\documentclass[11pt]{article}
\usepackage[margin=1in]{geometry}

\usepackage{setspace}
\usepackage{amsmath}
\usepackage{amsfonts}
\usepackage{enumerate}
\usepackage{float}
\usepackage{needspace}
\usepackage[utf8]{inputenc}
\usepackage{comment}
\usepackage{bm}
\usepackage{hyperref}
\usepackage{tocloft}
\usepackage{graphicx}
\usepackage{booktabs}
\usepackage{array}
\usepackage{multirow}
\usepackage{subcaption}
\usepackage{tikz}
\usetikzlibrary{calc}

\newlistof{appsec}{appsec}{Supplementary Appendix Contents}

\makeatletter
\newcommand{\listofappendixcontents}{
	\begingroup
	\let\float@listhead\@gobble
	\listof{appsec}{}
	\endgroup
}
\makeatother

\newcommand{\appsection}[1]{
	\section{#1}
	\addcontentsline{appsec}{section}{\protect\numberline{\thesection}#1}
}

\newcommand{\appsubsection}[1]{
	\subsection{#1}
	\addcontentsline{appsec}{subsection}{\protect\numberline{\thesubsection}#1}
}

\newcommand{\appsubsubsection}[1]{
	\subsubsection{#1}
	\addcontentsline{appsec}{subsubsection}{\protect\numberline{\thesubsubsection}#1}
}

\allowdisplaybreaks[4]

\usepackage[bottom]{footmisc}
\newtheorem{ass}{Assumption}

\newtheorem{prop}{Proposition}
\newtheorem{theorem}{Theorem}
\newtheorem{lem}{Lemma}

\renewcommand{\thesection}{\arabic{section}}
\renewcommand{\thesubsection}{\arabic{section}.\arabic{subsection}}
\renewcommand{\thesubsubsection}{\arabic{section}.\arabic{subsection}.\arabic{subsubsection}}

\newcolumntype{L}[1]{>{\raggedright\let\newline\\\arraybackslash\hspace{0pt}}m{#1}}
\newcolumntype{C}[1]{>{\centering\let\newline\\\arraybackslash\hspace{0pt}}m{#1}}

\usepackage{natbib}
\usepackage{bibunits}
\defaultbibliographystyle{aea}
\defaultbibliography{references}

\makeatletter
\AddToHook{env/bibunit/begin}{%
    \def\@extra@binfo{.\@bibunitname}%
    \let\@extra@b@citeb\@extra@binfo
}
\makeatother

\newcommand{\E}{\mathbb{E}}

\newcommand{\Var}{\mathbb{V}}
\newcommand{\Cov}{\mathbb{C}}

\newcommand{\Prob}{\mathbb{P}}
\newcommand{\supp}{\mathrm{supp}}

\makeatletter
\DeclareFontFamily{U}{mathx}{\hyphenchar\font45}
\DeclareFontShape{U}{mathx}{m}{n}{
	<-> mathx10
}{}
\DeclareSymbolFont{mathx}{U}{mathx}{m}{n}

\DeclareMathAccent{\widecheck}{0}{mathx}{"71}
\makeatother
\hypersetup{
    pdftitle={DynDisc},                 % title
    pdfauthor={Francesco Ruggieri},     % author
    pdfnewwindow=true,                  % links in new window
    colorlinks=true,                    % false: boxed links; true: colored links
    linkcolor=black,                    % color of internal links
    citecolor=blue,                     % color of links to bibliography
    filecolor=black,                    % color of file links
    urlcolor=blue                       % color of external links
}

\begin{document}
\begin{bibunit}

\title{Dynamic Regression Discontinuity: \\An Event-Study Approach\thanks{I am grateful to Alex Torgovitsky for his guidance and support. I also thank Scott Behmer, St\'ephane Bonhomme, Matias Cattaneo, Hazen Eckert, Nadav Kunievsky, Jonathan Roth, Noah Sobel-Lewin, and participants in the Econometrics Advising group at the University of Chicago for their feedback and comments.}}

\author{Francesco Ruggieri\thanks{University of Virginia, Department of Economics. Address: 248 McCormick Road, Charlottesville, VA 22903. Email: \href{mailto:ruggieri@virginia.edu}{ruggieri@virginia.edu}}}

\date{\today\thanks{An earlier version of this paper was circulated under the title ``Dynamic Regression Discontinuity: A Within-Design Approach'' and dated July 26, 2023. The present version supersedes and subsumes the contents of that draft.}}
\maketitle

\medskip

\medskip

\begin{abstract}
	I propose a novel argument to identify economically interpretable intertemporal treatment effects in dynamic regression discontinuity designs (RDDs). I develop a dynamic potential outcomes model and reformulate two assumptions from the difference-in-differences literature, no anticipation and common trends, to attain point identification of cutoff-specific impulse responses. Each time-specific parameter can be expressed as the sum of two static RDD contrasts, allowing for estimation with standard local polynomial methods and robust bias-corrected inference. I also propose a nonparametric approach to aggregate treatment effects across calendar time and treatment histories, and use a limited path independence restriction to reduce the number of history-specific estimands. I apply this method to estimate the dynamic effects of school district expenditure authorizations on housing prices in Wisconsin.
\end{abstract}

\medskip

\medskip

\begin{quote}
\small
\setstretch{1}
\end{quote}

\setcounter{page}{0}\thispagestyle{empty}
	
\newpage

\section{Introduction}\label{sec_intro}

Regression discontinuity designs (RDDs) are widely used to identify causal parameters when the probability of exposure to a treatment changes discontinuously at a known deterministic threshold. In a seminal contribution, \cite{cfr2010} extended the canonical design to multi-period settings in which treatment assignment is sequential and exhibits path dependence. Dynamic RDDs have proven popular in the empirical public finance literature that exploits local referenda to estimate the effect of increased government expenditure on various outcomes (\citealt{isen2014}; \citealt{hongzimmer2016}; \citealt{martorelletal2016}; \citealt{lang2018}; \citealt{langetal2018}; \citealt{abottetal2020}; \citealt{baron2022}; \citealt{rohlinetal2022}; \citealt{bilaschon2024}; \citealt{baronetal2024}).\footnote{Beyond applications to local referenda, dynamic RDDs have been used in settings in which treatment assignment is determined by federal financial aid eligibility rules (\citealt{darolia2013}), student test scores (\citealt{taylor2014}), shareholder vote shares (\citealt{cunatgineguadalupe2012}; \citealt{cunatgineguadalupe2016}; \citealt{ngwuzhaizhao2021}), and mayoral vote margins (\citealt{barbosaferreira2023}).} 

Despite their growing use in empirical research, dynamic RDDs remain methodologically understudied. This is particularly true in relation to the challenge posed by the identification of economically interpretable long-term causal parameters. Because units may be treated repeatedly, it is generally difficult to disentangle the effect of treatment received in a given period from the effects of treatments received in earlier or subsequent periods. In recent work, \cite{hsushen2024} address this challenge in a potential outcomes framework that allows for flexible patterns of path dependence and heterogeneity in treatment effects. To identify the intertemporal effects of treatment exposure in a given period, they impose a selection-on-observables assumption that exploits future realizations of the running variable and, potentially, additional covariates.

In this paper, I consider a dynamic model in which potential outcomes at any point in time depend on past, contemporaneous, and future treatment states (\citealt{robins1986}). I then propose an identification argument that does not require conditioning on observables and formalizes a connection between dynamic RDDs and event-study designs that has been recognized in recent empirical work (\citealt{bilaschon2024}). Specifically, I reformulate two assumptions from the difference-in-differences literature, no anticipation and common trends, and show that they are sufficient to point identify interpretable causal parameters. The resulting, time-specific estimands can be expressed as sums of two standard RD outcome contrasts and can be aggregated, for interpretability and exposition, using a set of convex weights. Finally, recognizing that the number of estimands grows exponentially with the number of time periods in the model, I reduce the dimensionality of the parameter space with a limited path independence assumption. This restriction takes advantage of the sparse and cyclical nature of treatment assignment that is typically true of empirical applications of dynamic RDDs. A more detailed comparison between the approach developed in this paper and that of \cite{hsushen2024} is provided in Section~\ref{sec_comparison_hsushen}.

For nonparametric estimation and statistical inference, I use local polynomial methods that are widely applied in empirical research. Specifically, I estimate time-specific parameters using local linear regression and aggregate them with convex sample weights (as in \citealt{callawaysantanna2021}). In the process, I incorporate standard bias correction techniques to address the distortion that arises when estimating cutoff-specific parameters with observations away from the cutoff. To quantify statistical uncertainty, I follow the method developed by \cite{cct2014} and construct nonparametric confidence intervals that account for the sampling variability of the estimated bias. Additionally, I derive bandwidth formulas that minimize the estimator's mean squared error and apply the rule-of-thumb adjustment proposed by \cite{ccf2020} to improve the coverage accuracy of confidence intervals. The accompanying package \texttt{dynrd} implements the estimation and inference procedures developed in this paper.\footnote{The package is available at \url{https://github.com/francescoruggieri/dynrd}.}

I implement the proposed method in the most common empirical application of dynamic RDDs: referenda through which school districts and other local jurisdictions authorize additional expenditures and bypass state-imposed fiscal policy constraints. I focus on Wisconsin, where such referenda are routinely held to finance both operational and capital spending (\citealt{baron2022}). The outcome of interest is housing prices in the years following referendum approval, a margin that has long been studied in public finance as a measure of how potential homebuyers value the tradeoff between improved public services and higher property taxes (see \citealt{rossyinger1999} for a review of the earlier literature).

The approach developed in this paper is similar in spirit to solutions to extrapolate effects away from the threshold in static designs using only features of the joint distribution of the outcome, the running variable, and one or more cutoffs (\citealt{donglewbel2015}, \citealt{bertanhaimbens2020}, \citealt{cattaneoetal2021}). The proposed approach offers several advantages. First, it is applicable to settings in which additional covariates may not be available or fail to satisfy prerequisites for validity, such as observed characteristics affected by earlier treatment. This concern may be especially salient in dynamic settings in which units are repeatedly subject to treatment assignment. Second, the plausibility of the main identifying assumption, local common trends, can be assessed by comparing pre-treatment outcome changes across groups defined by subsequent treatment paths at the cutoff. Third, the method is easy to implement because each target parameter is the sum of two RD outcome contrasts, which can be estimated using standard local polynomial methods.

This paper contributes to three related strands of the econometric literature. First, it proposes a novel approach to point identify intertemporal treatment effects in dynamic RDDs (\citealt{cfr2010}; \citealt{hsushen2024}). In this respect, it contributes more broadly to the literature on regression discontinuity designs (see \citealt{cattaneotitiunik2022} for a recent review). With regard to identification, the local common trends assumption has a conceptual link with the constant bias restriction that \cite{cattaneoetal2021} use for extrapolating average treatment effects in RDDs with multiple cutoffs. The local common trends assumption also relates to the difference-in-discontinuities design of \cite{grembinannicinitroiano2016}, whose identification argument requires the outcome discontinuity absent the policy of interest to be constant over time. In addition, the convex aggregation of heterogeneous effects is similar in spirit to the analysis by \cite{cattaneoetal2016}, who characterize pooled RD estimands as weighted averages of cutoff-specific treatment effects. With regard to statistical inference, this paper builds on the robust bias correction approach developed by \cite{cct2014}. In particular, the proposed estimator, like the fuzzy RD estimator, is a nonlinear function of several local polynomial estimators at the cutoff. This shared nonlinear structure motivates both the first-order linearization used for bias-corrected inference and the bandwidth adjustment recommended by \cite{ccf2020} for fuzzy RD estimators.

Second, this paper adds to the broad literature on the identification of economically interpretable dynamic effects under treatment effect heterogeneity (\citealt{dcdh2020}, \citealt{goodmanbacon2021}, \citealt{callawaysantanna2021}, \citealt{sunabraham2021}, \citealt{dcdh_intertemp2024}, \citealt{callawaybaconsantanna2024}, \citealt{dcdhvazquez2024}) and path dependence in treatment assignment (\citealt{heckmanhumphriesveramendi2016}). Third, it contributes to the growing literature on the causal interpretation of standard estimands in the presence of multiple treatments (\citealt{dcdh_multi2023}, \citealt{pgphullkolesar2022}) or multiple instrumental variables (\citealt{mogstadtorgowalters2021}, \citealt{magnetorgohole2025}) when treatment effects are allowed to be stochastic.

The remainder of the paper is organized as follows. In Section \ref{sec:background}, I introduce the main identification challenge and discuss previous approaches to address it. In Section \ref{sec_twoperiod}, I develop a simple two-period regression discontinuity design and illustrate the core of my identification strategy. In Section \ref{sec_multiperiod}, I generalize the model to a setting with multiple time periods. In Section \ref{sec_aggreg}, I discuss the aggregation of time-specific estimands and introduce a limited path independence restriction to reduce the number of parameters. I also describe how to assess the plausibility of local common trends using comparisons of pre-treatment outcome changes. In Section \ref{sec_estim}, I illustrate how to use standard local polynomial methods to estimate the target parameters and construct nonparametric confidence intervals around them. In Section \ref{sec:application}, I apply my method to estimate the average effects of school district expenditure authorizations on housing prices in Wisconsin. Section \ref{sec_concl} concludes.

\section{Background}\label{sec:background}

In sharp regression discontinuity designs, a treatment is assigned based on whether an observed running variable exceeds a known deterministic threshold. For example, \cite{lee2008} exploits U.S. House elections to estimate the average effect of a party's candidate narrowly winning an election on the party's subsequent victory in future elections for the same seat. Dynamic RDDs extend this framework by distinguishing between current and future treatment states, which depend on whether the running variable crosses the cutoff in each period.

In the context of \cite{lee2008}, the outcome of a House election in year $t$ determines political representation in both $t$ and $t+1$, allowing for the estimation of short-term treatment effects.\footnote{This example assumes that House elections occur at the beginning of a calendar year and that outcomes are observed at the end of it.} Estimating the average effect of winning an election in year $t$ on outcomes in years $t+2$ and beyond introduces an additional challenge. Because House elections occur every two years, a party that wins a seat in year $t$ may lose it in year $t+2$, and a party that loses in year $t$ may regain control. This dynamic turnover complicates the identification of interpretable long-term causal parameters. The key challenge lies in disentangling the long-term effects of the year-$t$ election outcome from the short-term effects of the year-$t+2$ election outcome. More broadly, the recurrence of elections introduces dynamic fuzziness, as future treatment states may not align with initial treatment assignments. Figure \ref{fig:dynfuzz} provides a stylized representation of the main identification challenge in dynamic RDDs.

\begin{figure}[!t]
	\begin{center}
		\caption{Stylized Representation of the Identification Challenge in Dynamic RDDs}\label{fig:dynfuzz}
		\vspace{1mm}
			\resizebox{0.95\textwidth}{!}{
			\begin{tikzpicture}[inner sep=2pt]
				% Bottom row: Outcomes at t, t+1, t+2, t+3
				\node (O1) at (0,0) {Outcome at \(t\)};
				\node (O2) at (4,0) {Outcome at \(t+1\)};
				\node (O3) at (8,0) {Outcome at \(t+2\)};
				\node (O4) at (12,0) {Outcome at \(t+3\)};
				
				% Left block (above O1 and O2): Election and Treatment at t
				\node (E1) at ($(O1)!0.5!(O2) + (0,3)$) {Election at \(t\)};
				\node (T1) at ($(O1)!0.5!(O2) + (0,2)$) {Treatment at \(t\)};
				
				% Right block (above O3 and O4): Election and Treatment at t+2
				\node (E2) at ($(O3)!0.5!(O4) + (0,3)$) {Election at \(t+2\)};
				\node (T2) at ($(O3)!0.5!(O4) + (0,2)$) {Treatment at \(t+2\)};
				
				% Arrows from elections to treatments
				\draw[->] (E1) -- (T1);
				\draw[->] (E2) -- (T2);
				
				% Arrows from Treatment at t to Outcomes
				\draw[->] (T1) -- (O1);
				\draw[->] (T1) -- (O2);
				\draw[->, dashed, bend left=15] (T1) to (O3);
				\draw[->, dashed, bend left=15] (T1) to (O4);
				
				% Arrows from Treatment at t+2 to Outcomes
				\draw[->] (T2) -- (O3);
				\draw[->] (T2) -- (O4);
			\end{tikzpicture}
			}
	\end{center}
\vspace{-1mm}
\begin{footnotesize}
	\begin{spacing}{1}
		\noindent
		\textsc{Notes}: This figure offers a stylized depiction of the identification challenge in dynamic regression discontinuity designs, drawing on the U.S. House election setting from \cite{lee2008}. Outcomes in years $t+2$ and $t+3$ may reflect both elections, whether party control changes or the same party wins twice. The identification challenge lies in disentangling the effect of the initial victory while holding subsequent party control fixed.
	\end{spacing}
\end{footnotesize}
\end{figure}

Since the seminal contribution by \cite{cfr2010}, dynamic regression discontinuity designs have been widely applied in local public finance. Numerous empirical studies leverage school district referenda on property tax hikes to estimate the effects of increased school expenditures on outcomes such as housing prices and student test scores. Similar to the stylized example above, dynamic fuzziness arises because school districts hold repeated referenda to authorize additional spending. A ballot measure rejected in year $t$ may be resubmitted and approved in year $t+1$. Comparing later outcomes between districts that narrowly approved or rejected the year-$t$ referendum may therefore capture the effects of spending authorized in both years.

Most empirical studies employing a dynamic regression discontinuity design adopt the ``one-step'' approach proposed by \cite{cfr2010}. This method relies on the following dynamic two-way fixed effects specification:
\begin{align}\label{eq:onestep}
	Y_{it} = \alpha_i + \beta_t + \sum_{s=0}^{\overline{s}} \left ( \theta_s D_{i,t-s} + \gamma_s Q_{i,t-s} + p_g \left ( \delta_s, R_{i,t-s} \right ) \right ) + U_{it}
\end{align}
where $D$ is a binary treatment indicator, $Q$ is an indicator for whether a referendum is held, and $p_g \left ( \cdot \right )$ represents a $g$th-degree polynomial of the approval vote share $R$. This specification is typically estimated without bandwidth restrictions, allowing researchers to flexibly control for both current and past treatment states, as well as for current and past realizations of the running variable (\citealt{darolia2013}; \citealt{hongzimmer2016}; \citealt{martorelletal2016}; \citealt{baronetal2024}).

More recently, empirical researchers have proposed modifications to the ``one-step'' approach introduced by \cite{cfr2010}. \cite{baron2022} extends the estimating equation in \eqref{eq:onestep} by including lead treatment indicators, allowing $s$ to take negative values. Similarly, \cite{bilaschon2024} incorporate leads but estimate the resulting linear regression in a ``stacked'' sample, where each component corresponds to a cohort-specific subsample. Here, a cohort is defined as a calendar year in which referenda occur.\footnote{\cite{bilaschon2024} restrict each cohort to include only treated jurisdictions and untreated jurisdictions that held a referendum in the same year but did not approve any ballot measure in the subsequent ten years.} Their stacked specification pools treatment effect coefficients across cohorts while allowing fixed effects to vary by cohort. Departing from \cite{cfr2010}, \cite{hsushen2024} develop a dynamic potential outcomes model that accounts for heterogeneity in treatment effects and path dependence in treatment assignments. To identify interpretable long-term effects, the authors impose a selection-on-observables assumption that exploits future approval vote shares and, possibly, additional covariates. More details on their framework are provided in the following section.

\section{A Two-Period Regression Discontinuity Design}\label{sec_twoperiod}

In this section, I present the core of my identification strategy within a simple two-period model. Sections \ref{sec_multiperiod} and \ref{sec_aggreg} extend this framework to accommodate an arbitrary number of time periods.

\subsection{Model Setup}\label{sec_model}

Consider a dynamic setting in which time periods are indexed by $t \in \left \{ 1, 2 \right \}$. Let $D_t \in \left \{ 0,1 \right \}$ denote a binary treatment that is assigned at the beginning of period $t$. The treatment is determined by whether the realization of an observed running variable $R_t \in \mathbb{R}$ exceeds a known deterministic constant $c \in \mathbb{R}$: $D_t \equiv \mathbb{I} \left [ R_t \geq c \right ]$. Let $Y_t \in \mathbb{R}$ be an outcome that is observed at the end of period $t$.

In the canonical static setting, potential outcomes depend on a single treatment state, and the observed outcome satisfies the switching equation $Y_t = \sum_{d \in \left \{ 0,1 \right \}} \mathbb{I} \left [ D_t = d \right ] Y_t \left ( d \right )$. In a two-period model with repeated treatment assignment, the potential outcome in period 1 may reflect the treatment state in period 2 due to anticipation and, symmetrically, the potential outcome in period 2 may be affected by the treatment state in period 1 due to path dependence. Thus, following \cite{robins1986}, it is natural to model potential outcomes as depending on the path of past, contemporaneous, and future treatment states at any point in time. Specifically, let the period-$t$ potential outcome be $Y_t \left ( d_1, d_2 \right )$, where $d_1, d_2 \in \left \{ 0,1 \right \}$. Without further assumptions, the evolution of treatment states is unrestricted. In fact, a unit may be repeatedly assigned the treatment, never receive it, or be exposed to it in only one period.

As in \cite{hsushen2024}, I allow for path dependence in treatment assignment by modeling potential treatments as functions of the history of treatment states. Specifically, let $D_1$ and $D_2 \left ( d_1 \right )$ denote the first- and second-period potential treatments, respectively.\footnote{A more flexible setting could allow for anticipation in treatment assignment by letting the potential treatment in period 1 depend on the treatment state in period 2, i.e., $D_1 \left ( d_2 \right )$. For simplicity, I abstract from this scenario and allow for anticipation only in potential outcomes.} In the canonical application of dynamic RDDs to local referenda, it is useful to interpret potential treatments as characterizing jurisdiction ``types''. For example, a school district in which voters repeatedly reject a bond for school construction ($D_1 = 0$ and $D_2 \left ( 0 \right ) = D_2 \left ( 1 \right ) = 0$) is likely more fiscally conservative than one in which increased public expenditure is approved at the first round of elections ($D_1 = 1$) or at the second round ($D_1 = 0$ and $D_2 \left ( 0 \right ) = 1$).

In this two-period model, the observed outcome in period $t$ is related to potential outcomes as
\begin{align}\label{potout}
	Y_t = \sum_{\left ( d_1, d_2 \right ) \in \left \{ 0,1 \right \}^2} \mathbb{I} \left [ D_1 = d_1, D_2 \left ( d_1 \right ) = d_2 \right ] Y_t \left ( d_1, d_2 \right )
\end{align}

\subsection{Target Parameters}

Having outlined features of the dynamic potential outcomes model, it is natural to define target parameters. As in static RDDs, all causal parameters of interest are local to the nonstochastic threshold above which the treatment is assigned (\citealt{hahnetal2001}). Without loss, the analysis will focus on target parameters implied by the first-period discontinuity. Letting $\tau \in \left \{ 0,1 \right \}$ denote the number of leads, consider a class of threshold-specific Average Treatment Effects (ATEs),
\begin{align}
	\mathrm{ATE}_{1,\tau} \left ( c, d_2, d_2' \right ) \equiv \E \left [ Y_{1+\tau} \left ( 1,d_2 \right ) - Y_{1+\tau} \left ( 0,d_2' \right ) | R_1 = c \right ]
\end{align}
where $d_2, d_2' \in \left \{ 0,1 \right \}$. Each of these average effects is constructed as the average contrast between a potential outcome that is treated in the first period and a potential outcome that is not. The choice of a second-period treatment state naturally gives rise to causal parameters with a different interpretation.

Since \cite{cfr2010}, the empirical public finance literature that studies the effects of increased public expenditure authorized by local referenda has focused on a class of Average Direct Treatment Effects that sets $d_2 = d_2' = 0$,
\begin{align}
	\mathrm{ADTE}_{1,\tau} \left ( c \right ) \equiv \E \left [ Y_{1+\tau} \left ( 1,0 \right ) - Y_{1+\tau} \left ( 0,0 \right ) | R_1 = c \right ]
\end{align}
Hereafter, I will refer to $\mathrm{ADTE}_{1,0} \left ( c \right )$ and $\mathrm{ADTE}_{1,1} \left ( c \right )$ as the instantaneous and cumulative Average Direct Treatment Effects, respectively. These target parameters can be interpreted as tracing out the impulse response of the outcome to the first-period discontinuity. Economically, they correspond to the thought experiment of measuring the cutoff-specific average effect of first-period treatment assignment in the counterfactual scenario in which exposure to the treatment is set to zero thereafter. For notational simplicity, I henceforth omit the dependence on $c$ and write $\mathrm{ADTE}_{1,\tau}$ in place of $\mathrm{ADTE}_{1,\tau} \left ( c \right )$.

\subsection{What Do Standard RD Estimands Identify?}

As discussed in \cite{hsushen2024}, standard RD estimands do not, in general, identify ADTEs. This subsection revisits this result within the potential outcomes framework of Section \ref{sec_model}, which allows outcomes to depend on future treatment assignments. First, the (sharp) RD estimand implied by the first-period discontinuity is defined as
\begin{align}
	\mathrm{RD}_{Y_t} \left ( c \right ) \equiv \lim_{r \downarrow c} \E \left [ Y_{t} | R_1 = r \right ] - \lim_{r \uparrow c} \E \left [ Y_{t} | R_1 = r \right ]
\end{align}
\cite{hahnetal2001} shows that continuity in average potential outcomes at the cutoff is sufficient to point identify the Average Treatment Effect at the threshold. It is natural to generalize this assumption to a dynamic setting.
\begin{ass}[Continuity at the Cutoff]\label{ass_cont0} For any $t \in \left \{ 1,2 \right \}$ and $d_1, d_2 \in \left \{ 0,1 \right \}$,
	\begin{align*}
		\E \left [ D_{2} \left ( d_1 \right ) | R_1 = r \right ] \quad \text{ and } \quad \E \left [ Y_{t} \left ( d_1, d_2 \right )  | R_1 = r, D_{2} \left ( d_1 \right ) \right ]
	\end{align*}
	are continuous at $r=c$.
\end{ass}
Assumption \ref{ass_cont0} restricts dynamic sorting around the threshold. First, it requires the probability of any jurisdiction ``type'', as defined by the potential treatment $D_2 \left ( d_1 \right )$, not to vary abruptly at the first-period cutoff. Second, within each jurisdiction ``type'', it rules out discontinuities in average potential outcomes arising from systematic differences in observed and unobserved characteristics on either side of the first-period cutoff. Under this assumption, the following identification argument holds.

\begin{prop}[Standard RD Estimand]\label{prop_rd}
	Suppose that Assumption \ref{ass_cont0} holds. Then
	\begin{align*}
		\mathrm{RD}_{Y_t} \left ( c \right ) & = \sum_{d_2 \in \left \{ 0,1 \right \}} \E \left [ Y_t \left ( 1,d_2 \right )  | R_1 = c, D_2 \left ( 1 \right ) = d_2 \right ] \times \Prob \left ( D_2 \left ( 1 \right ) = d_2 | R_1 = c \right ) \\
		& - \sum_{d_2 \in \left \{ 0,1 \right \}} \E \left [ Y_t \left ( 0,d_2 \right )  | R_1 = c, D_2 \left ( 0 \right ) = d_2 \right ] \times \Prob \left ( D_2 \left ( 0 \right ) = d_2 | R_1 = c \right )
	\end{align*}
\end{prop}
Without further assumptions, the sharp regression discontinuity estimand identifies the contrast between two convex combinations of conditional average potential outcomes over second-period treatment states. As discussed in \cite{cfr2010} and similarly in \cite{heckmanhumphriesveramendi2016}, this target parameter may be interpreted as the ``intent-to-treat'' implied by the first-period discontinuity. Because treatment assignment exhibits path dependence and outcomes reflect past and future treatment states, the intent-to-treat is generally hard to interpret. In the next two sections, I specify additional sufficient restrictions to identify the instantaneous and cumulative Average Direct Treatment Effects implied by the first-period discontinuity. These parameters trace out the impulse response of an outcome to the initial round of treatment assignment.

\subsection{Identification of the Instantaneous ADTE}\label{sec_ident_adte0}

In the potential outcomes framework of Section \ref{sec_model}, first-period outcomes may depend on the second-period treatment state. Therefore, as Proposition \ref{prop_rd} shows, Assumption \ref{ass_cont0} alone is insufficient to identify $\E \left [ Y_1 \left ( d_1, 0 \right ) | R_1 = c, D_2 \left ( d_1 \right ) = 1 \right ]$ for $d_1 \in \left \{ 0,1 \right \}$. This counterfactual expectation is the average first-period outcome under the $\left ( d_1,0 \right )$ path among units who would be treated in period 2 if their period-1 treatment state were $d_1$.

To identify this counterfactual expectation, it is sufficient to impose that, within strata defined by potential second-period treatment states, the average first-period potential outcome be invariant to the second-period treatment assignment.
\begin{ass}[Local No Anticipation] \label{ass_noant} For any $d_1, d_2 \in \left \{ 0,1 \right \}$,
	\begin{align*}
		\E \left [ Y_{1} \left ( d_1, 0 \right ) | R_1 = c, D_{2} \left ( d_1 \right ) = d_2 \right ] = \E \left [ Y_{1} \left ( d_1, 1 \right ) | R_1 = c, D_{2} \left ( d_1 \right ) = d_2 \right ]
	\end{align*}
\end{ass}
Assumption \ref{ass_noant} is a cutoff-specific version of the canonical no anticipation assumption required to point identify the Average Treatment Effect on the Treated in difference-in-differences designs (\citealt{abbringvanderberg2003}). Although this restriction may appear to limit the applicability of the argument to forward-looking outcome variables, it should be noted that the lack of anticipatory behavior need hold only at the first-period cutoff. In the application to the effect of increased school spending on housing prices, Assumption \ref{ass_noant} does not preclude home values from reflecting the prospect of a subsequent referendum. Instead, it requires that, within each potential second-period treatment stratum at the cutoff, average home values following the initial referendum be invariant to whether a subsequent bond proposition is ultimately approved.

Under Assumption \ref{ass_noant}, the first-period conditional mean potential outcomes in Proposition \ref{prop_rd} can be evaluated at $d_2=0$, yielding the following identification result.

\begin{prop}[Identification of the Instantaneous ADTE]\label{prop_apte0}
	\sloppy Suppose that Assumptions \ref{ass_cont0} and \ref{ass_noant} hold. Then
	\begin{align*}
		\textsc{ADTE}_{1,0} = \lim_{r \downarrow c} \E \left [ Y_1 | R_1 = r \right ] - \lim_{r \uparrow c} \E \left [ Y_1 | R_1 = r \right ] \equiv \mathrm{RD}_{Y_1} \left ( c \right )
	\end{align*}
\end{prop}
The result follows directly from Proposition \ref{prop_rd}. Under no anticipation, the first-period conditional mean potential outcomes are invariant to the second-period treatment state. Thus, the standard static RD estimand identifies the instantaneous ADTE.

\subsection{Identification of the Cumulative ADTE}\label{sec_ident_adte1}

\sloppy Consider, instead, the goal of identifying the cumulative Average Direct Treatment Effect, $\E \left [ Y_2 \left ( 1,0 \right ) | R_1 = c \right ] - \E \left [ Y_2 \left ( 0,0 \right ) | R_1 = c \right ]$. By an application of the Law of Iterated Expectations, the first term can be expressed as
\begin{align}\label{apte11}
	\E \left [ Y_2 \left ( 1,0 \right ) | R_1 = c \right ] & = \overbrace{\E \left [ Y_2 \left ( 1,0 \right ) | R_1 = c, D_2 \left ( 1 \right ) = 0 \right ]}^{\text{identified}} \times \overbrace{\Prob \left ( D_2 \left ( 1 \right ) = 0 | R_1 = c \right )}^{\text{identified}} \nonumber \\
	& + \underbrace{\E \left [ Y_2 \left ( 1,0 \right ) | R_1 = c, D_2 \left ( 1 \right ) = 1 \right ]}_{\text{counterfactual}} \times \underbrace{\Prob \left ( D_2 \left ( 1 \right ) = 1 | R_1 = c \right )}_{\text{identified}}
\end{align}
Notice that both probabilities are identified under Assumption \ref{ass_cont0}. In fact, for $d_2 \in \left \{ 0,1 \right \}$,
\begin{align}
	\Prob \left ( D_2 \left ( 1 \right ) = d_2 | R_1 = c \right ) = \lim_{r \downarrow c} \Prob \left ( D_{2} = d_2 | R_1 = r \right )
\end{align}
Similarly, the first conditional expectation in equation \eqref{apte11} is identified as
\begin{align}
	\E \left [ Y_2 \left ( 1,0 \right ) | R_1 = c, D_2 \left ( 1 \right ) = 0 \right ] = \lim_{r \downarrow c} \E \left [ Y_2 | R_1 = r, D_2 = 0 \right ]
\end{align}
On the other hand, $\E \left [ Y_2 \left ( 1,0 \right ) | R_1 = c, D_2 \left ( 1 \right ) = 1 \right ]$ is counterfactual because $Y_2 \left ( 1,0 \right )$ is unobserved among units who would sort into the treated arm in $t=2$ after being treated in $t=1$. Differently from the previous section, Assumptions \ref{ass_cont0} and \ref{ass_noant} alone is not sufficient to identify the counterfactual mean because the outcome is measured in $t=2$. However, point identification of the target parameter may be achieved by also assuming that the expected growth in untreated-at-2 potential outcomes is independent of the potential second-period treatment state at the cutoff.
\begin{ass}[Local Common Trends] \label{ass_ct0} For $d_1 \in \left \{ 0,1 \right \}$,
	\begin{align*}
		\E \left [ \Delta_{1,2} \left ( d_1, 0 \right ) | R_1 = c, D_{2} \left ( d_1 \right ) = 0 \right ] = \E \left [ \Delta_{1,2} \left ( d_1, 0 \right ) | R_1 = c, D_{2} \left ( d_1 \right ) = 1 \right ]
	\end{align*}
	where $\Delta_{1,2} \left ( d_1, 0 \right ) \equiv Y_{2} \left ( d_1,0 \right ) - Y_1 \left ( d_1, 0 \right )$.
\end{ass}
Assumption \ref{ass_ct0} is a cutoff-specific version of the common trends assumption required to point identify the Average Treatment Effect on the Treated in difference-in-differences designs (\citealt{abadie2005}). As in that setting, groups (``treatment'' and ``control'') are implied by paths of potential treatment states. Differently from that setting, the treatment is available also in the first period, implying that there are four, not two, groups or ``types''. As a consequence, the parallel trends restriction also applies to units who are potentially treated in the first period. In the classic application of dynamic RDDs to local public finance, Assumption \ref{ass_ct0} entails that, at the cutoff implied by the first-period referendum, the expected growth in potential test scores absent a second referendum approval is independent of a jurisdiction's ``type'', i.e., it is independent of whether a jurisdiction would approve or reject a school bond referendum in the second period. A more detailed discussion on the economic content of this restriction is presented in Section \ref{sec_localct}.

Under Assumptions \ref{ass_cont0}, \ref{ass_noant}, and \ref{ass_ct0}, the counterfactual conditional mean is identified as
\begin{align}
	\E \left [ Y_{2} \left ( 1,0 \right ) | R_1 = c, D_{2} \left ( 1 \right )=1 \right ] & = \lim_{r \downarrow c} \E \left [ Y_1 | R_1 = r, D_{2}=1 \right ] \notag \\
	& + \lim_{r \downarrow c} \E \left [ Y_{2} - Y_1 | R_1 = r, D_{2}=0 \right ]
\end{align}
As in the canonical difference-in-differences design, the counterfactual term is identified as the sum of a ``level'' and a ``growth'' term. In this case, the former measures the average observed first-period outcome among units who later sort into the treated arm. The latter captures the average observed outcome growth among units who sort into the untreated arm in the second period. Combining results yields the following proposition.

\begin{prop}[Identification of the Cumulative ADTE]\label{prop_apte1}
	Suppose that Assumptions \ref{ass_cont0}, \ref{ass_noant}, and \ref{ass_ct0} hold. Further assume that $\Prob \left ( D_2 \left ( d_1 \right ) = 0 | R_1 = c \right ) > 0$ for $d_1 \in \left \{ 0,1 \right \}$. Then
	\begin{align*}
		\textsc{ADTE}_{1,1} & = \lim_{r \downarrow c} \E \left [ Y_1 | R_1 = r \right ] - \lim_{r \uparrow c} \E \left [ Y_1 | R_1 = r \right ] \\
		& + \lim_{r \downarrow c} \E \left [ Y_2 - Y_1 | R_1 = r, D_2 =0 \right ] - \lim_{r \uparrow c} \E \left [ Y_2 - Y_1 | R_1 = r, D_2 =0 \right ]
	\end{align*}
\end{prop}
This proposition entails that an interpretable long-term causal parameter determined by the first-period discontinuity is identified by the sum of two standard sharp regression discontinuity estimands, namely the RD estimand that uses $Y_1$ as an outcome and the RD estimand that uses $Y_2 - Y_1$ as an outcome in the subpopulation implied by $D_2 = 0$.

In difference-in-differences designs, the common trends assumption is sometimes deemed more plausible conditional on predetermined covariates (\citealt{bakeretal2026}). Assumption \ref{ass_ct0} could likewise be imposed conditional on these covariates. Under corresponding conditional continuity and no anticipation restrictions, as well as overlap conditions, the identification argument proposed in this section would yield conditional ADTEs. The unconditional target parameter would then be obtained by averaging these effects over the distribution of covariates at the cutoff. Estimation could build on outcome regression (\citealt{heckmanichimuratodd1997}), inverse probability weighting (\citealt{abadie2005}), or doubly robust methods (\citealt{santannazhao2020}). A formal treatment of this extension is left for future research.

\subsection{Relation to \cite{hsushen2024}}\label{sec_comparison_hsushen}

The fundamental difference between this paper and \cite{hsushen2024} lies in the assumption used to identify the counterfactual expectation $\E \left [ Y_2 \left ( 1,0 \right ) | R_1 = c, D_2 \left ( 1 \right ) = 1 \right ]$. \cite{hsushen2024} introduces a potential indicator $Q_2 \left ( d_1 \right )$, which denotes whether a round of treatment assignment would occur in the second period if the first-period treatment state were $d_1 \in \left \{ 0,1 \right \}$. When $Q_2 \left ( d_1 \right ) = 1$, a corresponding potential second-period running variable $R_2 \left ( d_1 \right )$ is defined. Their key identifying assumption states that, if a second-period treatment assignment occurs, the potential outcome $Y_2 \left ( 1,0 \right )$ is mean independent of $R_2 \left ( d_1 \right )$ in a neighborhood of the first-period cutoff. This condition is further generalized to incorporate covariates $X$, such that, for any $r \in \supp \left ( R_2 \right )$,
\begin{align}
	& \E \left [ Y_2 \left ( d_1 ,0 \right ) | R_1 = c, Q_2 \left ( d_1 \right ) = 1, X = x, R_2 \left ( d_1 \right ) = r \right ] \notag \\
	& = \E \left [ Y_2 \left ( d_1 ,0 \right ) | R_1 = c, Q_2 \left ( d_1 \right ) = 1, X = x \right ]
\end{align}
This assumption has the appealing feature of restricting comparisons to units for which treatment assignment occurs in the second period as well. However, it may be vulnerable to violations if unobserved determinants of the second-period outcome also influence the second-period vote share. For instance, in the canonical application examining the effect of school expenditure referenda on student test scores, unobserved parental valuation of education may drive both support for a second-round referendum and student performance—e.g., because more involved parents are more likely to vote in favor of school funding (\citealt{poterba1997}) and also assist their children academically (\citealt{guryanhurstkearney2008}). Notably, such a violation can occur even if the unobservable is time-invariant and the first-round approval margin is idiosyncratically close to the cutoff.

Another distinction between the identification strategy developed in this paper and that in \cite{hsushen2024} pertains to the role of anticipation in the potential outcomes framework. The dynamic model presented here allows potential outcomes at any given time to depend on past, contemporaneous, and future treatment states. In a setting with two periods, this implies that potential outcomes in the first period are denoted as $Y_1 \left ( d_1, d_2 \right )$, while those in the second period are given by $Y_2 \left ( d_1, d_2 \right )$. By contrast, the setup considered by \cite{hsushen2024} restricts the sequence of treatment states to end in the period in which the outcome is realized, with first-period outcomes modeled as $Y_1 \left ( d_1 \right )$ and second-period outcomes as $Y_2 \left ( d_1, d_2 \right )$. In empirical applications where the outcome of interest may respond to anticipated future treatments---such as settings involving housing prices or other forward-looking variables commonly analyzed in public finance (e.g., \citealt{cfr2010} and \citealt{bilaschon2024})---it may be useful to explicitly state the role of anticipation in the underlying model.

Finally, the approach developed in this paper is related in spirit to event-study-type specifications commonly employed by empirical researchers in dynamic RDD applications. The inclusion of leads in the estimating equation \eqref{eq:onestep} by \cite{abottetal2020} and \cite{baron2022}, together with the cohort-by-cohort identification strategy based on ``clean controls'' proposed by \cite{bilaschon2024}, suggests that a conceptual link between dynamic RDDs and difference-in-differences designs has already been implicitly recognized in the literature. This paper formalizes that connection.

\subsection{Relation to \cite{cfr2010}}\label{sec_comparison_cfr}

As discussed in Section \ref{sec:background}, most empirical applications of dynamic regression discontinuity designs have relied on the ``one-step'' approach proposed by \cite{cfr2010}. In the same article, the authors also propose a recursive approach to identify impulse-response-like parameters implied by treatment assignment at a focal discontinuity. Although this approach has been less widely adopted than its ``one-step'' counterpart (\citealt{isen2014}; \citealt{taylor2014}; \citealt{enamietal2021}; \citealt{rohlinetal2022}), it is useful to reconcile its identifying assumptions with those proposed in this paper. Building on \cite{hsushen2024}, this section reformulates the recursive identification argument within the more general dynamic potential outcomes framework outlined in Section \ref{sec_model}, which allows potential outcomes to depend on future treatment assignments.

Equation (4) of \cite{cfr2010} assumes that treatment effects depend exclusively on the time elapsed since treatment assignment, thereby ruling out heterogeneity across units and dependence on previous treatment assignments. In the two-period model considered here, these restrictions imply $Y_1 \left ( 1,0 \right ) - Y_1 \left ( 0,0 \right ) = Y_2 \left ( 0,1 \right ) - Y_2 \left ( 0,0 \right ) = Y_2 \left ( 1,1 \right ) - Y_2 \left ( 1,0 \right ) = k_0$ and $Y_2 \left ( 1,0 \right ) - Y_2 \left ( 0,0 \right ) = k_1$ almost surely, where $k_0$ and $k_1$ are deterministic constants. The specification also rules out anticipatory effects, as outcomes depend only on current and past rounds of treatment assignment.\footnote{\cite{cfr2010} relax this restriction in Section IV.E by allowing housing prices to respond to expectations of future treatment assignments.} In the potential outcomes model of this paper, this restriction implies $Y_1 \left ( d_1, 0 \right ) = Y_1 \left ( d_1, 1 \right )$ almost surely for $d_1 \in \left \{ 0,1 \right \}$.

Under Assumption \ref{ass_cont0}, these restrictions yield the following recursive identification result:
\begin{equation}\label{eq_recursive}
	\mathrm{ADTE}_{1,0} = \text{RD}_{Y_1} \qquad \mathrm{ADTE}_{1,1} = \text{RD}_{Y_2} - \text{RD}_{Y_1} \text{RD}_{D_2}
\end{equation}
In Lemma 2.1, \cite{hsushen2024} show that the recursive identification argument remains valid under heterogeneous treatment effects, subject to assumptions on their expectations and on selection into subsequent treatment. Their result can be restated as follows.

\begin{lem}[Recursive Identification of ADTEs]\label{lemma_recursive}
	\sloppy Suppose that Assumptions \ref{ass_cont0} and \ref{ass_noant} hold. Further assume that average instantaneous treatment effects are homogeneous across periods and treatment histories,
	\begin{align*}
		& \E \left [ Y_1 \left ( 1,0 \right ) - Y_1 \left ( 0,0 \right ) | R_1 = c \right ] \\
		& = \E \left [ Y_2 \left ( 0,1 \right ) - Y_2 \left ( 0,0 \right ) | R_1 = c \right ] = \E \left [ Y_2 \left ( 1,1 \right ) - Y_2 \left ( 1,0 \right ) | R_1 = c \right ],
	\end{align*}
	and that the instantaneous effects of second-period treatment are mean independent of second-period treatment assignment,
	\begin{equation*}
		\E \left [ Y_2 \left ( d_1,1 \right ) - Y_2 \left ( d_1,0 \right ) | R_1 = c, D_2 \left ( d_1 \right ) \right ] = \E \left [ Y_2 \left ( d_1,1 \right ) - Y_2 \left ( d_1,0 \right ) | R_1 = c \right ]
	\end{equation*}
	for $d_1 \in \left \{ 0,1 \right \}$. Then $\mathrm{ADTE}_{1,0}$ and $\mathrm{ADTE}_{1,1}$ are identified as in equation \eqref{eq_recursive}.
\end{lem}
The proof of this lemma is presented in Supplementary Appendix \ref{app_lemma_recursive}.

The identification argument illustrated in Sections \ref{sec_ident_adte0} and \ref{sec_ident_adte1} differs from the recursive approach of \cite{cfr2010} in two respects. First, it does not require average instantaneous treatment effects to be equal across periods and treatment histories. Second, it does not impose mean independence between the instantaneous effects of second-period treatment and the potential second-period treatment state $D_2 \left ( d_1 \right )$. Instead, Assumption \ref{ass_ct0} requires that potential outcomes absent second-period treatment exhibit common trends across groups implied by $D_2 \left ( d_1 \right )$, conditional on the first-period running variable being at the cutoff, for each $d_1 \in \left \{ 0,1 \right \}$. Thus, the two approaches rely on distinct, non-nested identifying restrictions.

Local common trends may be more plausible than the restrictions underlying recursive identification when agents select into subsequent treatments according to their expected benefits. For example, school districts with larger potential housing price gains from additional investments may be more likely to approve a subsequent referendum, thereby violating the mean independence condition in Lemma \ref{lemma_recursive}. This mechanism remains compatible with local common trends if these gains are unrelated to potential housing price growth absent further approval, as illustrated in Section \ref{sec_localct}. Moreover, earlier investments may change the housing price gains from additional school spending, implying that average instantaneous effects would differ across treatment histories.

\subsection{The Economic Content of Local Common Trends}\label{sec_localct}

Recent research has explored the economic content of the conventional parallel trends assumption in difference-in-differences designs, with particular attention to the relationship between dynamic selection into treatment and the evolution of potential outcomes (\citealt{marxtamertang2024}; \citealt{ghanemsantannawuthrich2026}).

Assumption \ref{ass_ct0} accommodates several forms of dynamic selection into treatment. In particular, local common trends allows for path dependence in treatment assignment and does \textit{not} require subsequent treatment to be independent of persistent unobserved unit characteristics or heterogeneous effects of subsequent treatment. Moreover, the restriction is imposed separately for each initial treatment state and only at the first-period cutoff, allowing potential outcome trends to differ across initial treatment states and for realizations of the running variable away from the cutoff. The examples below illustrate economic mechanisms under which this restriction may be credible.

First, consider the canonical application of dynamic RDD to local government referenda in which voters decide whether to increase taxation to finance additional public services, such as the renovation, expansion, or construction of school facilities. For simplicity, let the outcome $Y$ be local housing prices. Let $U$ denote unobserved determinants of $D_2 \left ( d_1 \right )$, such as political preferences and residents' valuation of school quality. Similarly, let $V$ denote unobserved determinants of potential housing price growth absent second-period referendum approval, such as shocks to local housing demand and economic conditions.

\textbf{Path Dependence in Treatment Assignment} \quad A local government may be more likely to submit another ballot proposition following the rejection of an initial referendum, particularly when the margin of defeat is narrow. Consequently, $D_2 \left ( d_1 \right )$ may depend on $R_1 - c$. Such dependence on the first-period running variable does not, by itself, violate local common trends, since Assumption \ref{ass_ct0} is imposed only at $R_1 = c$. In other words, Assumption \ref{ass_ct0} allows local government officials to learn about residents' preferences from the initial referendum margin and strategically decide whether to submit the same or a revised proposition in a subsequent period.

\textbf{Selection on Persistent Characteristics} \quad Consider a scenario in which, following an initial referendum, districts with a higher willingness to pay for public education are more likely to approve a referendum in the second period. Formally, suppose that $D_{2} \left ( d_1 \right ) = \mathbb{I} \left [ \alpha d_1 + U \geq 0 \right ]$, where linearity is assumed only for expositional purposes. Although $U$ may affect potential housing price \textit{levels} in both periods, suppose that it does not systematically affect their \textit{growth} between periods. Specifically, let $\Delta_{1,2} \left ( d_1, 0 \right ) = \gamma d_1 + V$, with $\E \left [ V | U, R_1 = c \right ] = \E \left [ V | R_1 = c \right ]$. In this case, Assumption \ref{ass_ct0} is satisfied. This may be plausible when persistent differences in residents' willingness to pay for public education are already capitalized into housing prices and contribute equally to potential price levels over time. Instead, potential price changes absent subsequent referendum approval reflect common housing market movements or local economic shocks unrelated to these preferences at the cutoff.

\textbf{Selection on Expected Gains from Treatment} \quad Consider a scenario in which districts are more likely to approve a second referendum when they anticipate larger gains in housing prices from additional school spending. Formally, suppose that \sloppy $D_{2} \left ( d_1 \right ) = \mathbb{I} \left [ Y_{2} \left ( d_1, 1 \right ) - Y_{2} \left ( d_1, 0 \right ) \geq 0 \right ]$, where, for simplicity, districts are assumed to know their potential treatment gains. This selection mechanism is compatible with local common trends if these gains are not systematically related to housing price growth absent second-period referendum approval. For example, let $\Delta_{1,2} \left ( d_1, 0 \right ) = \gamma d_1 + V$, with $\E \left [ V | Y_{2} \left ( d_1, 1 \right ) - Y_{2} \left ( d_1, 0 \right ), R_1 = c \right ] = \E \left [ V | R_1 = c \right ]$. Under this restriction, subsequent treatment assignment may depend on heterogeneous treatment effects without violating Assumption \ref{ass_ct0}. This may be plausible when the expected housing price gains from referendum approval reflect perceived differences in how effectively proposed projects would improve school services. Such differences may influence subsequent approval without being systematically related to housing price growth when the projects are not implemented.

The same reasoning applies to other empirical applications. For example, \cite{barbosaferreira2023} use a dynamic RDD to study the effects of winning a mayoral election in Brazil on public employment and wages among party affiliates. Specifically, let the outcome be the share of municipal employees affiliated with a given party. Assumption \ref{ass_ct0} concerns how this share would change if the party did not win the subsequent election. At the focal electoral cutoff, the expected change must be equal between parties that would subsequently win and those that would lose, separately for each initial election result. This restriction may be plausible when the strength of a party's local organization influences its subsequent electoral success but contributes equally to potential employment shares in both periods. Thus, unobserved differences in local party organization do not affect how employment shares would change in the absence of a subsequent electoral victory.

\section{Generalization to Multiple Leads}\label{sec_multiperiod}

This section generalizes the potential outcomes framework and identifying assumptions of the two-period model to an arbitrary number of time periods, indexed by $t \in \left \{ 1, 2, \dots, \overline{t} \right \}$.

First, potential outcomes at any point in time may depend on the path of past, contemporaneous, and future treatment states. Specifically, for $\left \{ d_j \right \}_{j=1}^{\overline{t}}$ with $d_j \in \left \{ 0,1 \right \}$ for all $j$, the period-$t$ potential outcome is denoted as $Y_t \left ( d_1, \dots, d_{\overline{t}} \right )$. Second, to allow for path dependence in treatment assignment, the period-$t$ potential treatment is modeled as $D_t \left ( d_1, \dots, d_{t-1} \right )$. To simplify notation, define $P \left ( d_1, \dots, d_t \right )$ as an indicator that equals one if the sequence of potential treatment assignments following initial treatment state $d_1$ coincides with the path $\left ( d_1, \dots, d_t \right )$:
\begin{align}\label{path_indicator}
	P \left ( d_1, \dots, d_{t} \right ) = \mathbb{I} \left [ D_2 \left ( d_1 \right ) = d_2, D_3 \left ( d_1, d_2 \right ) = d_3, \dots, D_{t} \left ( d_1, \dots, d_{t-1} \right ) = d_{t} \right ]
\end{align}
Thus, as a generalization of equation \eqref{potout}, the relationship between potential and observed outcomes can be characterized as
\begin{align}\label{potout_gen2}
	Y_t = \sum_{\left ( d_1, \dots, d_{\overline{t}} \right ) \in \left \{ 0,1 \right \}^{\overline{t}}} \mathbb{I} \left [ D_1 = d_1 \right ] P \left ( d_1, \dots, d_{\overline{t}} \right ) Y_t \left (  d_1, \dots, d_{\overline{t}} \right )
\end{align}
The target parameter of interest is a class of Average Direct Treatment Effects implied by the first-period discontinuity. Letting $0_{\overline{t}-1}$ denote a vector of $\overline{t}-1$ zeros and $\tau \in \left \{ 0, \dots, \overline{t}-1 \right \}$, the $\tau$-period-ahead ADTE at the first-period cutoff is defined as
\begin{align}\label{class_apte_mult}
	\mathrm{ADTE}_{1,\tau} \equiv \E \left [ Y_{1+\tau} \left ( 1, 0_{\overline{t}-1} \right ) - Y_{1+\tau} \left ( 0, 0_{\overline{t}-1} \right ) | R_1 = c \right ]
\end{align}
These parameters trace out the impulse response function of the outcome $Y$ to the initial treatment assignment determined by the threshold-crossing rule $D_1 \equiv \mathbb{I} \left [ R_1 \geq c \right ]$. Because subsequent treatment assignments may depend on the initial treatment state, the all-zero continuation need not coincide with the realized treatment path. Consequently, standard RD estimands do not generally identify ADTEs, as they may also reflect the contribution of subsequent treatment assignments. The first step in extending the identification arguments of the two-period model is to generalize Assumption \ref{ass_cont0} to arbitrary treatment paths.

\begin{ass}[Continuity at the Cutoff]\label{ass_cont1} For any $t \in \left \{ 1, \dots, \overline{t} \right \}$, any $\left \{ d_j \right \}_{j=1}^{\overline{t}}$ with $d_j \in \left \{ 0,1 \right \}$ for all $j$, and any $\left \{ d'_j \right \}_{j=2}^{\overline{t}}$ with $d'_j \in \left \{ 0,1 \right \}$ for all $j$,
	\begin{equation*}
		\E \left [ P \left ( d_1, \dots, d_{\overline{t}} \right ) | R_1 = r \right ] \quad \text{ and } \quad \E \left [ Y_{t} \left ( d_1, d_2, \dots, d_{\overline{t}} \right ) | R_1 = r, P \left ( d_1, d'_2, \dots, d'_{\overline{t}} \right ) = 1 \right ]
	\end{equation*}
	are continuous at $r=c$.
\end{ass}
Under Assumption \ref{ass_cont1}, the standard RD estimand admits the following decomposition.

\begin{prop}[Standard RD Estimand]\label{prop_rd1}
	Suppose that Assumption \ref{ass_cont1} holds. Then the estimand $\mathrm{RD}_{Y_t} \left ( c \right ) \equiv \lim_{r \downarrow c} \E \left [ Y_t | R_1 = r \right ] - \lim_{r \uparrow c} \E \left [ Y_t | R_1 = r \right ]$ identifies
	\begin{align*}
		& \sum_{\left ( d_2, \dots, d_{\overline{t}} \right ) \in \left \{ 0,1 \right \}^{\overline{t}-1}} \E \left [ Y_t \left ( 1,d_2, \dots, d_{\overline{t}} \right )  | R_1 = c, P \left ( 1, d_2, \dots, d_{\overline{t}} \right ) = 1 \right ] \times \E \left [ P \left ( 1, d_2, \dots, d_{\overline{t}} \right ) | R_1 = c \right ] \\
		- \ & \sum_{\left ( d_2, \dots, d_{\overline{t}} \right ) \in \left \{ 0,1 \right \}^{\overline{t}-1}} \E \left [ Y_t \left ( 0, d_2, \dots, d_{\overline{t}} \right )  | R_1 = c, P \left ( 0, d_2, \dots, d_{\overline{t}} \right ) = 1 \right ] \times \E \left [ P \left ( 0, d_2, \dots, d_{\overline{t}} \right ) | R_1 = c \right ]
	\end{align*}
\end{prop}
Since the indicators $P \left ( d_1, \dots, d_{\overline{t}} \right )$ partition the possible subsequent treatment paths for each initial treatment state, their conditional expectations at the cutoff are probabilities that sum to one. Thus, a standard RD estimand identifies the contrast between two convex combinations of conditional mean potential outcomes, where the initial treatment state is held fixed within each combination and subsequent treatment paths are weighted by their probabilities at the cutoff.

In the remainder of this section, I first consider identification of the instantaneous ADTE and then examine long-term effects. The following assumption generalizes the no anticipation restriction of the two-period model to an arbitrary number of time periods.

\begin{ass}[Local No Anticipation] \label{ass_noant1} For any $t < \overline{t}$, any $\left \{ d_j \right \}_{j=1}^{\overline{t}}$ with $d_j \in \left \{ 0,1 \right \}$ for all $j$, any $\left \{ d'_j \right \}_{j=t+1}^{\overline{t}}$ with $d'_j \in \left \{ 0,1 \right \}$ for all $j$, and any $\left \{ d''_j \right \}_{j=t+1}^{\overline{t}}$ with $d''_j \in \left \{ 0,1 \right \}$ for all $j$,
	\begin{align*}
		& \E \left [ Y_{t} \left ( d_1, \dots, d_t, d_{t+1}, \dots, d_{\overline{t}} \right ) \big | R_1 = c, P \left ( d_1, \dots, d_t, d''_{t+1}, \dots,  d''_{\overline{t}} \right ) = 1 \right ] \\
		= \ & \E \left [ Y_{t} \left ( d_1, \dots, d_t, d'_{t+1}, \dots, d'_{\overline{t}} \right ) \big | R_1 = c, P \left ( d_1, \dots, d_t, d''_{t+1}, \dots,  d''_{\overline{t}} \right ) = 1 \right ]
	\end{align*}
\end{ass}
Under Assumptions \ref{ass_cont1} and \ref{ass_noant1}, the first-period RD estimand identifies the instantaneous ADTE.

\begin{prop}[Identification of the Instantaneous ADTE]\label{prop_apte0_gen}
	\sloppy Suppose that Assumptions \ref{ass_cont1} and \ref{ass_noant1} hold. Then
	\begin{align*}
		\textsc{ADTE}_{1,0} = \lim_{r \downarrow c} \E \left [ Y_1 | R_1 = r \right ] - \lim_{r \uparrow c} \E \left [ Y_1 | R_1 = r \right ] \equiv \mathrm{RD}_1 \left ( c \right )
	\end{align*}
\end{prop}
To identify long-term ADTEs, it is sufficient to generalize the common trends assumption of the two-period model to an arbitrary number of time periods.

\begin{ass}[Local Common Trends] \label{ass_ct1} For any $t \in \left \{ 2, \dots, \overline{t} \right \}$ and $\left \{ d_j \right \}_{j=1}^{\overline{t}}$ with $d_j \in \left \{ 0,1 \right \}$,
	\begin{align*}
		& \E \left [ \Delta_{1,t} \left ( d_1, 0_{\overline{t}-1} \right ) | R_1 = c, P \left ( d_1, 0_{\overline{t}-1} \right ) = 1 \right ] = \E \left [ \Delta_{1,t} \left ( d_1, 0_{\overline{t}-1} \right ) | R_1 = c, P \left ( d_1, d_2, \dots, d_{\overline{t}} \right ) = 1 \right ]
	\end{align*}
with $\Delta_{1,t} \left ( d_1, 0_{\overline{t}-1} \right ) \equiv Y_{t} \left ( d_1, 0_{\overline{t}-1} \right ) - Y_1 \left ( d_1, 0_{\overline{t}-1} \right )$.
\end{ass}
For any initial treatment state $d_1$, $\Delta_{1,t} \left ( d_1, 0_{\overline{t}-1} \right )$ measures the change in the potential outcome between the first and a subsequent period when all treatment states after the first are set to zero. Assumption \ref{ass_ct1} requires the conditional mean of this change at the first-period cutoff to be invariant across potential subsequent treatment paths. Together with continuity and no anticipation, this restriction is sufficient to point identify all long-term ADTEs.

\begin{prop}[Identification of the Cumulative ADTE]\label{prop_apte1_gen}
	Suppose that Assumptions \ref{ass_cont1}, \ref{ass_noant1}, and \ref{ass_ct1} hold. Further assume that $\E \left [ P \left ( d_1, 0_{\overline{t}-1} \right ) | R_1 = c \right ] > 0$ for $d_1 \in \left \{ 0,1 \right \}$. Then, for $\tau \in \left \{ 1, \dots, \overline{t}-1 \right \}$,
	\begin{align*}
		\mathrm{ADTE}_{1,\tau} & = \lim_{r \downarrow c} \E \left [ Y_{1} | R_1 = r \right ] - \lim_{r \uparrow c} \E \left [ Y_{1} | R_1 = r \right ] \\
		& + \lim_{r \downarrow c} \E \left [ Y_{1+\tau} - Y_{1} \bigg | R_1 = r, \bigcap_{s=2}^{1+\tau} \left \{ D_{s} = 0 \right \} \right ] - \lim_{r \uparrow c} \E \left [ Y_{1+\tau} - Y_{1} \bigg | R_1 = r, \bigcap_{s=2}^{1+\tau} \left \{ D_{s} = 0 \right \} \right ]
	\end{align*}
\end{prop}
Proposition \ref{prop_apte1_gen} establishes that continuity, no anticipation, and common trends are sufficient to identify any long-term ADTE as the sum of two regression discontinuity estimands at the first-period cutoff. The first estimand measures the average outcome contrast immediately after the first round of treatment assignment. The second compares average outcome growth between periods 1 and $1+\tau$ on either side of the cutoff, among units who do not receive additional treatment during that interval.

\section{Aggregation of Time-Specific Parameters}\label{sec_aggreg}

The identification arguments presented so far have focused on causal parameters implied by the first-period discontinuity. In practice, the focal round of treatment assignment may occur in a later period and agents may be exposed to multiple rounds of treatment.

To accommodate this setting, let $H_{t-1} \in \left \{ 0,1 \right \}^{t-1}$ denote the history of treatment states up to and excluding period $t$. Because past treatments may affect subsequent potential outcomes, a causal parameter that isolates the effect of period-$t$ treatment assignment must compare potential outcomes under identical prior and subsequent treatment states. Accordingly, for a given history $h_{t-1} \in \supp \left ( H_{t-1} \right )$, the Average Direct Treatment Effect at the period-$t$ cutoff is defined as
\begin{align}
	\mathrm{ADTE}_{t,\tau} \left ( h_{t-1} \right ) \equiv \E \left [ Y_{t+\tau} \left ( h_{t-1}, 1, 0_{\overline{t}-t} \right ) - Y_{t+\tau} \left ( h_{t-1}, 0, 0_{\overline{t}-t} \right ) | H_{t-1} = h_{t-1}, R_t = c \right ]
\end{align}
for any $\tau \in \left \{ 0, 1, \dots, \overline{t} - t \right \}$.\footnote{Although the conditioning set need not include the event $H_{t-1} = h_{t-1}$, an empirical researcher is unlikely to be interested in the average effect of a treatment subsequent to a treatment history $h_{t-1}$ in the subpopulation implied by a \textit{different} treatment history $h'_{t-1}$. For practical purposes, it is natural for \textit{one} treatment history to both be an argument of potential outcomes and belong to the conditioning set.} By construction, the two potential outcomes differ only in their period-$t$ treatment state, so this class of impulse response parameters generalizes equation \eqref{class_apte_mult}.

Because each $\mathrm{ADTE}_{t,\tau} \left ( h_{t-1} \right )$ is specific to a treatment history, focal period of treatment assignment, and horizon, these parameters can serve as building blocks to construct aggregate policy-relevant measures. This approach parallels \cite{callawaysantanna2021}, which proposes several approaches to aggregate cohort- and time-specific causal parameters in difference-in-differences designs with staggered adoption of an absorbing treatment. Similarly, \cite{bojinovetal2021} aggregates unit-time-specific causal parameters in sequentially randomized experiments. In dynamic RDDs, a natural first step is to aggregate over treatment histories for a given focal period and horizon. The resulting parameter is defined as
\begin{align}
	\mathrm{ADTE}_{t,\tau} \equiv \sum_{h_{t-1} \in \supp \left ( H_{t-1} \right )} \mathrm{ADTE}_{t,\tau} \left ( h_{t-1} \right ) \times \Prob \left ( H_{t-1} = h_{t-1} | R_t = c \right )
\end{align}
The parameter $\mathrm{ADTE}_{t,\tau}$ averages history-specific Average Direct Treatment Effects over the distribution of treatment histories at the period-$t$ cutoff. It can thus be interpreted as the average impulse response of the outcome to the period-$t$ discontinuity, $\tau$ periods after treatment assignment.

In addition, an empirical researcher may be interested in aggregating time-specific parameters into a single summary measure of the $\tau$-period-ahead effect of treatment. Assuming a balanced panel, each term of the aggregate impulse response function is defined as the equally weighted average of the corresponding time-specific effects:
\begin{align}\label{eq:aggr_reltime}
	\mathrm{ADTE}_{\tau} \equiv \frac{1}{\overline{t}-\tau} \sum_{t=1}^{\overline{t}-\tau} \mathrm{ADTE}_{t,\tau}
\end{align}

\subsection{Dimension Reduction via Limited Path Independence}\label{sec:markov}

Conditional versions of Assumptions \ref{ass_cont1}, \ref{ass_noant1}, and \ref{ass_ct1} identify each $\mathrm{ADTE}_{t,\tau} \left ( h_{t-1} \right )$. In observational settings, however, it is generally infeasible to estimate one target parameter for each possible treatment history, focal calendar time, and number of leads. For instance, in the canonical application of dynamic RDDs to the effects of local referenda, units are infrequently subject to treatment assignment because jurisdictions typically hold new elections only once funds approved by a previously approved proposition are close to expiration. As a consequence, only a limited number of local governments participates in treatment assignment every year and any given jurisdiction does so in a cyclical fashion. In addition, a narrowly rejected referendum is generally submitted to voters again only once or twice briefly after the initial attempt, implying that the identification challenge posed by repeated treatment assignment is particularly salient shortly after an initial referendum. To summarize, in the local public finance application of dynamic RDDs, treatment assignment is sparse and cyclical.

Motivated by this feature of treatment assignment, I impose a limited path independence assumption to reduce the number of target parameters. This assumption is similar in spirit to that in \cite{hsushen2024}. For a given focal period and horizon, I assume that the ADTE is equal across treatment histories whose last $k$ states are zero.

\begin{ass}[Limited Path Independence] \label{ass_mark} Let $k \in \mathbb{N}_{+}$ be a deterministic constant.
For any $t \in \left \{ k+1,\dots,\overline{t} \right \}$, any $\tau \in \left \{ 0,1,\dots,\overline{t}-t \right \}$, and any histories $h_{t-1},h'_{t-1} \in \supp \left ( H_{t-1} \right )$ satisfying $h_{t-1} = \left ( h_{t-k-1},0_k \right )$ and $h'_{t-1} = \left ( h'_{t-k-1},0_k \right )$,
	\begin{align*}
    	\mathrm{ADTE}_{t,\tau} \left ( h_{t-1} \right ) = \mathrm{ADTE}_{t,\tau} \left ( h'_{t-1} \right )
	\end{align*}
\end{ass}
The constant $k$ is to be determined by the researcher depending on the empirical application.\footnote{In a similar spirit, \cite{callawaysantanna2021} consider a limited treatment anticipation assumption, based on which the researcher must specify the number of time periods the outcome variable may reflect agents' knowledge of (or expectations over) future exposure to the treatment.} The choice of a smaller $k$ requires ADTEs to be invariant across a larger set of treatment histories and therefore imposes a stronger restriction. A larger $k$ weakens this restriction, but reduces the number of units eligible for pooling.

Assumption \ref{ass_mark} makes the ADTE common across treatment histories ending in $0_k$. To estimate this common effect by pooling observations, I impose continuity, local no anticipation, and local common trends within the subpopulation of units satisfying $D_{t-k} = \dots = D_{t-1} = 0$. These three identifying restrictions apply directly to the pooled population, allowing earlier treatment histories to vary across units.

\subsection{Assessing the Plausibility of Local Common Trends}\label{sec_check_ct}

The population restriction motivated by Assumption \ref{ass_mark} also makes it possible to assess the plausibility of local common trends, as in difference-in-differences and event-study designs. Specifically, restricting the population to units untreated for $k$ periods before the focal round of treatment assignment provides pre-treatment outcomes. In this section, I compare changes in average pre-treatment outcomes across groups defined by subsequent treatment paths, separately on each side of the focal cutoff.

For ease of exposition, consider a four-period model ($\overline{t}=4$) in which all units are untreated until $t=3$, when a round of treatment assignment occurs. Let the target parameter be
\begin{align}
	\mathrm{ADTE}_{3,1} \left ( 0_2 \right ) \equiv \E \left [ Y_{4} \left ( 0_2,1,0 \right ) - Y_{4} \left ( 0_2,0,0 \right ) | H_2 = 0_2, R_3 = c \right ]
\end{align}
This is the cutoff-specific, one-period-ahead average effect of the treatment assigned in $t=3$, conditional on a never-treated path in the first two periods. Because, by construction, $D_1 = D_2 = 0$, the history of treatment states $H_2 = 0_2$ is observed. To identify $\mathrm{ADTE}_{3,1} \left ( 0_2 \right )$, Assumption \ref{ass_ct1} restricts $\E \left [ Y_4 \left ( 0_2,d_3,0 \right ) - Y_3 \left ( 0_2,d_3,0 \right ) | H_2 = 0_2, R_3 = c, P \left ( 0_2, d_3, 0 \right ) = 1 \right ]$ to be equal to $\E \left [ Y_4 \left ( 0_2,d_3,0 \right ) - Y_3 \left ( 0_2,d_3,0 \right ) | H_2 = 0_2, R_3 = c, P \left ( 0_2, d_3, 1 \right ) = 1 \right ]$ for both $d_3 = 0$ and $d_3 = 1$. The identifying restriction pertains to the growth in potential outcomes between periods 3 and 4. To assess its plausibility, I examine whether the groups defined by $D_4 \left ( 0_2,d_3 \right )$ also exhibit common trends between periods 1 and 2. Specifically, I consider the following pre-period restriction:
\begin{align}\label{eq:null_simple}
	& \E \left [ Y_2 \left ( 0_2,d_3,0 \right ) - Y_1 \left ( 0_2,d_3,0 \right ) | H_2 = 0_2, R_3 = c, P \left ( 0_2, d_3, 0 \right ) = 1 \right ] \nonumber \\
	= \ & \E \left [ Y_2 \left ( 0_2,d_3,0 \right ) - Y_1 \left ( 0_2,d_3,0 \right ) | H_2 = 0_2, R_3 = c, P \left ( 0_2, d_3, 1 \right ) = 1 \right ]
\end{align}
for $d_3 \in \left \{ 0,1 \right \}$. Suppose that continuity and local no anticipation hold for the relevant pre-period outcomes at the period-3 cutoff. Then, for $d_3=1$, the restriction in equation \eqref{eq:null_simple} implies the following equality of upper-limit conditional means:
\begin{align}
	\lim_{r \downarrow c} \E \left [ Y_2 - Y_1 | H_2 = 0_2, R_3 = r, D_4 = 0 \right ] = \lim_{r \downarrow c} \E \left [ Y_2 - Y_1 | H_2 = 0_2, R_3 = r, D_4 = 1 \right ]
\end{align}
For $d_3=0$, the corresponding restriction uses lower-limit conditional means.

These equalities naturally give rise to testable restrictions aimed at assessing the plausibility of Assumption \ref{ass_ct1}. Specifically, an empirical researcher may provide a more compelling argument in favor of the local common trends restriction by failing to reject the equality of average pre-treatment outcome changes at the cutoff in all available ``pre-periods''. In the canonical difference-in-differences design, it is common practice to test the equality of pre-period outcome trends across groups implied by whether units are assigned to the ``treatment'' or ``control'' arm. In a dynamic RDD, groups are implicitly determined by the treatment path subsequent to the focal time period. Unless more restrictions are placed on the evolution of treatment states, the number of groups or ``types'' will in general exceed two, thereby increasing the number of testable equalities.

To illustrate these alternatives, I extend the preceding example to $\overline{t}=5$ periods and consider a horizon of $\tau=2$. The focal round of treatment assignment remains in $t=3$. The subsequent treatment path is now $\left ( D_4, D_5 \right )$, while the pre-treatment outcome change remains $\Delta_{1,2} \equiv Y_2-Y_1$.

\subsubsection{Aggregate Comparison}

A natural starting point is to compare units with an all-zero subsequent treatment path with units treated in at least one subsequent period. This approach pools all nonzero paths irrespective of treatment timing or frequency. The null hypothesis requires average pre-treatment outcome growth to be equal between these two groups on each side of the cutoff. In the example above, the joint null hypothesis imposes the following upper-limit equality and its lower-limit counterpart:
\begin{align}
	& \lim_{r \downarrow c} \E \left [ \Delta_{1,2} \big | H_2 = 0_2, R_3 = r, \left ( D_4 , D_5 \right ) = \left ( 0, 0 \right ) \right ] \notag \\
	& = \lim_{r \downarrow c} \E \left [ \Delta_{1,2} \big | H_2 = 0_2, R_3 = r, \left ( D_4 , D_5 \right ) \neq \left ( 0, 0 \right ) \right ]
\end{align}
The joint null hypothesis consists of two restrictions, one for each side of the cutoff.

\subsubsection{Comparisons with Individual Treatment Paths}

The pooling approach may conceal differences in pre-treatment outcome growth across individual paths.
In particular, positive and negative differences relative to the all-zero group may offset. At the other extreme,
an alternative approach is therefore to compare the all-zero subsequent treatment path separately with each nonzero path. In the example above, the joint null hypothesis imposes the following equality for each $\left ( d_4,d_5 \right ) \in
\left \{ \left ( 1,0 \right ),\left ( 0,1 \right ), \left ( 1,1 \right ) \right \}$, together with the corresponding lower-limit equalities:
\begin{align}
	& \lim_{r \downarrow c} \E \left [ \Delta_{1,2} \big | H_2 = 0_2, R_3 = r, \left ( D_4 , D_5 \right ) = \left ( 0, 0 \right ) \right ] \notag \\
	& = \lim_{r \downarrow c} \E \left [ \Delta_{1,2} \big | H_2 = 0_2, R_3 = r, \left ( D_4 , D_5 \right ) = \left ( d_4, d_5 \right ) \right ]
\end{align}
For a horizon of $\tau$ periods, there are up to $2^\tau-1$ nonzero treatment paths. Comparing each path with the
all-zero path yields up to $2 \left ( 2^\tau-1 \right )$ scalar restrictions. In the example above, the joint null hypothesis contains six restrictions.

\subsubsection{Intermediate Groupings}

A limitation of the approach based on comparing individual treatment paths is that few observations might remain available for each comparison, thereby reducing the precision of the estimated contrasts. An intermediate approach consists of grouping nonzero paths according to economically relevant features. I illustrate two possibilities: treatment \textit{frequency} and treatment \textit{timing}.

A researcher may group paths based on the number of subsequent treated periods. This comparison is motivated by concerns that factors influencing treatment frequency may also affect potential outcome growth absent subsequent treatment. In the example above, pooling paths with exactly one treated period yields the following upper-limit equality and its lower-limit counterpart:
\begin{align}
	& \lim_{r \downarrow c} \E \left [ \Delta_{1,2} \big | H_2 = 0_2, R_3 = r, \left ( D_4 , D_5 \right ) = \left ( 0, 0 \right ) \right ] \notag \\
	& = \lim_{r \downarrow c} \E \left [ \Delta_{1,2} \big | H_2 = 0_2, R_3 = r, \left ( D_4 , D_5 \right ) \in \left \{ \left ( 0, 1 \right ), \left ( 1, 0 \right ) \right \} \right ]
\end{align}

Alternatively, a researcher may group paths by the timing of the first subsequent treatment. This comparison is motivated by concerns that factors influencing early subsequent treatment may also affect potential outcome growth absent subsequent treatment. In the example above, pooling paths with treatment in period 4 yields the following upper-limit equality and its lower-limit counterpart:
\begin{align}
	& \lim_{r \downarrow c} \E \left [ \Delta_{1,2} \big | H_2 = 0_2, R_3 = r, \left ( D_4 , D_5 \right ) = \left ( 0, 0 \right ) \right ] \notag \\
	& = \lim_{r \downarrow c} \E \left [ \Delta_{1,2} \big | H_2 = 0_2, R_3 = r, \left ( D_4 , D_5 \right ) \in \left \{ \left ( 1, 0 \right ), \left ( 1, 1 \right ) \right \} \right ]
\end{align}

\subsubsection{Full Generalization}

The comparisons discussed above extend to arbitrary focal periods and subsequent treatment horizons. Within the subpopulation of units satisfying $D_{t-1} = \dots = D_{t-k} = 0$, the tests may be conducted for any pair of pre-treatment periods $u$ and $v$ such that $t-k \leq u < v < t$. Supplementary Appendix \ref{app_testct} develops statistical theory for assessing the plausibility of local common trends.

\section{Estimation and Statistical Inference}\label{sec_estim}

In this section, I illustrate that standard local polynomial methods are applicable to the estimation of Average Direct Treatment Effects in a dynamic RDD. Hereafter, I do not explicitly condition on the history of treatment states $H_{t-1} = h_{t-1}$ to keep notation concise. Then, by an application of the Law Iterated Expectations and the properties of limits, the target estimand from Proposition \ref{prop_apte1_gen} can be equivalently expressed as
\begin{align}
	\mathrm{ADTE}_{t,\tau} & = \lim_{r \downarrow c} \E \left [ Y_{t} | R_t = r \right ] - \lim_{r \uparrow c} \E \left [ Y_{t} | R_t = r \right ] \notag \\
	& + \frac{\lim_{r \downarrow c} \E \left [ \left ( Y_{t+\tau} - Y_t \right ) \prod_{s=1}^{\tau} \left ( 1 - D_{t+s} \right ) | R_t = r \right ]}{\lim_{r \downarrow c} \E \left [ \prod_{s=1}^{\tau} \left ( 1 - D_{t+s} \right ) | R_t = r \right ]} \notag \\
	& - \frac{\lim_{r \uparrow c} \E \left [ \left ( Y_{t+\tau} - Y_t \right ) \prod_{s=1}^{\tau} \left ( 1 - D_{t+s} \right ) | R_t = r \right ]}{\lim_{r \uparrow c} \E \left [ \prod_{s=1}^{\tau} \left ( 1 - D_{t+s} \right ) | R_t = r \right ]} \label{apte_gen_lie}
\end{align}
In this expression, the parameter of interest is a nonlinear function of cutoff-specific conditional means, all of which can be nonparametrically estimated in the full sample, rather than in subsamples implied by realizations of $\left \{ D_{t+s} \right \}_{s=1}^{\tau}$. This feature will prove convenient for the purpose of constructing robust nonparametric confidence intervals. To keep notation compact, I also define the following random variables:
\begin{align}
	R \equiv R_t \qquad Y \equiv Y_t \qquad W \equiv \left ( Y_{t+\tau} - Y_t \right ) \prod_{s=1}^{\tau} \left ( 1 - D_{t+s} \right ) \qquad D \equiv \prod_{s=1}^{\tau} \left ( 1 - D_{t+s} \right )
\end{align}
For any pair of random variables $A$ and $B$, let $\mu_{A} \left ( r \right ) \equiv \E \left [ A | R = r \right ]$, $\sigma^2_{AA} \left ( r \right ) \equiv \Var \left [ A | R = r \right ]$, and $\sigma^2_{AB} \left ( r \right ) \equiv \Cov \left [ A , B | R = r \right ]$. Furthermore, let $\mu_{A+} \equiv \lim_{r \downarrow c} \mu_{A} \left ( r \right )$ and $\mu_{A-} \equiv \lim_{r \uparrow c} \mu_{A} \left ( r \right )$. Then the parameter of interest can be written as
\begin{align}\label{eq_estimand}
	\theta \equiv \mu_{Y+} - \mu_{Y-} + \frac{\mu_{W+}}{\mu_{D+}} - \frac{\mu_{W-}}{\mu_{D-}}
\end{align}
Prior to discussing statistical inference on $\theta$, Assumption \ref{ass_moments} outlines regularity conditions
on the running variable and the conditional moments of $Y$, $W$, and $D$. These conditions build on Assumptions
1 and 3 of \cite{cct2014}.

\begin{ass} \label{ass_moments}
	Let $\delta \in \mathbb{N}$. For some $\kappa_0 \in \mathbb{R}_{++}$, the following conditions hold in a neighborhood of the cutoff $\left ( c-\kappa_0, c+\kappa_0 \right )$.
	\begin{enumerate}[(a)]
		\item The density  $f_R \left ( r \right )$ is continuous and bounded away from zero.
		\item $\E \left [ Y^4 | R=r \right ]$ and $\E \left [ W^4 | R=r \right ]$ are bounded.
		\item $\mu_{Y} \left ( r \right )$, $\mu_{W} \left ( r \right )$, $\mu_{D} \left ( r \right )$ are $\delta$ times continuously differentiable separately on $\left ( c-\kappa_0,c \right )$ and $\left ( c,c+\kappa_0 \right )$. These functions and their derivatives through order $\delta$ have finite one-sided limits at $c$.
		\item $\sigma^2_{YY} \left ( r \right )$, $\sigma^2_{WW} \left ( r \right )$, $\sigma^2_{DD} \left ( r \right )$ are continuous separately on either side of $c$, have finite one-sided limits at $c$, and are bounded away from zero.
		\item $\sigma^2_{YW} \left ( r \right )$, $\sigma^2_{YD} \left ( r \right )$, $\sigma^2_{WD} \left ( r \right )$ are continuous separately on either side of $c$ and have finite one-sided limits at $c$.
		\item The one-sided limits at $c$ of the conditional covariance matrix of $\left [ Y,W,D \right ]'$ given $R=r$ are positive definite.
	\end{enumerate}
\end{ass}
Consider a random sample $\left \{ \left [ R_{i}, Y_{i}, W_{i}, D_{i} \right ]' \right \}_{i=1}^{n}$. Given a positive bandwidth $h_n$ and a positive real-valued constant $\kappa$, let $\mathcal{R} \left ( h_n \right ) = \left [ c - \kappa h_n, c + \kappa h_n \right ]$ be a discontinuity window implied by realizations of the running variable around the first-period cutoff. Moreover, let $\mathcal{R}^{-} \left ( h_n \right ) = \left [ c - \kappa h_n, c \right )$ and $\mathcal{R}^{+} \left ( h_n \right ) = \left [ c, c + \kappa h_n \right ]$ indicate the left and right discontinuity windows, respectively. Throughout this section, I employ standard local polynomial estimators to approximate the value of several conditional means on either side of the cutoff. Specifically, for any random variable $A$, I consider local linear estimators defined as
\begin{align}
	\left [ \widehat{\mu}^{(0)}_{A+,1} \left ( h_n \right ), \widehat{\mu}^{(1)}_{A+,1} \left ( h_n \right ) \right ]' & \equiv \arg \min_{b_0,b_1 \in \mathbb{R}} \sum_{i : R_{i} \in \mathcal{R}^{+} \left ( h_n \right )} \left ( A_{i} - b_0 - b_1 \left ( R_{i} - c \right ) \right )^2 k_{h_n} \left ( R_{i} - c \right ) \label{loclinplus} \\
	\left [ \widehat{\mu}^{(0)}_{A-,1} \left ( h_n \right ), \widehat{\mu}^{(1)}_{A-,1} \left ( h_n \right ) \right ]' & \equiv \arg \min_{b_0,b_1 \in \mathbb{R}} \sum_{i : R_{i} \in \mathcal{R}^{-} \left ( h_n \right )} \left ( A_{i} - b_0 - b_1 \left ( R_{i} - c \right ) \right )^2 k_{h_n} \left ( R_{i} - c \right ) \label{loclinminus}
\end{align}
where $k_h \left ( u \right ) \equiv k \left ( |u| \slash h \right ) \slash h$, with $k \left ( \cdot \right )$ denoting a kernel function. Mirroring Assumption 2 in \cite{cct2014}, Assumption \ref{ass_kernel} below reports minimal conditions for the validity of $k \left ( \cdot \right )$.

\begin{ass}\label{ass_kernel}
	For $\kappa \in \mathbb{R}_{++}$, the kernel function $k \left ( \cdot \right ): \left [ 0,\kappa \right ] \to \mathbb{R}$ is bounded and nonnegative, zero outside its support, and positive and continuous on $\left ( 0, \kappa \right )$.
\end{ass}
In the notation of equations \eqref{loclinplus} and \eqref{loclinminus}, the subscript $p$ in $\widehat{\mu}^{(\nu)}_{+,p}$ indicates the chosen polynomial degree in the regression specification, while the superscript $\nu$ denotes the order of the derivative of interest. For example, in standard sharp designs, the estimated coefficient of interest is $\widehat{\mu}^{(0)}_{+,1} \left ( h_n \right ) - \widehat{\mu}^{(0)}_{-,1} \left ( h_n \right )$, namely the difference between two intercepts in a local linear regression specification. In this paper's dynamic design, the estimator of interest replaces the conditional means in equation \eqref{apte_gen_lie} with their sample counterparts. As a matter of fact, a natural estimator for $\theta$ is
\begin{align}\label{estimator}
	\widehat{\theta} \left ( h_n \right ) \equiv \widehat{\mu}^{(0)}_{Y+,1} \left ( h_n \right ) - \widehat{\mu}^{(0)}_{Y-,1} \left ( h_n \right ) + \frac{\widehat{\mu}^{(0)}_{W+,1} \left ( h_n \right )}{\widehat{\mu}^{(0)}_{D+,1} \left ( h_n \right )} - \frac{\widehat{\mu}^{(0)}_{W-,1} \left ( h_n \right )}{\widehat{\mu}^{(0)}_{D-,1} \left ( h_n \right )}
\end{align}
As in any regression discontinuity design, the estimator in equation \eqref{estimator} is a function of the width of the discontinuity window $h_n$. The choice of a larger bandwidth may lead to more precise estimates, but also increase their bias relative to the true intercept at the cutoff. Concerns about bias are particularly salient when adopting the prevalent data-driven approach to bandwidth selection, namely choosing one that minimizes the mean squared error of the estimator (\citealt{ik2012}). A natural solution is to first estimate the bias using a pilot bandwidth $b_n$ and then subtract the estimated bias from the point estimate of interest. \cite{cct2014} show how to account for the statistical uncertainty stemming from the first step and construct robust nonparametric confidence intervals for canonical estimands in sharp and fuzzy regression discontinuity and kink designs. \cite{ccf2018} demonstrate that robust bias correction can yield more accurate confidence interval coverage than undersmoothing. \cite{ccf2022} provide further motivation for robust bias correction by establishing optimal rates at which the coverage error of robust bias-corrected local polynomial confidence intervals converges to zero.

In this section, I follow the approach developed by \cite{cct2014} to construct robust nonparametric confidence intervals for the target estimand in equation \eqref{apte_gen_lie}. The choice of notation and the structure of proofs largely mirror those in \cite{cct2014supp}. Similarly to standard estimators in fuzzy regression discontinuity and kink designs, the estimator in equation \eqref{estimator} is nonlinear. Thus, to obtain the bias expression in equation \eqref{bias}, I linearize $\widehat{\theta} \left ( h_n \right )$ with a first-order Taylor expansion. The resulting estimator is defined as $\widetilde{\theta} \left ( h_n \right ) \equiv \widehat{\mu}^{(0)}_{Y+,1} \left ( h_n \right ) - \widehat{\mu}^{(0)}_{Y-,1} \left ( h_n \right ) + \widetilde{\eta}_{+} \left ( h_n \right ) - \widetilde{\eta}_{-} \left ( h_n \right )$, with
\begin{align}
	\widetilde{\eta}_{+} \left ( h_n \right ) & \equiv \frac{1}{\mu_{D+}} \left ( \widehat{\mu}^{(0)}_{W+,1} \left ( h_n \right ) - \mu_{W+} \right ) - \frac{\mu_{W+}}{\mu_{D+}^2} \left ( \widehat{\mu}^{(0)}_{D+,1} \left ( h_n \right ) - \mu_{D+} \right ) \label{eq:etaplus}\\
	\widetilde{\eta}_{-} \left ( h_n \right ) & \equiv \frac{1}{\mu_{D-}} \left ( \widehat{\mu}^{(0)}_{W-,1} \left ( h_n \right ) - \mu_{W-} \right ) - \frac{\mu_{W-}}{\mu_{D-}^2} \left ( \widehat{\mu}^{(0)}_{D-,1} \left ( h_n \right ) - \mu_{D-} \right ) \label{eq:etaminus}
\end{align}
Letting $\widehat{\textsc{B}} \left ( h_n, b_n \right )$ denote the estimated bias of $\widetilde{\theta} \left ( h_n \right )$, the bias-corrected estimator of $\theta$ is
\begin{align}\label{eq_thetabc}
	\widehat{\theta}^{\textsc{bc}} \left ( h_n, b_n \right ) \equiv \widehat{\theta} \left ( h_n \right ) - h^2_n \widehat{\textsc{B}} \left ( h_n, b_n \right )
\end{align}
with
\begin{align}
	\widehat{\textsc{B}} \left ( h_n, b_n \right ) & \equiv \left [ \frac{\widehat{\mu}^{(2)}_{Y+,2} \left ( b_n \right )}{2} + \frac{1}{\widehat{\mu}^{(0)}_{D+,1} \left ( h_n \right )} \frac{\widehat{\mu}^{(2)}_{W+,2} \left ( b_n \right )}{2} - \frac{\widehat{\mu}^{(0)}_{W+,1} \left ( h_n \right )}{\left ( \widehat{\mu}^{(0)}_{D+,1} \left ( h_n \right ) \right )^2} \frac{\widehat{\mu}^{(2)}_{D+,2} \left ( b_n \right )}{2} \right ] \mathcal{B}_{+} \left ( h_n \right ) \notag \\
	& - \left [ \frac{\widehat{\mu}^{(2)}_{Y-,2} \left ( b_n \right )}{2} + \frac{1}{\widehat{\mu}^{(0)}_{D-,1} \left ( h_n \right )} \frac{\widehat{\mu}^{(2)}_{W-,2} \left ( b_n \right )}{2} - \frac{\widehat{\mu}^{(0)}_{W-,1} \left ( h_n \right )}{\left ( \widehat{\mu}^{(0)}_{D-,1} \left ( h_n \right ) \right )^2} \frac{\widehat{\mu}^{(2)}_{D-,2} \left ( b_n \right )}{2} \right ] \mathcal{B}_{-} \left ( h_n \right ) \label{bias}
\end{align}
where $\mathcal{B}_{+} \left ( h_n \right )$ and $\mathcal{B}_{-} \left ( h_n \right )$ are observed and asymptotically bounded quantities defined in Supplementary Appendix \ref{app_def_bias}. Proposition \ref{prop_confint} below establishes the asymptotic distribution of the bias-corrected estimator $\widehat{\theta}^{\textsc{bc}} \left ( h_n, b_n \right )$.

\begin{prop}[Asymptotic Distribution]\label{prop_confint}
	Suppose that Assumptions \ref{ass_moments} and \ref{ass_kernel} hold with $\delta \geq 3$. Let $\mu_{D+} \neq 0$ and $\mu_{D-} \neq 0$. If $n \min \left \{ h^5_n, b^5_n \right \} \max \left \{ h^2_n, b^2_n \right \} \to 0$ and $n \min \left \{ h_n, b_n \right \} \to \infty$,
		\begin{align*}
			T^{\textsc{rbc}} \left ( h_n , b_n \right ) \equiv \frac{\widehat{\theta}^{\textsc{bc}} \left ( h_n, b_n \right ) - \theta}{\sqrt{\textsc{V}^{\textsc{bc}} \left ( h_n, b_n \right )}} \stackrel{d}{\to} \mathcal{N} \left ( 0,1 \right )
		\end{align*}
	provided that $h_n \to 0$ and $\kappa b_n < \kappa_0$. The exact expression for $\textsc{V}^{\textsc{bc}} \left ( h_n, b_n \right )$ is provided in Supplementary Appendix \ref{app_prop_confint}.
\end{prop}

\subsection{Bandwidth Selection}\label{sec:bandwidth}

Following \cite{cct2014}, I select a main bandwidth $h_n$ to estimate the parameter of interest and a pilot bandwidth $b_n$ to estimate its bias. I first derive bandwidths that minimize the mean squared error (MSE) of the relevant linearized estimators. As discussed by \cite{ccf2020}, MSE-optimal bandwidths can yield valid robust bias-corrected inference, but generally do not minimize confidence interval coverage error. I therefore also apply their rule-of-thumb adjustment to the main bandwidth.

Two distinct conditional MSE criteria are considered. First, the main bandwidth minimizes the leading terms of the conditional MSE of the linearized estimator $\widetilde{\theta} \left ( h_n \right )$, which combines intercepts estimated with local linear regressions.\footnote{In general, $\widetilde{\theta}_{\nu,p} \left ( h_n \right )$ denotes the linearized estimator involving $\nu$th-order derivatives in local polynomial specifications of order $p$. Thus, $\widetilde{\theta} \left ( h_n \right )$ corresponds to $\widetilde{\theta}_{0,1} \left ( h_n \right )$.} Second, the pilot bandwidth minimizes the leading terms of the conditional MSE of the estimator of the leading bias of $\widetilde{\theta} \left ( h_n \right )$. This bias estimator combines second-order derivatives estimated with local quadratic regressions but retains the population coefficients from the original linearization. Its definition is provided in Supplementary Appendix \ref{app_prop_bandwidth}. The following proposition reports both MSE expansions and the corresponding optimal bandwidths.

\begin{prop}\label{prop_optbandwidth}
	Suppose that Assumptions \ref{ass_moments} and \ref{ass_kernel} hold with $\delta \geq 3$. Let $\mu_{D+} \neq 0$ and $\mu_{D-} \neq 0$. Suppose that $h_n \to 0$, $n h_n \to \infty$, $b_n \to 0$, and $n b_n \to \infty$. The conditional mean squared error of the linearized estimator $\widetilde{\theta} \left ( h_n \right )$ is
	\begin{align*}
		\text{MSE} \left ( \widetilde{\theta} \left ( h_n \right ) \right ) & = h^4_{n} \left ( \dot{\textsc{B}}^2_{h} + o_p \left ( 1 \right ) \right ) + \frac{1}{n h_n} \left ( \dot{\textsc{V}}_{h} + o_p \left ( 1 \right ) \right )
	\end{align*}
	If $\dot{\textsc{B}}_h \neq 0$, the MSE-optimal main bandwidth is
	\begin{align*}
		h_{\textsc{mse}} = \left ( \frac{\dot{\textsc{V}}_{h}}{4 \dot{\textsc{B}}^2_{h}} \right )^{1 \slash 5} n^{-1 \slash 5}
	\end{align*}
	Let $\widetilde{\textsc{B}}_h \left ( b_n \right )$ denote the estimator of $\dot{\textsc{B}}_h$. The conditional mean squared error of $\widetilde{\textsc{B}}_h \left ( b_n \right )$ is
	\begin{align*}
		\text{MSE} \left ( \widetilde{\textsc{B}}_h \left ( b_n \right ) \right ) & = b_n^2 \left ( \dot{\textsc{B}}_b^2 + o_p \left ( 1 \right ) \right ) + \frac{1}{n b_n^5} \left ( \dot{\textsc{V}}_b + o_p \left ( 1 \right ) \right )
	\end{align*}
	If $\dot{\textsc{B}}_b \neq 0$, the MSE-optimal pilot bandwidth is
	\begin{align*}
		b_{\textsc{mse}} & = \left ( \frac{5 \dot{\textsc{V}}_b}{2 \dot{\textsc{B}}_b^2} \right )^{1 \slash 7} n^{-1 \slash 7}
	\end{align*}
	Exact expressions for $\widetilde{\textsc{B}}_h \left ( b_n \right )$, $\dot{\textsc{B}}_{h}$, $\dot{\textsc{V}}_{h}$, $\dot{\textsc{B}}_{b}$, and $\dot{\textsc{V}}_{b}$ are provided in Supplementary Appendix \ref{app_prop_bandwidth}.
\end{prop}
The MSE-optimal bandwidths specified in Proposition \ref{prop_optbandwidth} depend on unknown population quantities and, hence, are not directly implementable. In practice, I follow the steps outlined in Section S.2.6 of \cite{cct2014supp} and employ analogous direct plug-in bandwidth selectors. For inference, I multiply the resulting main bandwidth by $n^{-1\slash20}$, while retaining the MSE-optimal pilot bandwidth to estimate bias. This rule-of-thumb adjustment follows the recommendation for fuzzy RD estimators in Section 5.1 of \cite{ccf2020}. The application of this approach here is motivated by the nonlinear structure shared by fuzzy RD estimators and the estimator considered in this paper.

\subsection{Standard Errors}\label{sec:stderr}

The variance of the bias-corrected estimator $\textsc{V}^{\textsc{bc}} \left ( h_n, b_n \right )$ in Proposition \ref{prop_confint} is asymptotic. To compute standard errors and conduct statistical inference on the parameter of interest, it is natural to replace unknown population moments with their sample analogues. First, note that the linearized estimator in equations \eqref{eq:etaplus} and \eqref{eq:etaminus} is defined as a linear combination of estimated intercepts, where the constant coefficients are unknown population means. I replace these means with their corresponding estimates computed within each discontinuity half-window. Second, population variances and covariances are unknown. For any pair of random variables $A$ and $B$, define these second-order moments as $\sigma^2_{AB+} \equiv \lim_{r \downarrow c} \Cov \left [ A,B | R=r \right ]$ and $\sigma^2_{AB-} \equiv \lim_{r \uparrow c} \Cov \left [ A,B | R=r \right ]$. To estimate these moments, I follow the approach in \cite{abadieimbens2006} and \cite{cct2014}, employing a nearest-neighbor method that computes each sample covariance using a fixed number of observations. Specifically, letting $j^*$ denote this tuning parameter,
\begin{align}
	\widehat{\sigma}^2_{AB+} \left ( R_i \right ) & \equiv \mathbb{I} \left [ R_i \geq c \right ] \frac{j^*}{j^*+1} \left ( A_i - \sum_{j \in \mathcal{J}^{+}_{i}} \frac{A_{j}}{j^*} \right ) \left ( B_i - \sum_{j \in \mathcal{J}^{+}_{i}} \frac{B_{j}}{j^*} \right ) \\
	\widehat{\sigma}^2_{AB-} \left ( R_i \right ) & \equiv \mathbb{I} \left [ R_i < c \right ] \frac{j^*}{j^*+1} \left ( A_i - \sum_{j \in \mathcal{J}^{-}_{i}} \frac{A_{j}}{j^*} \right ) \left ( B_i - \sum_{j \in \mathcal{J}^{-}_{i}} \frac{B_{j}}{j^*} \right )
\end{align}
where $\mathcal{J}^{+}_{i}$ and $\mathcal{J}^{-}_{i}$ denote the sets of the $j^*$ closest units to unit $i$ in the right and left discontinuity half-windows, respectively. Once all of the first- and second-order moments are estimated, the resulting estimated variance $\widehat{\textsc{V}}^{\textsc{bc}} \left ( h_n, b_n \right )$ can be used to compute the standard error of $\widehat{\theta}^{\textsc{bc}} \left ( h_n, b_n \right )$ and construct confidence intervals for the target parameter.

Supplementary Appendix \ref{sec_simulation} reports the results of a Monte Carlo simulation that assesses the performance of the proposed estimators and confidence intervals, as well as the size and power of the tests used to assess the plausibility of local common trends.

\section{Empirical Application}\label{sec:application}

In this section, I apply the proposed method to estimate the intertemporal effects of changes in school district expenditures on housing prices in Wisconsin. Due to property tax limits imposed by state governments, local jurisdictions often face constraints in their fiscal policy decisions \citep{lincolntaxlimits}. To secure additional funding, school districts may hold referenda seeking voter authorization for property tax increases designated to finance specific expenditure initiatives.

Housing prices are chosen as the outcome variable for two primary reasons. First, research in local public finance has long recognized that the extent to which property tax changes are capitalized into housing prices is an informative measure of the efficiency of local public goods provision (\citealt{oates1969}; \citealt{brueckner1982}; \citealt{cushing1984}; \citealt{barrowrouse2004}; \citealt{figliolucas2004}; \citealt{cfr2010}; \citealt{bilaschon2024}). Evidence of positive capitalization following an expenditure increase is typically interpreted as an indication that prospective homebuyers value the associated improvement in public service provision more than the change in the tax burden required to fund it. Second, many capital investment projects approved through school referenda unfold over multiple years, and their long-term implications may not be fully internalized by potential buyers at the time of the vote. This feature underscores the importance of identifying interpretable dynamic treatment effects over extended horizons.

\subsection{Data}

Wisconsin school districts routinely hold referenda to authorize both operational and capital expenditures (\citealt{baron2022}). The Wisconsin Department of Public Instruction collects and publishes comprehensive data on all such local referenda held in the state since 1990 (\citealt{wiscreferenda}). For each referendum, I define the running variable as the difference between the share of votes in favor and the 50 percent approval threshold.

To construct the outcome of interest, I follow an approach similar to that of \cite{bilaschon2024}. Specifically, I rely on a repeat-sales house price index developed by \cite{contatlarson2024}, which covers Census tracts located within Core-Based Statistical Areas\footnote{The term “Core-Based Statistical Area” refers collectively to both Metropolitan Statistical Areas and Micropolitan Statistical Areas (\citealt{censusglossary}).} in the United States from 1989 to 2021. The index is normalized to 100 in 1989 for all tracts. It therefore measures changes in housing prices within each tract, but cannot be used to establish differences in housing price levels across tracts. To allow for level comparisons across school districts, I incorporate data on the average value of owner-occupied single-family homes at the Census tract level, as reported in the U.S. Census Bureau's 2000 Decennial Census.\footnote{This variable was not collected in the 2010 Decennial Census. I harmonize the 2000 Census values to 2010 tract boundaries using Census tract relationship files and housing unit weights.} For each tract, I compute a calibration factor as the ratio of the 2000 Census home value to the 2000 value of the house price index from \cite{contatlarson2024}, and apply this factor to the full time series of the index. The resulting measure of housing prices allows for both cross-sectional and intertemporal comparisons.

Next, I aggregate tract-level housing values to elementary, secondary, and unified school districts using 2010 Census housing unit counts and Census block-to-school-district assignments.\footnote{I use 2010 tract and school district boundaries because the house price index constructed by \cite{contatlarson2024} is based on 2010 Census tracts.} When a tract spans multiple districts of the same type, I weight its housing value in each district by the number of housing units located in the portion of the tract within that district.\footnote{I retain districts only if tracts with an available house price index account for at least 80 percent of the district's housing units in the 2010 Census.} I then take the natural logarithm of each district's unit-weighted average housing value.

\begin{figure}[!t]
	\begin{center}
		\caption{Distribution of the Approval Vote Share Margin}\label{fig_runningvar}
		\includegraphics[width=0.8\textwidth]{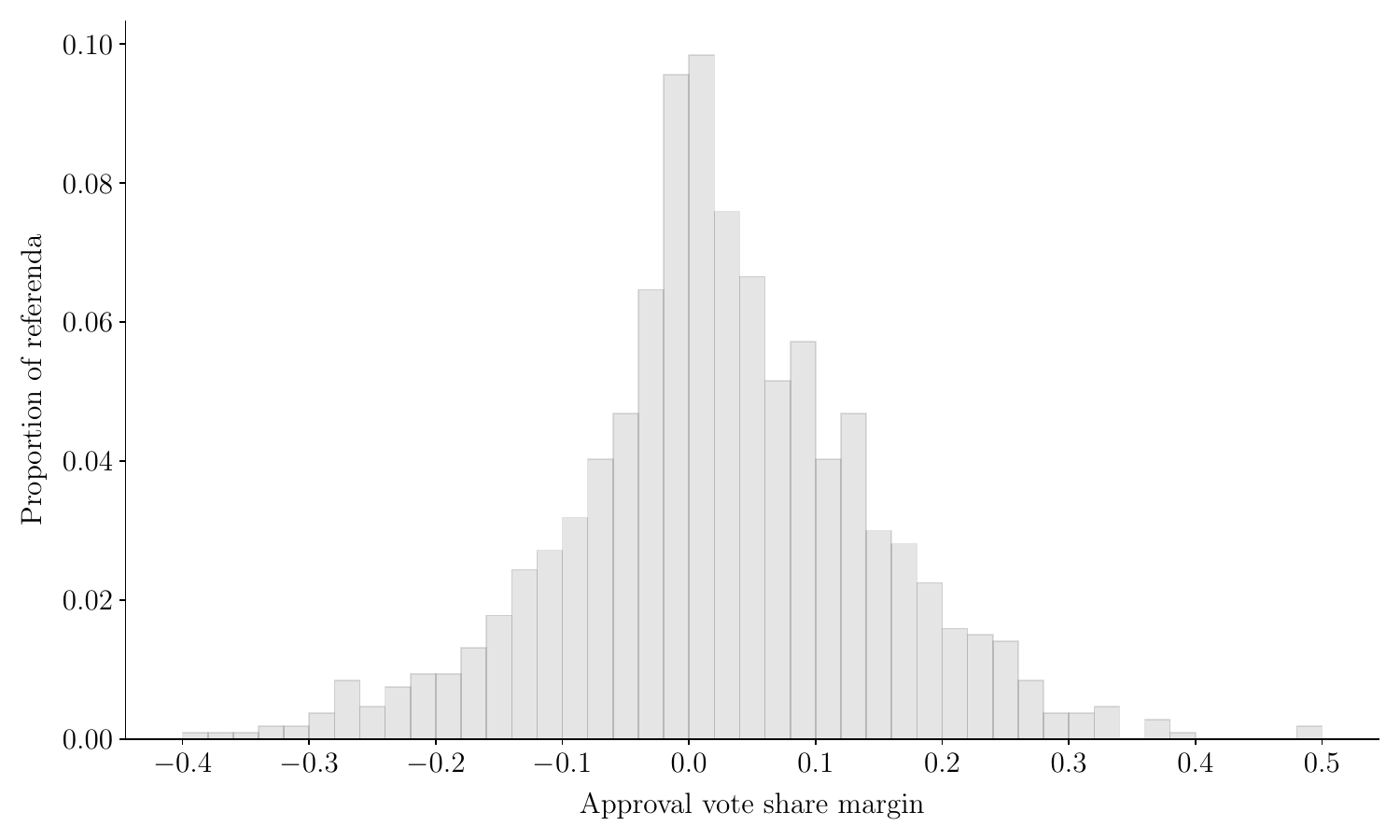}
	\end{center}
	\vspace{-2mm}
	\begin{footnotesize}
		\begin{spacing}{1}
			\noindent
			\textsc{Notes}: This figure displays a histogram of the approval vote share margin, defined as the difference between the share of votes in favor of the proposed expenditure measure and the 50 percent approval threshold, for 1,067 referenda held by Wisconsin school districts between 1993 and 2015.
		\end{spacing}
	\end{footnotesize}
\end{figure}

In the potential outcome model of Section \ref{sec_multiperiod}, $Y_t$ is measured at the end of period $t$. Consistent with this timing, I use the housing value observed in the calendar year following the referendum as the outcome at relative year zero.\footnote{This timing is particularly relevant for referenda held during November general elections.} To construct a balanced panel in event time, I restrict the sample to referenda with outcomes observed in all relative years from $-5$ to 5.\footnote{In addition, I follow \cite{baron2022} in retaining referenda with the smallest absolute margin whenever a school district holds more than one referendum in the same year.} The matched sample includes 1,067 referenda, of which 58.8 percent were approved. The average approval vote share margin is 2.4 percentage points, with an average of 9.8 percentage points among approved referenda and $-8.3$ percentage points among those that were rejected. Figure \ref{fig_runningvar} displays a histogram of the approval vote share margin. To assess the validity of the design, I test for discontinuities in the density of the running variable at the cutoff using the local polynomial density estimators developed by \cite{cattaneojanssonma2020}. Reassuringly, the null hypothesis of continuity of the density at the cutoff is not rejected ($p$-value = 0.91).

\subsection{Results}

\begin{figure}[!t]
	\caption{Estimated $\mathrm{ADTE}$s of Referendum Approval on Housing Prices}\label{fig_eventplot}
	\vspace{2mm}
	\begin{minipage}[b]{0.495\textwidth}
		\begin{center}
			\includegraphics[width=\textwidth]{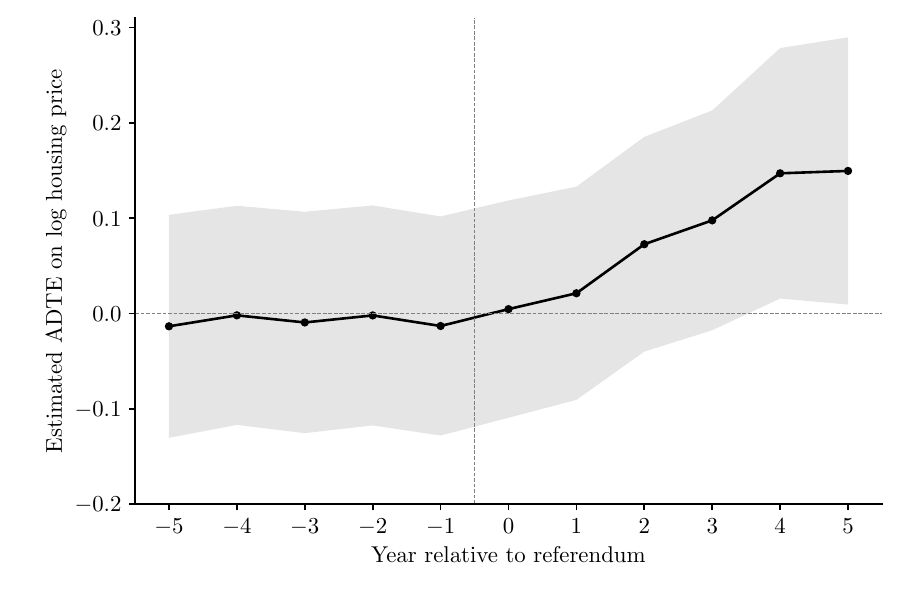}
			\subcaption{Aggregated Cohort-Specific Estimates}
		\end{center}
	\end{minipage}
	\hfill
	\begin{minipage}[b]{0.495\textwidth}
		\begin{center}
			\includegraphics[width=\textwidth]{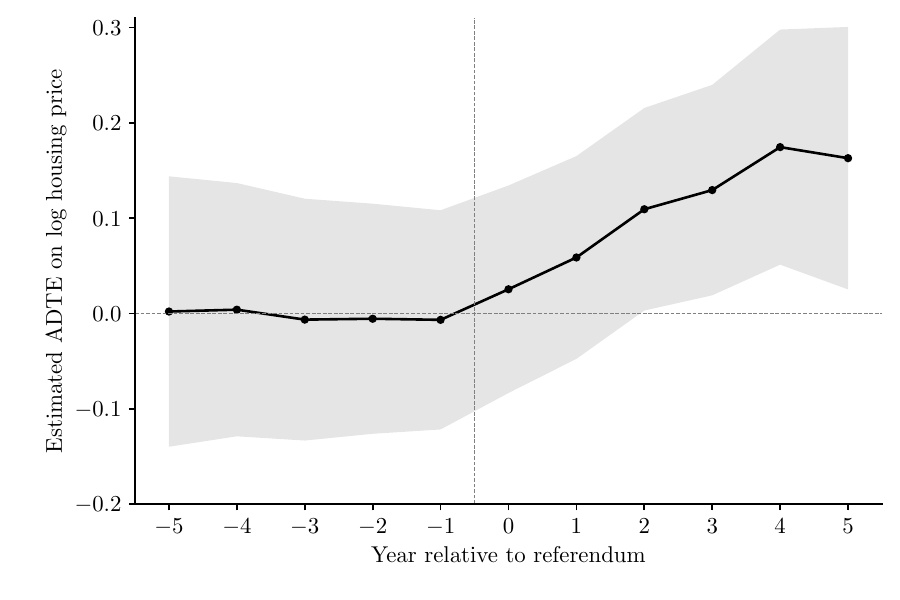}
			\subcaption{Pooled Estimates}
		\end{center}
	\end{minipage}
	\vspace{-4mm}
	\begin{footnotesize}
		\begin{spacing}{1}
			\noindent
			\textsc{Notes:} 
			This figure displays estimates of average direct treatment effects ($\mathrm{ADTE}$s) of the approval of school district expenditure referenda on housing prices in Wisconsin between 1995 and 2015. Estimates for relative years 1 through 5 correspond to the estimand stated in equation~\eqref{apte_gen_lie}. Estimates for relative years -5 through 0 are based on standard (i.e., static) local linear regression discontinuity estimators; those for years -5 to -1 serve as placebo tests. Panel~(a) reports aggregated estimates of cohort-specific effects, where cohorts are defined as groups of adjacent years. Aggregated $\mathrm{ADTE}$s are computed as weighted averages of cohort-specific $\mathrm{ADTE}$s, with weights proportional to each cohort's share of the sample. Panel~(b) displays estimates from a pooled specification in which a single parameter is estimated for each relative year, using data from all cohorts combined. Shaded gray areas denote 90 percent confidence intervals. Standard errors are computed using the nearest-neighbor method described in Section~\ref{sec:stderr}, with tuning parameter $j^* = 3$. In Panel~(a), standard errors are aggregated using the Delta method.
		\end{spacing}
	\end{footnotesize}
\end{figure}

I leverage Assumption~\ref{ass_mark} by defining a cohort as the set of jurisdictions that held a referendum in a given year and did not approve a ballot measure in any of the preceding $k=5$ years. Given the limited number of referenda in individual years, estimating a separate parameter for each cohort would be inefficient. To address this constraint, I group adjacent referendum years into three intervals that broadly follow the national housing cycle. These intervals cover the expansion preceding the financial crisis (1995-2006), the housing downturn (2007-2011), and the subsequent recovery (2012-2015) (\citealt{glicklansingmolitor2015}). To quantify statistical uncertainty, I compute standard errors using the nearest-neighbor method described in Section~\ref{sec:stderr}, setting the tuning parameter to $j^* = 3$. This choice corresponds to the default setting in the \texttt{rdrobust} package (\citealt{rdrobust2014}, \citealt{rdrobust2015}, \citealt{rdrobust2017}). Cohort-specific standard errors are then aggregated using the Delta method, as described in Supplementary Appendix \ref{appsec_aggrconfint}.

\begin{figure}[!t]
	\caption{Estimated $\mathrm{ADTE}$ of Referendum Approval on Housing Prices at $\tau = 5$}\label{fig_rdplot}
	\vspace{1mm}
	\begin{minipage}[b]{0.495\textwidth}
		\begin{center}
			\includegraphics[width=\textwidth]{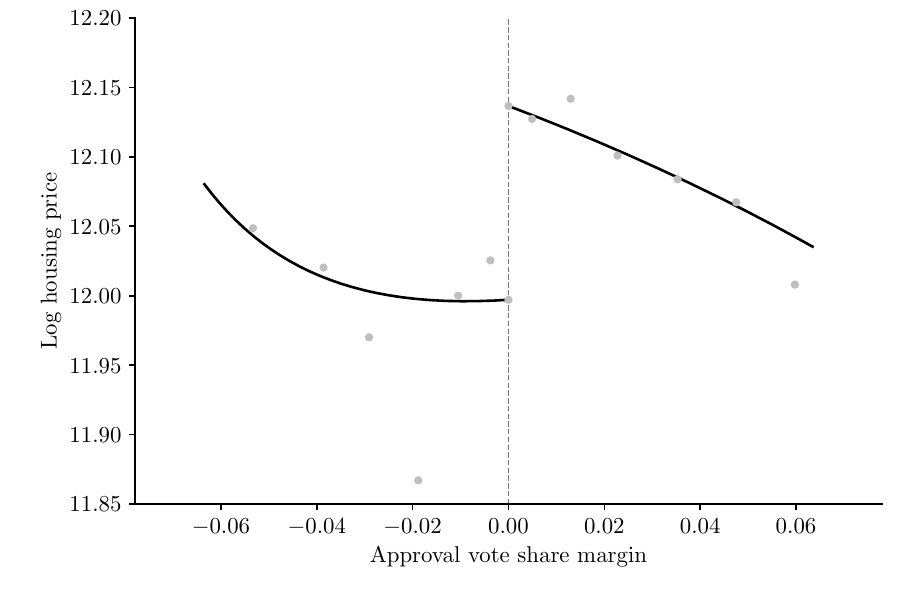}
			\subcaption{Aggregated Cohort-Specific Estimates}
		\end{center}
	\end{minipage}
	\hfill
	\begin{minipage}[b]{0.495\textwidth}
		\begin{center}
			\includegraphics[width=\textwidth]{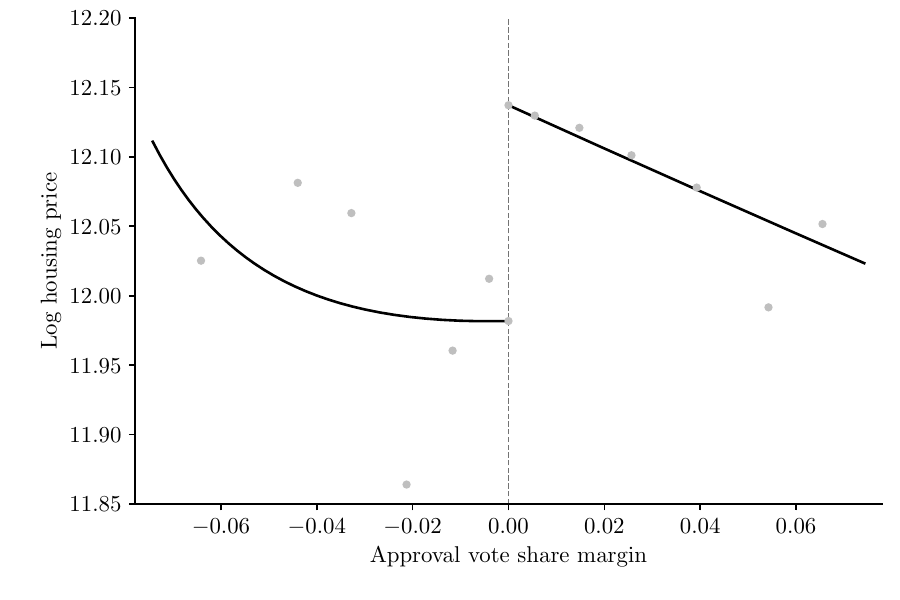}
			\subcaption{Pooled Estimates}
		\end{center}
	\end{minipage}
	\vspace{-4mm}
	\begin{footnotesize}
		\begin{spacing}{1}
			\noindent
			\textsc{Notes:} 
			This figure displays estimates of the five-year-ahead Average Direct Treatment Effect ($\mathrm{ADTE}_{5}$) of the approval of school district expenditure referenda on housing prices in Wisconsin between 1995 and 2015. Black curves combine local linear fitted values for $Y$, $W$, and $D$. At the cutoff, their limits are $\widehat{\mu}^{(0)}_{Y-,1} + \widehat{\mu}^{(0)}_{W-,1} \slash \widehat{\mu}^{(0)}_{D-,1}$ on the left and $\widehat{\mu}^{(0)}_{Y+,1} + \widehat{\mu}^{(0)}_{W+,1} \slash \widehat{\mu}^{(0)}_{D+,1}$ on the right. Away from the cutoff, the same calculation uses the fitted values evaluated at each approval vote share margin. The difference between the two limits at the cutoff equals the estimated $\mathrm{ADTE}_{5}$ prior to bias correction. Panel~(a) averages cohort-specific curves, with weights equal to each cohort's share of the estimation sample. Panel~(b) uses a single set of local linear regressions estimated from all cohorts combined. Gray dots away from the cutoff summarize observations in quantile bins on each side. Dots at the cutoff mark the fitted limits. In Panel~(b), each dot is computed as the bin mean of $Y$ plus the ratio of the bin means of $W$ and $D$. In Panel~(a), this calculation is performed separately within each cohort and bin, and the resulting values are averaged using the cohort weights.
		\end{spacing}
	\end{footnotesize}
\end{figure}

Figure~\ref{fig_eventplot} presents event-study-style plots that show average direct treatment effects for each relative year, based on two specifications: one that aggregates cohort-specific estimates, and another that pools observations across all cohorts. The two approaches yield comparable results in both magnitude and precision. As a placebo exercise, I estimate differences in pre-referendum housing prices at the approval cutoff using standard static local linear regression discontinuity estimators. Reassuringly, the estimates are close to zero and not statistically significant, suggesting no systematic differences in housing prices between jurisdictions that later approved versus rejected referenda. On average, the approval of school expenditure measures leads to cumulative increases in housing prices of approximately 0.15 log points after five years. These estimates are consistent with recent findings in the literature (\citealt{bilaschon2024}).

Figure~\ref{fig_rdplot} provides a graphical representation of the estimator in equation~\eqref{estimator} at horizon $\tau=5$. Specifically, the figure plots estimates of $\mu_Y \left ( r \right ) + \mu_W \left ( r \right ) \slash \mu_D \left ( r \right )$, constructed from separate local linear regressions for $Y$, $W$, and $D$ on each side of the approval cutoff. The difference between the right and left fitted limits at the cutoff yields the estimated $\mathrm{ADTE}_{5}$ prior to bias correction. Panel~(a) averages the cohort-specific curves using cohort sample shares as weights, while Panel~(b) uses the pooled sample. Both specifications show a positive discontinuity at the cutoff, indicating that referendum approval increased housing prices five years later.

\begin{table}[!t]
	\setstretch{1}
	\begin{center}
		\caption{Checks on the Plausibility of the Common Trends Assumption}\label{table_wi_testct}
		\begingroup
\renewcommand{\arraystretch}{1.4}
\setlength{\tabcolsep}{5pt}
\begin{tabular}{>{\centering\arraybackslash}m{2.3cm}*{3}{>{\centering\arraybackslash}m{1.8cm}}}
\hline\hline
\multirow{2}{2.3cm}[-1pt]{\centering \makebox[\linewidth][c]{Pre-Period} Pair $(u,v)$} & \multicolumn{3}{c}{Type of Comparison Group} \\
\cline{2-4}
 & \makebox[\linewidth][c]{Aggregate} & \makebox[\linewidth][c]{Frequency} & Timing \\
\midrule
\addlinespace[2pt]
\multicolumn{4}{@{}l@{}}{\textit{Panel A: Four-Year Differences}} \\
\addlinespace[2pt]
$(-5,-1)$ & $0.978$ & $0.336$ & $0.994$ \\
\midrule
\addlinespace[2pt]
\multicolumn{4}{@{}l@{}}{\textit{Panel B: Three-Year Differences}} \\
\addlinespace[2pt]
$(-5,-2)$ & $0.850$ & $0.338$ & $0.844$ \\
$(-4,-1)$ & $0.815$ & $0.278$ & $0.821$ \\
\midrule
\addlinespace[2pt]
\multicolumn{4}{@{}l@{}}{\textit{Panel C: Two-Year Differences}} \\
\addlinespace[2pt]
$(-5,-3)$ & $0.331$ & $0.271$ & $0.584$ \\
$(-4,-2)$ & $0.837$ & $0.207$ & $0.781$ \\
$(-3,-1)$ & $0.621$ & $0.483$ & $0.881$ \\
\midrule
\addlinespace[2pt]
\multicolumn{4}{@{}l@{}}{\textit{Panel D: One-Year Differences}} \\
\addlinespace[2pt]
$(-5,-4)$ & $0.123$ & $0.339$ & $0.177$ \\
$(-4,-3)$ & $0.973$ & $0.138$ & $0.303$ \\
$(-3,-2)$ & $0.543$ & $0.392$ & $0.685$ \\
$(-2,-1)$ & $0.810$ & $0.679$ & $0.809$ \\
\hline\hline
\end{tabular}
\endgroup

	\end{center}
	\begin{footnotesize}
		\begin{spacing}{1}
			\noindent
			\textsc{Notes:} This table reports $p$-values of joint Wald tests comparing the average pre-treatment outcome change $Y_{g+v} - Y_{g+u}$ across groups defined by subsequent treatment paths. Each row corresponds to a pair of relative years $(u,v)$, with $-5 \leq u < v \leq -1$. All tests use the pooled sample. The reference group follows the all-zero path $D_{g+1} = \dots = D_{g+5} = 0$. Aggregate comparisons pool all nonzero paths into a single comparison group. Frequency-based comparisons divide these paths into two groups: exactly one treatment and at least two treatments between $g+1$ and $g+5$. Timing-based comparisons divide these paths according to whether the first subsequent treatment occurs between $g+1$ and $g+3$ or between $g+4$ and $g+5$. Each test jointly imposes equality of average outcome changes between the reference group and each comparison group at both cutoff limits. Aggregate tests therefore have two restrictions, while frequency- and timing-based tests have four.
		\end{spacing}
	\end{footnotesize}
\end{table}

As a final validation exercise, I assess the plausibility of the local common trends assumption, as discussed in Section~\ref{sec_check_ct}. The estimation sample conditions on $D_{g-5} = \dots = D_{g-1} = 0$. Thus, I examine all ten pairs of relative years $-5 \leq u < v \leq -1$ and compare average pre-treatment outcome changes $Y_{g+v} - Y_{g+u}$ across groups defined by subsequent treatment paths. The all-zero subsequent treatment path serves as the reference group. The comparison groups distinguish whether subsequent treatment occurs, how often it occurs, and when it first occurs. Table~\ref{table_wi_testct} reports the $p$-values of joint Wald tests estimated from the pooled sample. Each test jointly evaluates the restrictions for all comparison groups on both sides of the cutoff. None of the joint null hypotheses is rejected at conventional significance levels, providing suggestive support for the plausibility of local common trends in this setting.

\section{Conclusion}\label{sec_concl}

This paper proposed a novel argument to point identify economically interpretable intertemporal treatment effects in dynamic regression discontinuity designs. I developed a dynamic potential outcomes model and reformulated two assumptions from the difference-in-differences literature, no anticipation and common trends, to identify impulse response parameters at the cutoff. The estimand of each target parameter can be expressed as the sum of two static RD outcome contrasts, allowing for nonparametric estimation with standard local polynomial methods. I derived robust bias-corrected confidence intervals and bandwidth formulas that minimize the estimator's mean squared error. I also proposed a nonparametric approach to aggregate treatment effects across focal periods and prior treatment histories, relying on a limited path independence restriction to reduce the dimensionality of the parameter space. I applied the proposed method to study the dynamic effects of school district expenditure referenda on housing prices in Wisconsin. The results indicate that, on average, referendum approval is positively capitalized into housing prices.

% Proofs

\section*{Proofs}

\subsection*{Propositions \ref{prop_rd} and \ref{prop_rd1}}

Consider the right-hand limit entering $\mathrm{RD}_{Y_t}$:
\begin{align}
	& \lim_{r \downarrow c} \E \left [ Y_t | R_1 = r \right ] \notag \\
	& = \lim_{r \downarrow c} \sum_{\left ( d_2, \dots, d_{\overline{t}} \right )} \E \left [ P \left ( 1, d_2, \dots, d_{\overline{t}} \right ) Y_t \left ( 1,d_2, \dots, d_{\overline{t}} \right )  | R_1 = r \right ] \\
	& = \lim_{r \downarrow c} \sum_{\left ( d_2, \dots, d_{\overline{t}} \right )} \E \left [ Y_t \left ( 1,d_2, \dots, d_{\overline{t}} \right )  | R_1 = r, P \left ( 1, d_2, \dots, d_{\overline{t}} \right ) = 1 \right ] \E \left [ P \left ( 1, d_2, \dots, d_{\overline{t}} \right ) | R_1 = r \right ] \\
	& = \sum_{\left ( d_2, \dots, d_{\overline{t}} \right )} \E \left [ Y_t \left ( 1,d_2, \dots, d_{\overline{t}} \right )  | R_1 = c, P \left ( 1, d_2, \dots, d_{\overline{t}} \right ) = 1 \right ] \E \left [ P \left ( 1, d_2, \dots, d_{\overline{t}} \right ) | R_1 = c \right ]
\end{align}
The first equality exploits the threshold-crossing definition $D_1 \equiv \mathbb{I} \left [ R_1 \geq c \right ]$ and the switching equation \eqref{potout_gen2}: conditional on $R_1 = r > c$, all terms associated with $d_1=0$ vanish. The second equality follows from an application of the Law of Iterated Expectations. The third equality applies Assumption \ref{ass_cont1} and the algebraic properties of limits. A symmetric argument identifies the left limit. Taking the difference yields the expression in Proposition \ref{prop_rd1}. Setting $\overline t=2$, the path indicator becomes $P \left ( d_1,d_2 \right ) = \mathbb{I} \left [ D_2 \left (d_1 \right ) = d_2 \right ]$ and Assumption \ref{ass_cont1} reduces to Assumption \ref{ass_cont0}. The same argument therefore proves Proposition \ref{prop_rd}.

\subsection*{Propositions \ref{prop_apte0} and \ref{prop_apte0_gen}}

Letting $t=1$, the preceding proof gives
\begin{align}
	& \lim_{r \downarrow c} \E \left [ Y_1 | R_1 = r \right ] \notag \\
	& = \sum_{\left ( d_2, \dots, d_{\overline{t}} \right )} \E \left [ Y_1 \left ( 1,d_2, \dots, d_{\overline{t}} \right )  | R_1 = c, P \left ( 1, d_2, \dots, d_{\overline{t}} \right ) = 1 \right ] \E \left [ P \left ( 1, d_2, \dots, d_{\overline{t}} \right ) | R_1 = c \right ] \\
	& = \sum_{\left ( d_2, \dots, d_{\overline{t}} \right )} \E \left [ Y_1 \left ( 1, 0_{\overline{t}-1} \right )  | R_1 = c, P \left ( 1, d_2, \dots, d_{\overline{t}} \right ) = 1 \right ] \E \left [ P \left ( 1, d_2, \dots, d_{\overline{t}} \right ) | R_1 = c \right ] \\
	& = \E \left [ Y_1 \left ( 1, 0_{\overline{t}-1} \right )  | R_1 = c \right ]
\end{align}
The second equality applies the local no anticipation restriction in Assumption \ref{ass_noant1}. The third equality follows from the Law of Iterated Expectations, since the path indicators partition the possible continuations for a fixed initial treatment state. By a symmetric argument, the left limit identifies $\E \left [ Y_1 \left ( 0, 0_{\overline{t}-1} \right )  | R_1 = c \right ]$. Thus, $\mathrm{ADTE}_{1,0} = \mathrm{RD}_{Y_1}$, which proves Proposition \ref{prop_apte0_gen}. Setting $\overline{t} = 2$, Assumptions \ref{ass_noant1} and \ref{ass_cont1} reduce to Assumptions \ref{ass_noant} and \ref{ass_cont0}, respectively, yielding Proposition \ref{prop_apte0}.

\subsection*{Propositions \ref{prop_apte1} and \ref{prop_apte1_gen}}

For $\tau \in \left \{ 1, \dots, \overline{t}-1 \right \}$, fix $t=1+\tau$. Let $\Delta_{1,t} \equiv Y_t - Y_1$ and $\Delta_{1,t} \left ( d_1, \dots, d_{\overline{t}} \right ) \equiv Y_t \left ( d_1, \dots, d_{\overline{t}} \right ) - Y_1 \left ( d_1, \dots, d_{\overline{t}} \right )$ for any $\left ( d_1, \dots, d_{\overline{t}} \right )$. By an application of the Law of Iterated Expectations and Assumption \ref{ass_ct1}, the first term in $\mathrm{ADTE}_{1,\tau}$ can be expressed as
\begin{align}
	\E \left [ Y_t \left ( 1,0_{\overline t-1} \right ) | R_1=c \right ] = \E \left [ Y_1 \left ( 1,0_{\overline t-1} \right ) | R_1 = c \right ] + \E \left [ \Delta_{1,t}(1,0_{\overline t-1}) | R_1=c, P \left ( 1,0_{\overline t-1} \right ) = 1 \right ]
\end{align}
Assumption \ref{ass_ct1} makes the mean of $\Delta_{1,t} \left ( 1,0_{\overline t-1} \right )$ at $R_1 = c$ invariant across potential continuation paths. Thus, its mean conditional only on $R_1 = c$ is equal to its mean in the stratum defined by $P \left ( 1,0_{\overline t-1} \right ) = 1$. As shown in the preceding proof, the first term is identified by $\lim_{r \downarrow c} \E \left [ Y_1 | R_1 = r \right ]$. To identify the second term, consider
\begin{align}
	& \lim_{r \downarrow c} \E \left [ \Delta_{1,t} | R_1=r, D_2=0, \dots, D_t=0 \right ] \notag \\
	& = \sum_{(d_{t+1},\dots,d_{\overline t})} \E \left [ \Delta_{1,t} \left ( 1, 0_{t-1}, d_{t+1}, \dots, d_{\overline t} \right ) | R_1=c, P \left ( 1, 0_{t-1}, d_{t+1}, \dots, d_{\overline t} \right ) = 1 \right ] \notag \\
	& \times \frac{ \E \left [ P \left ( 1, 0_{t-1}, d_{t+1}, \dots, d_{\overline t} \right ) | R_1 = c \right ]}{
	\E \left [ P \left ( 1,0_{t-1} \right ) | R_1 = c \right ]} \\
	& = \sum_{(d_{t+1},\dots,d_{\overline t})} \E \left [ \Delta_{1,t} \left ( 1,0_{\overline{t}-1} \right ) | R_1=c, P \left ( 1, 0_{t-1}, d_{t+1}, \dots, d_{\overline t} \right ) = 1 \right ] \notag \\
	& \times \frac{ \E \left [ P \left ( 1, 0_{t-1}, d_{t+1}, \dots, d_{\overline t} \right ) | R_1 = c \right ]}{
	\E \left [ P \left ( 1,0_{t-1} \right ) | R_1 = c \right ]} \\
	& = \E \left [ \Delta_{1,t} \left ( 1,0_{\overline t-1} \right ) | R_1=c, P \left ( 1,0_{\overline t-1} \right ) = 1 \right ]
\end{align}
where the sums range over $ \left ( d_{t+1}, \dots, d_{\overline t} \right ) \in \left \{ 0,1 \right \}^{\overline t-t}$. The first equality follows from the threshold-crossing definition of the treatment, the switching equation \eqref{potout_gen2}, the Law of Iterated Expectations, and Assumption \ref{ass_cont1}. Each ratio of probabilities measures the share, at $R_1 = c$, of the continuation path $\left ( d_{t+1}, \dots, d_{\overline{t}} \right )$ among units whose potential treatment path satisfies $P \left ( 1,0_{t-1} \right ) = 1$. The second equality applies Assumption \ref{ass_noant1} separately to $Y_t$ and $Y_1$, holding the conditioning path fixed. The third equality invokes Assumption \ref{ass_ct1} and the fact that the probability ratios sum to one. When $t = \overline{t}$, the sum contains a single term corresponding to the empty continuation and thus the second equality is the identity. Combining these identification results,
\begin{align}
	\E \left [ Y_t \left ( 1,0_{\overline t-1} \right ) | R_1=c \right ] = \lim_{r\downarrow c} \E \left [ Y_1 | R_1=r \right ] + \lim_{r\downarrow c} \E \left [ Y_t - Y_1 | R_1=r, D_2=0, \dots, D_t=0 \right ]
\end{align}
By a symmetric argument, the corresponding left limits identify $\E \left [ Y_t \left ( 0, 0_{\overline t-1} \right ) | R_1 = c \right ]$. Subtracting the left limits from the right limits yields Proposition \ref{prop_apte1_gen}. Setting $\overline t=2$, Assumptions \ref{ass_cont1}, \ref{ass_noant1}, and \ref{ass_ct1} reduce to Assumptions \ref{ass_cont0}, \ref{ass_noant}, and \ref{ass_ct0}, respectively, yielding Proposition \ref{prop_apte1}.

\subsection*{Proposition \ref{prop_confint}}

The proof closely adapts the argument of Theorem A.2 in \cite{cct2014supp}. First, I linearize $\widehat{\theta} \left ( h_n \right )$ and estimate its leading bias, thereby obtaining an expression for $\widehat{\theta}^{\textsc{bc}} \left ( h_n, b_n \right )$. Second, I compute the variance of $\widetilde{\theta}^{\textsc{bc}} \left ( h_n, b_n \right )$, the linearized version of the bias-corrected estimator. This variance accounts for the variance of the estimated bias and its covariance with the original estimator. Under the stated bandwidth conditions, the remaining bias and the linearization remainders are negligible relative to the standard deviation of the linearized bias-corrected estimator. The result follows by specializing parts (V) and (D) of Theorem \ref{thm_asydist_bias} in the Supplemental Appendix to $\nu=0$, $p=1$, and $q=2$. Section \ref{app_prop_confint} provides the corresponding expression for the variance.

\subsection*{Proposition \ref{prop_optbandwidth}}

The proof closely adapts the MSE argument of Lemma 2 in \cite{cct2014}. First, I derive the conditional bias and variance of $\widetilde{\theta} \left ( h_n \right )$, the linearized estimator of the parameter of interest. The conditional MSE is obtained as the sum of squared bias and variance. Minimizing its leading terms yields the MSE-optimal main bandwidth. Second, I derive the conditional bias and variance of $\widetilde{\textsc{B}}_h \left ( b_n \right )$. This estimator combines estimates of second-order derivatives from local quadratic regressions, retaining the population coefficients from the original linearization. Minimizing the leading terms of its conditional MSE yields the MSE-optimal pilot bandwidth. The main bandwidth result follows by specializing Theorem \ref{thm_optbandwidth} in the Supplementary Appendix to $\nu=0$ and $p=1$. The pilot bandwidth result follows by combining the bias and covariance expansions for the individual estimators of the second-order derivatives. Section \ref{app_prop_bandwidth} provides the corresponding derivations and population coefficients.

% References
\begin{spacing}{1} 
	\def\bibfont{\small}
	\putbib
\end{spacing}
\clearpage
\end{bibunit}

% Supplementary Appendix
\begin{bibunit}

\clearpage

\setcounter{page}{1}
\renewcommand{\thepage}{S.\arabic{page}}

\appendix

%\pagenumbering{arabic}
\renewcommand{\thesubsection}{\Alph{section}.\arabic{subsection}}
\renewcommand{\thesubsubsection}{\Alph{section}.\arabic{subsection}.\arabic{subsubsection}}

\setcounter{figure}{0} \renewcommand{\thefigure}{\Alph{section}.\arabic{figure}}
\setcounter{table}{0} \renewcommand{\thetable}{\Alph{section}.\arabic{table}}
\numberwithin{equation}{section}
\setcounter{theorem}{0} \renewcommand{\thetheorem}{\Alph{section}.\arabic{theorem}}

\phantomsection
\addcontentsline{toc}{section}{Supplementary Appendix}

\begin{spacing}{1}
\begin{center}
	{\LARGE Supplementary Appendix to\\[0.5em]
	``Dynamic Regression Discontinuity: An Event-Study Approach''\par}
	\vspace{2em}
	{\large Francesco Ruggieri \par}
	\vspace{2em}
	{\large \today \par}
\end{center}
\end{spacing}
\listofappendixcontents

\appsection{Identification}

\appsubsection{Proof of Lemma \ref{lemma_recursive}}\label{app_lemma_recursive}

Identification of $\mathrm{ADTE}_{1,0}$ follows directly from Proposition \ref{prop_apte0}. Since Assumptions \ref{ass_cont0} and \ref{ass_noant} hold,
\begin{equation}
	\mathrm{ADTE}_{1,0} = \text{RD}_{Y_{1}} \equiv \lim_{r \downarrow c} \E \left [ Y_1 | R_1 = r \right ] - \lim_{r \uparrow c} \E \left [ Y_1 | R_1 = r \right ]
\end{equation}
To establish identification of $\mathrm{ADTE}_{1,1}$, consider the first summation in the expression for $\mathrm{RD}_{Y_2}$ provided by Proposition \ref{prop_rd}:
\begin{align}
	& \sum_{d_2 \in \left \{ 0,1 \right \}} \E \left [ Y_2 \left ( 1,d_2 \right )  | R_1 = c, D_2 \left ( 1 \right ) = d_2 \right ] \times \Prob \left ( D_2 \left ( 1 \right ) = d_2 | R_1 = c \right ) \notag \\
	& = \sum_{d_2 \in \left \{ 0,1 \right \}} \E \left [ Y_2 \left ( 1,d_2 \right ) \mathbb{I} \left [ D_2 \left ( 1 \right ) = d_2 \right ] | R_1 = c \right ] \\
	& = \E \left [ Y_2 \left ( 1,0 \right ) \left ( 1 - D_2 \left ( 1 \right ) \right ) | R_1 = c \right ] + \E \left [ Y_2 \left ( 1,1 \right ) D_2 \left ( 1 \right ) | R_1 = c \right ] \\
	& = \E \left [ Y_2 \left ( 1,0 \right ) | R_1 = c \right ] + \E \left [ \left ( Y_2 \left ( 1,1 \right ) - Y_2 \left ( 1,0 \right ) \right ) D_2 \left ( 1 \right ) | R_1 = c \right ] \\
	& = \E \left [ Y_2 \left ( 1,0 \right ) | R_1 = c \right ] \\
	& + \E \left [ Y_2 \left ( 1,1 \right ) - Y_2 \left ( 1,0 \right ) | R_1 = c, D_2 \left ( 1 \right ) = 1 \right ] \times \Prob \left ( D_2 \left ( 1 \right ) = 1 | R_1 = c \right ) \notag \\
	& = \E \left [ Y_2 \left ( 1,0 \right ) | R_1 = c \right ] + \E \left [ Y_2 \left ( 1,1 \right ) - Y_2 \left ( 1,0 \right ) | R_1 = c \right ] \times \Prob \left ( D_2 \left ( 1 \right ) = 1 | R_1 = c \right )
\end{align}
The first and fourth equalities follow from the Law of Iterated Expectations, the second exploits the binary nature of $D_2 \left ( 1 \right )$, and the third rearranges terms. The final equality follows from the mean independence assumption on instantaneous second-period treatment effects. By a symmetric argument,
\begin{align}
	& \sum_{d_2 \in \left \{ 0,1 \right \}} \E \left [ Y_2 \left ( 0,d_2 \right )  | R_1 = c, D_2 \left ( 0 \right ) = d_2 \right ] \times \Prob \left ( D_2 \left ( 0 \right ) = d_2 | R_1 = c \right ) \notag \\
	& = \E \left [ Y_2 \left ( 0,0 \right ) | R_1 = c \right ] + \E \left [ Y_2 \left ( 0,1 \right ) - Y_2 \left ( 0,0 \right ) | R_1 = c \right ] \times \Prob \left ( D_2 \left ( 0 \right ) = 1 | R_1 = c \right )
\end{align}
Combining these expressions yields
\begin{align}
	\mathrm{RD}_{Y_2} & = \E \left [ Y_2 \left ( 1,0 \right ) - Y_2 \left ( 0,0 \right ) | R_1 = c \right ] \\
	& + \E \left [ Y_2 \left ( 1,1 \right ) - Y_2 \left ( 1,0 \right ) | R_1 = c \right ] \times \Prob \left ( D_2 \left ( 1 \right ) = 1 | R_1 = c \right ) \\
	& - \E \left [ Y_2 \left ( 0,1 \right ) - Y_2 \left ( 0,0 \right ) | R_1 = c \right ] \times \Prob \left ( D_2 \left ( 0 \right ) = 1 | R_1 = c \right )
\end{align}
Then, by the definition of $\mathrm{ADTE}_{1,1}$ and the assumption that all instantaneous average treatment effects are equal to each other, 
\begin{align}
	\mathrm{RD}_{Y_2} & = \mathrm{ADTE}_{1,1} + \E \left [ Y_1 \left ( 1,0 \right ) - Y_1 \left ( 0,0 \right ) | R_1 = c \right ] \times \E \left [ D_2 \left ( 1 \right ) - D_2 \left ( 0 \right ) | R_1 = c \right ] \\
	& = \mathrm{ADTE}_{1,1} + \mathrm{ADTE}_{1,0} \times \E \left [ D_2 \left ( 1 \right ) - D_2 \left ( 0 \right ) | R_1 = c \right ] \\
	& = \mathrm{ADTE}_{1,1} + \mathrm{RD}_{Y_1} \mathrm{RD}_{D_2}
\end{align}
where the last equality follows from the identification of $\mathrm{ADTE}_{1,0}$ established above and the continuity of potential second-period treatment probabilities under Assumption \ref{ass_cont0}. Rearranging terms yields the desired result.

\appsection{Statistical Inference}\label{app_statinf}

In this section, I state and prove more general versions of Propositions \ref{prop_confint} and \ref{prop_optbandwidth}. These theorems apply to derivatives of arbitrary order of the conditional means in the estimator of interest $\widehat{\theta} \left ( h_n \right )$. In a similar fashion, I also develop statistical theory to assess the plausibility of Assumption \ref{ass_ct1}. The notation, intermediate propositions, and structure of proofs largely mirror those in \cite{cct2014supp}. 

\appsubsection{Notation}\label{appsec_notation}

For any pair of random variables $A$ and $B$, consider the following conditional moments:
\begin{align}
	\mu_{A} \left ( r \right ) \equiv \E \left [ A | R = r \right ] \qquad \sigma^2_{AA} \left ( r \right ) \equiv \Var \left [ A | R = r \right ] \qquad \sigma^2_{AB} \left ( r \right ) \equiv \Cov \left [ A,B | R=r \right ]
\end{align}
For any nonnegative integer $s$, the $s$th order derivative of the conditional mean with respect to the running variable is
\begin{align}
	\mu^{(s)}_{A} \left ( r \right ) \equiv \frac{\partial^s \mu_{A} \left ( r \right )}{\partial r^s}
\end{align}
Clearly, $\mu^{(0)}_{A} \left ( r \right ) = \mu_{A} \left ( r \right )$. Then define the following limits for the running variable to the deterministic cutoff $c$:
\begin{align}
	\mu^{(s)}_{A+} \equiv \lim_{r \downarrow c} \mu^{(s)}_{A} \left ( r \right ) \quad & \quad \mu^{(s)}_{A-} \equiv \lim_{r \uparrow c} \mu^{(s)}_{A} \left ( r \right ) \\
	\sigma^2_{AA+} \equiv \lim_{r \downarrow c} \sigma^2_{AA} \left ( r \right ) \quad & \quad \sigma^2_{AA-} \equiv \lim_{r \uparrow c} \sigma^{2}_{AA} \left ( r \right ) \\
	\sigma^2_{AB+} \equiv \lim_{r \downarrow c} \sigma^2_{AB} \left ( r \right ) \quad & \quad \sigma^2_{AB-} \equiv \lim_{r \uparrow c} \sigma^2_{AB} \left ( r \right )
\end{align}
Focusing on the conditional mean, let $0 \leq s \leq p$ and define
\begin{align}
	\beta_{A+,p} \equiv \left [ \mu_{A+}, \mu^{(1)}_{A+}, \frac{\mu^{(2)}_{A+}}{2}, \dots, \frac{\mu^{(p)}_{A+}}{p!} \right ]' \qquad \beta_{A-,p} \equiv \left [ \mu_{A-}, \mu^{(1)}_{A-}, \frac{\mu^{(2)}_{A-}}{2}, \dots, \frac{\mu^{(p)}_{A-}}{p!} \right ]'
\end{align}
so that
\begin{align}
	\mu^{(s)}_{A+} = s! e_{s}' \beta_{A+,p} \qquad \mu^{(s)}_{A-} = s! e_{s}' \beta_{A-,p}
\end{align}
where $e_s$ is the conformable $(s+1)$th unit vector. The estimand of interest is
\begin{align}
	\theta \equiv \mu^{(0)}_{Y+} - \mu^{(0)}_{Y-} + \frac{\mu^{(0)}_{W+}}{\mu^{(0)}_{D+}} - \frac{\mu^{(0)}_{W-}}{\mu^{(0)}_{D-}}
\end{align}
I consider the following generalization of the target parameter, with $\nu = 0$ corresponding to $\theta$:
\begin{align}
	\theta_{\nu} \equiv \mu^{(\nu)}_{Y+} - \mu^{(\nu)}_{Y-} + \frac{\mu^{(\nu)}_{W+}}{\mu^{(\nu)}_{D+}} - \frac{\mu^{(\nu)}_{W-}}{\mu^{(\nu)}_{D-}}
\end{align}
where $\nu$ is a nonnegative integer satisfying $0 \leq \nu \leq p$, and the denominator terms are nonzero. Let $h$ be a positive bandwidth sequence and let
\begin{align}
	k_h \left ( u \right ) = k \left ( \frac{|u|}{h} \right ) \Big \slash h
\end{align}
where $k \left ( \cdot \right )$ is the kernel function. I apply the kernel to the absolute distance from the cutoff, using the same weighting function on both sides. Then, given the polynomial basis $x_p \left ( r \right ) = \left [ 1, r, r^2, \dots, r^p \right ]'$, the local polynomial estimators of the conditional mean of $A$ are
\begin{align}
	\widehat{\beta}_{A+,p} \left ( h_n \right ) & \equiv \arg \min_{b \in \mathbb{R}^{p+1}} \sum_{i=1}^{n} \mathbb{I} \left [ R_i \geq c \right ] \left ( A_i - x_p \left ( R_i - c \right )' b \right )^2 k_{h_n} \left ( R_i - c \right ) \label{polest_plus} \\
	\widehat{\beta}_{A-,p} \left ( h_n \right ) & \equiv \arg \min_{b \in \mathbb{R}^{p+1}} \sum_{i=1}^{n} \mathbb{I} \left [ R_i < c \right ] \left ( A_i - x_p \left ( R_i - c \right )' b \right )^2 k_{h_n} \left ( R_i - c \right ) \label{polest_minus}
\end{align}
It follows that one can estimate the conditional $\nu$th derivative of a random variable $A$ with
\begin{align}
	\widehat{\mu}^{(\nu)}_{A+,p} \left ( h_n \right ) = \nu! e_{\nu}' \widehat{\beta}_{A+,p} \left ( h_n \right ) \qquad \widehat{\mu}^{(\nu)}_{A-,p} \left ( h_n \right ) = \nu! e_{\nu}' \widehat{\beta}_{A-,p} \left ( h_n \right )
\end{align}
As a consequence, a local polynomial estimator for the target parameter $\theta_\nu$ is
\begin{align}
	\widehat{\theta}_{\nu,p} \left ( h_n \right ) \equiv \widehat{\mu}^{(\nu)}_{Y+,p} \left ( h_n \right ) - \widehat{\mu}^{(\nu)}_{Y-,p} \left ( h_n \right ) + \frac{\widehat{\mu}^{(\nu)}_{W+,p} \left ( h_n \right )}{\widehat{\mu}^{(\nu)}_{D+,p} \left ( h_n \right )} - \frac{\widehat{\mu}^{(\nu)}_{W-,p} \left ( h_n \right )}{\widehat{\mu}^{(\nu)}_{D-,p} \left ( h_n \right )}
\end{align}
For any random variable $A$, define the data vectors
\begin{align}
	\bm{A} \equiv \left [ A_1, \dots, A_n \right ]' \qquad \bm{R} \equiv \left [ R_1, \dots, R_n \right ]'
\end{align}
Also define the vector of residuals
\begin{align}
	\bm{\varepsilon_{A}} \equiv \left [ \varepsilon_{A,1}, \dots, \varepsilon_{A,n} \right ]' \quad \text{ with }\quad \varepsilon_{A,i} \equiv A_i - \mu_A \left ( R_i \right )
\end{align}
Under independent sampling, the conditional covariance matrix of the residual vectors is
\begin{align}
	 \bm{\Sigma_{AB}} \equiv \E \left [ \bm{\varepsilon_{A}} \bm{\varepsilon_{B}}' | \bm{R} \right ] = \text{diag} \left ( \sigma^2_{AB} \left ( R_1 \right ), \dots, \sigma^2_{AB} \left ( R_n \right ) \right )
\end{align}
In addition, define the following intermediate data vectors and matrices:
\begin{align}
	\bm{X_{p} \left ( h_n \right )} & \equiv \left [ x_p \left ( \left ( R_1 - c \right ) \slash h_n \right ), \dots, x_p \left ( \left ( R_n - c \right ) \slash h_n \right ) \right ]' \\
	\bm{S_{q} \left ( h_n \right )} & \equiv \left [ \left ( \left ( R_1 - c \right ) \slash h_n \right )^q, \dots, \left ( \left ( R_n - c \right ) \slash h_n \right )^q \right ]' \\
	\bm{Z_{+} \left ( h_n \right )} & \equiv \text{diag} \left ( \mathbb{I} \left [ R_1 \geq c \right ] k_{h_n} \left ( R_1 - c \right ), \dots, \mathbb{I} \left [ R_n \geq c \right ] k_{h_n} \left ( R_n - c \right ) \right ) \\
	\bm{Z_{-} \left ( h_n \right )} & \equiv \text{diag} \left ( \mathbb{I} \left [ R_1 < c \right ] k_{h_n} \left ( R_1 - c \right ), \dots, \mathbb{I} \left [ R_n < c \right ] k_{h_n} \left ( R_n - c \right ) \right )
\end{align}
Finally, given a second positive bandwidth sequence $b_n$, define the scaled matrices
\begin{align}
	\Gamma_{+,p} \left ( h_n \right ) & \equiv \bm{X_{p} \left ( h_n \right )}' \bm{Z_{+} \left ( h_n \right )} \bm{X_{p} \left ( h_n \right )} \slash n \\
	\vartheta_{+,p,q} \left ( h_n \right ) & \equiv \bm{X_{p} \left ( h_n \right )}' \bm{Z_{+} \left ( h_n \right )} \bm{S_{q} \left ( h_n \right )} \slash n \\
	\Psi_{AB+,p,q} \left ( h_n,b_n \right ) & \equiv \bm{X_{p} \left ( h_n \right )}' \bm{Z_{+} \left ( h_n \right )} \bm{\Sigma_{AB}} \bm{Z_{+} \left ( b_n \right )} \bm{X_{q} \left ( b_n \right )} \slash n \\
	\Psi_{AB+,p} \left ( h_n \right ) & \equiv \Psi_{AB+,p,p} \left ( h_n,h_n \right )
\end{align}
The corresponding lower-side matrices are defined by replacing the $+$ subscripts with $-$ subscripts throughout. It follows that the vector of polynomial estimators in \eqref{polest_plus} and \eqref{polest_minus} can be expressed as
\begin{align}
	\widehat{\beta}_{A+,p} \left ( h_n \right ) & = H_p \left ( h_n \right ) \Gamma^{-1}_{+,p} \left ( h_n \right ) \bm{X_{p} \left ( h_n \right )}' \bm{Z_{+} \left ( h_n \right )} \bm{A} \slash n \\
	\widehat{\beta}_{A-,p} \left ( h_n \right ) & = H_p \left ( h_n \right ) \Gamma^{-1}_{-,p} \left ( h_n \right ) \bm{X_{p} \left ( h_n \right )}' \bm{Z_{-} \left ( h_n \right )} \bm{A} \slash n
\end{align}
with $H_p \left ( h_n \right ) \equiv \text{diag} \left ( 1, h_n^{-1}, h_n^{-2}, \dots, h_n^{-p} \right )$. Finally, define the following population kernel moments:
\begin{align}
	\Gamma_{p} & \equiv \int_{0}^{\infty} k \left ( u \right ) x_p \left ( u \right ) x_p \left ( u \right )' du \\
	\vartheta_{p,q} & \equiv \int_{0}^{\infty} k \left ( u \right ) u^q x_p \left ( u \right ) du \\
	\Psi_{p} & \equiv \int_{0}^{\infty} k \left ( u \right )^2 x_p \left ( u \right ) x_p \left ( u \right )' du
\end{align}

\appsubsection{Bias-Corrected Confidence Intervals}

In this section, I derive the asymptotic distribution of the bias-corrected estimator for the target parameter $\theta_{\nu}$. Following a similar approach to \cite{cct2014}, I first linearize the estimator $\widehat{\theta}_{\nu,p} \left ( h_n \right )$ and compute the asymptotic bias and variance of the linearized estimator $\widetilde{\theta}_{\nu,p} \left ( h_n \right )$. Then, I derive the asymptotic distribution of an estimator that subtracts the true (rather than estimated) bias from $\widehat{\theta}_{\nu,p} \left ( h_n \right )$. Before stating the main proposition, it is convenient to define a few recurring bias- and variance-related terms.

\appsubsubsection{Definitions of Bias Terms}\label{app_def_bias}

First, define two widely used bias terms:
\begin{align}
	\mathcal{B}_{+,\nu,p,q} \left ( h_n \right ) & \equiv \nu! e_{\nu}' \Gamma^{-1}_{+,p} \left ( h_n \right ) \vartheta_{+,p,q} \left ( h_n \right ) \\
	\mathcal{B}_{-,\nu,p,q} \left ( h_n \right ) & \equiv \nu! e_{\nu}' \Gamma^{-1}_{-,p} \left ( h_n \right ) \vartheta_{-,p,q} \left ( h_n \right )
\end{align}
If $n h_n \to \infty$ and $h_n \to 0$, Lemma S.A.1 in \cite{cct2014supp} gives
\begin{align}
	\mathcal{B}_{+,\nu,p,q} \left ( h_n \right ) = \nu! e_{\nu}' \Gamma^{-1}_{p} \vartheta_{p,q} + o_p \left ( 1 \right ) \qquad \mathcal{B}_{-,\nu,p,q} \left ( h_n \right ) = \left ( -1 \right )^{\nu+q} \nu! e_{\nu}' \Gamma^{-1}_{p} \vartheta_{p,q} + o_p \left ( 1 \right )
\end{align}
where the limits are finite constants determined by the kernel and polynomial orders. In addition, for any random variable $A$, define the following recurring bias terms:
\begin{align}
	\textsc{B}_{A+,\nu,p,q} \left ( h_n \right ) \equiv \frac{\mu^{(q)}_{A+}}{q!} \mathcal{B}_{+,\nu,p,q} \left ( h_n \right ) \qquad \textsc{B}_{A-,\nu,p,q} \left ( h_n \right ) \equiv \frac{\mu^{(q)}_{A-}}{q!} \mathcal{B}_{-,\nu,p,q} \left ( h_n \right )
\end{align}

\appsubsubsection{Definitions of Variance Terms}

For any pair of random variables $A$ and $B$, the conditional covariance between $\widehat{\mu}^{(\nu)}_{A+,p} \left ( h_n \right )$ and $\widehat{\mu}^{(\nu)}_{B+,p} \left ( h_n \right )$ is
\begin{align}
	\mathcal{V}_{AB+,\nu,p} \left ( h_n \right ) \equiv \Cov \left [ \widehat{\mu}^{(\nu)}_{A+,p} \left ( h_n \right ), \widehat{\mu}^{(\nu)}_{B+,p} \left ( h_n \right ) \Big | \bm{R} \right ] = \frac{\nu!^2}{n h_n^{2\nu}} e'_{\nu} \Gamma^{-1}_{+,p} \left ( h_n \right ) \Psi_{AB+,p} \left ( h_n \right ) \Gamma^{-1}_{+,p} \left ( h_n \right ) e_{\nu}
\end{align}
The equality follows from the matrix representation in Section \ref{appsec_notation}, using $e'_{\nu} H_p \left ( h_n \right ) = h_n^{-\nu} e'_{\nu}$ and the definition of $\Psi_{AB+,p} \left ( h_n \right )$. If $n h_n \to \infty$ and $h_n \to 0$, Lemma S.A.3 in \cite{cct2014supp} gives
\begin{align}
	\mathcal{V}_{AB+,\nu,p} \left ( h_n \right ) = \frac{1}{n h_n^{1+2\nu}} \left [ \frac{\sigma^2_{AB+}}{f_R \left ( c \right )} \nu!^2 e'_{\nu} \Gamma^{-1}_{p} \Psi_p \Gamma^{-1}_{p} e_{\nu} + o_p \left ( 1 \right ) \right ]
\end{align}
where $e'_{\nu} \Gamma^{-1}_{p} \Psi_p \Gamma^{-1}_{p} e_{\nu}$ is a finite constant determined by the kernel and polynomial orders. Define $\mathcal{V}_{AB-,\nu,p} \left ( h_n \right )$ analogously, replacing $+$ subscripts with $-$ subscripts throughout. Setting $B=A$ yields the corresponding conditional variances.

\appsubsubsection{Linearization of the Estimator}\label{app_linearization}

To study $\widehat{\theta}_{\nu,p} \left ( h_n \right )$, I first linearize the ratio of conditional mean estimators on each side of the cutoff. Define the following ratios of polynomial estimators on either side of the cutoff:
\begin{align}
	\widehat{\eta}_{+,\nu,p} \left ( h_n \right ) \equiv \frac{\widehat{\mu}^{(\nu)}_{W+,p} \left ( h_n \right )}{\widehat{\mu}^{(\nu)}_{D+,p} \left ( h_n \right )} \qquad 	\widehat{\eta}_{-,\nu,p} \left ( h_n \right ) \equiv \frac{\widehat{\mu}^{(\nu)}_{W-,p} \left ( h_n \right )}{\widehat{\mu}^{(\nu)}_{D-,p} \left ( h_n \right )}
\end{align}
Their population counterparts are
\begin{align}
	\eta_{+,\nu} \equiv \frac{\mu^{(\nu)}_{W+}}{\mu^{(\nu)}_{D+}} \qquad \eta_{-,\nu} \equiv \frac{\mu^{(\nu)}_{W-}}{\mu^{(\nu)}_{D-}}
\end{align}
To determine the asymptotic behavior of $\widehat{\eta}_{+,\nu,p} \left ( h_n \right )$ and $\widehat{\eta}_{-,\nu,p} \left ( h_n \right )$, I proceed as in \cite{cct2014} and linearize the ratio using a first-order Taylor expansion:
\begin{align}
	\widehat{\eta}_{+,\nu,p} \left ( h_n \right ) - \eta_{+,\nu} & \approx \frac{1}{\mu^{(\nu)}_{D+}} \left ( \widehat{\mu}^{(\nu)}_{W+,p} \left ( h_n \right ) - \mu^{(\nu)}_{W+} \right ) - \frac{\mu^{(\nu)}_{W+}}{\left ( \mu^{(\nu)}_{D+} \right )^2} \left ( \widehat{\mu}^{(\nu)}_{D+,p} \left ( h_n \right ) - \mu^{(\nu)}_{D+} \right )
\end{align}
The expansion for $\widehat{\eta}_{-,\nu,p} \left ( h_n \right ) - \eta_{-,\nu}$ follows by replacing the $+$ subscripts with $-$ subscripts throughout. Then, define a linear term and an exact remainder term:
\begin{align}
	\widetilde{\eta}_{+,\nu,p} \left ( h_n \right ) & \equiv \frac{1}{\mu^{(\nu)}_{D+}} \left ( \widehat{\mu}^{(\nu)}_{W+,p} \left ( h_n \right ) - \mu^{(\nu)}_{W+} \right ) - \frac{\mu^{(\nu)}_{W+}}{\left ( \mu^{(\nu)}_{D+} \right )^2} \left ( \widehat{\mu}^{(\nu)}_{D+,p} \left ( h_n \right ) - \mu^{(\nu)}_{D+} \right ) \\
	R_{+,n} & \equiv - \frac{\widehat{\mu}^{(\nu)}_{D+,p} \left ( h_n \right ) - \mu^{(\nu)}_{D+}}{\widehat{\mu}^{(\nu)}_{D+,p} \left ( h_n \right )} \widetilde{\eta}_{+,\nu,p} \left ( h_n \right )
\end{align}
Define $\widetilde{\eta}_{-,\nu,p} \left ( h_n \right )$ and $R_{-,n}$ symmetrically. Clearly,
\begin{align}
	\widehat{\eta}_{+,\nu,p} \left ( h_n \right ) - \eta_{+,\nu} & = \widetilde{\eta}_{+,\nu,p} \left ( h_n \right ) + R_{+,n} \\
	\widehat{\eta}_{-,\nu,p} \left ( h_n \right ) - \eta_{-,\nu} & = \widetilde{\eta}_{-,\nu,p} \left ( h_n \right ) + R_{-,n}
\end{align}
Finally, define
\begin{align}
	\widetilde{\eta}_{\nu,p} \left ( h_n \right ) \equiv \widetilde{\eta}_{+,\nu,p} \left ( h_n \right ) - \widetilde{\eta}_{-,\nu,p} \left ( h_n \right ) \qquad R_{n} \equiv R_{+,n} - R_{-,n} 
\end{align}
The linearized estimator for $\theta_\nu$ is
\begin{align}
	\widetilde{\theta}_{\nu,p} \left ( h_n \right ) \equiv \widehat{\mu}^{(\nu)}_{Y+,p} \left ( h_n \right ) - \widehat{\mu}^{(\nu)}_{Y-,p} \left ( h_n \right ) + \widetilde{\eta}_{\nu,p} \left ( h_n \right )
\end{align}
To conclude, the relationship between $\widehat{\theta}_{\nu,p} \left ( h_n \right )$ and $\widetilde{\theta}_{\nu,p} \left ( h_n \right )$ is 
\begin{align}
	\widehat{\theta}_{\nu,p} \left ( h_n \right ) - \theta_\nu = \widetilde{\theta}_{\nu,p} \left ( h_n \right ) - \left ( \mu^{(\nu)}_{Y+} - \mu^{(\nu)}_{Y-} \right ) + R_n
\end{align}

\appsubsubsection{General Theorem: Statement}

\begin{theorem}\label{thm_asydist}
	Suppose that Assumptions \ref{ass_moments} and \ref{ass_kernel} hold with $\delta \geq p+2$ and $n h_n \to \infty$.
	\begin{enumerate}[(R)]
		\item If $h_n \to 0$ and $n h_n^{1+2\nu} \to \infty$, then
		\begin{align*}
			R_n = O_p \left ( h_n^{2 (p+1-\nu)} + \frac{1}{n h_n^{1+2\nu}} \right )
		\end{align*}
	\end{enumerate}
	
	\begin{enumerate}[(B)]
		\item If $h_n \to 0$, then
		\begin{align*}
			\E \left [ \widetilde{\theta}_{\nu,p} \left ( h_n \right ) \Big | \bm{R} \right ] & = \mu^{(\nu)}_{Y+} - \mu^{(\nu)}_{Y-} + h^{p+1-\nu}_{n} \textsc{B}_{\nu,p,p+1} \left ( h_n \right )\\
			& + h^{p+2-\nu}_{n} \textsc{B}_{\nu,p,p+2} \left ( h_n \right ) + o_p \left ( h^{p+2-\nu}_{n} \right )
		\end{align*}
		where
		\begin{align*}
			\textsc{B}_{\nu,p,q} \left ( h_n \right ) \equiv \textsc{B}_{Y,\nu,p,q} \left ( h_n \right ) + \widetilde{\textsc{B}}_{\nu,p,q} \left ( h_n \right )
		\end{align*}
		with
		\begin{align*}
			\textsc{B}_{Y,\nu,p,q} \left ( h_n \right ) & \equiv \textsc{B}_{Y+,\nu,p,q} \left ( h_n \right ) - \textsc{B}_{Y-,\nu,p,q} \left ( h_n \right ) \\
			\widetilde{\textsc{B}}_{\nu,p,q} \left ( h_n \right ) &  \equiv \widetilde{\textsc{B}}_{+,\nu,p,q} \left ( h_n \right ) - \widetilde{\textsc{B}}_{-,\nu,p,q} \left ( h_n \right )
		\end{align*}
		and
		\begin{align*}
			\widetilde{\textsc{B}}_{+,\nu,p,q} \left ( h_n \right ) & \equiv \frac{1}{\mu^{(\nu)}_{D+}} \textsc{B}_{W+,\nu,p,q} \left ( h_n \right ) - \frac{\mu^{(\nu)}_{W+}}{\left ( \mu^{(\nu)}_{D+} \right )^2} \textsc{B}_{D+,\nu,p,q} \left ( h_n \right ) \\
			\widetilde{\textsc{B}}_{-,\nu,p,q} \left ( h_n \right ) & \equiv \frac{1}{\mu^{(\nu)}_{D-}} \textsc{B}_{W-,\nu,p,q} \left ( h_n \right ) - \frac{\mu^{(\nu)}_{W-}}{\left ( \mu^{(\nu)}_{D-} \right )^2} \textsc{B}_{D-,\nu,p,q} \left ( h_n \right )
		\end{align*}
	\end{enumerate}
	
	\begin{enumerate}[(V)]
		\item For any $h_n > 0$,
		\begin{align*}
			\textsc{V}_{\nu,p} \left ( h_n \right ) \equiv \Var \left [ \widetilde{\theta}_{\nu,p} \left ( h_n \right ) \Big | \bm{R} \right ] = \textsc{V}_{Y,\nu,p} \left ( h_n \right ) + \widetilde{\textsc{V}}_{\nu,p} \left ( h_n \right ) + 2 \widetilde{\textsc{V}}_{Y,\nu,p} \left ( h_n \right )
		\end{align*}
		where
		\begin{align*}
			\textsc{V}_{Y,\nu,p} \left ( h_n \right ) & \equiv \mathcal{V}_{YY+,\nu,p} \left ( h_n \right ) + \mathcal{V}_{YY-,\nu,p} \left ( h_n \right ) \\
			\widetilde{\textsc{V}}_{\nu,p} \left ( h_n \right ) & \equiv \widetilde{\textsc{V}}_{+,\nu,p} \left ( h_n \right ) + \widetilde{\textsc{V}}_{-,\nu,p} \left ( h_n \right ) \\
			\widetilde{\textsc{V}}_{Y,\nu,p} \left ( h_n \right ) & \equiv \widetilde{\textsc{V}}_{Y+,\nu,p} \left ( h_n \right ) + \widetilde{\textsc{V}}_{Y-,\nu,p} \left ( h_n \right )
		\end{align*}
		with 
		\begin{align*}
			\widetilde{\textsc{V}}_{+,\nu,p} \left ( h_n \right ) & \equiv \frac{1}{\left ( \mu^{(\nu)}_{D+} \right )^2} \mathcal{V}_{WW+,\nu,p} \left ( h_n \right ) + \frac{\left ( \mu^{(\nu)}_{W+} \right )^2}{\left ( \mu^{(\nu)}_{D+} \right )^4} \mathcal{V}_{DD+,\nu,p} \left ( h_n \right ) - \frac{2 \mu^{(\nu)}_{W+}}{\left ( \mu^{(\nu)}_{D+} \right )^3} \mathcal{V}_{WD+,\nu,p} \left ( h_n \right ) \\
			\widetilde{\textsc{V}}_{-,\nu,p} \left ( h_n \right ) & \equiv \frac{1}{\left ( \mu^{(\nu)}_{D-} \right )^2} \mathcal{V}_{WW-,\nu,p} \left ( h_n \right ) + \frac{\left ( \mu^{(\nu)}_{W-} \right )^2}{\left ( \mu^{(\nu)}_{D-} \right )^4} \mathcal{V}_{DD-,\nu,p} \left ( h_n \right ) - \frac{2 \mu^{(\nu)}_{W-}}{\left ( \mu^{(\nu)}_{D-} \right )^3} \mathcal{V}_{WD-,\nu,p} \left ( h_n \right )
		\end{align*}
		and 
		\begin{align*}
			\widetilde{\textsc{V}}_{Y+,\nu,p} \left ( h_n \right ) & \equiv \frac{1}{\mu^{(\nu)}_{D+}} \mathcal{V}_{YW+,\nu,p} \left ( h_n \right ) - \frac{\mu^{(\nu)}_{W+}}{\left ( \mu^{(\nu)}_{D+} \right )^2} \mathcal{V}_{YD+,\nu,p} \left ( h_n \right ) \\
			\widetilde{\textsc{V}}_{Y-,\nu,p} \left ( h_n \right ) & \equiv \frac{1}{\mu^{(\nu)}_{D-}} \mathcal{V}_{YW-,\nu,p} \left ( h_n \right ) - \frac{\mu^{(\nu)}_{W-}}{\left ( \mu^{(\nu)}_{D-} \right )^2} \mathcal{V}_{YD-,\nu,p} \left ( h_n \right )
		\end{align*}
	\end{enumerate}
	
	\begin{enumerate}[(D)]
		\item If $n h_n^{2p+5} \to 0$ and $n h_n^{1+2\nu} \to \infty$, then
		\begin{align*}
			T^{\textsc{bc}}_{\nu,p} \left ( h_n \right ) \equiv \frac{\widehat{\theta}_{\nu,p} \left ( h_n \right ) - \theta_\nu - h^{p+1-\nu}_{n} \textsc{B}_{\nu,p,p+1} \left ( h_n \right )}{\sqrt{\textsc{V}_{\nu,p} \left ( h_n \right )}} \stackrel{d}{\to} \mathcal{N} \left ( 0,1 \right ) 
		\end{align*}
	\end{enumerate}
	
\end{theorem}

\appsubsubsection{General Theorem: Proof}
	
	\begin{enumerate}[(R)]
		\item Consider $R_{+,n}$ and let $A \in \left \{ W,D \right \}$. By parts (B) and (V) of Lemma S.A.3 in \cite{cct2014supp}, together with the conditional Chebyshev inequality,
		\begin{align*}
			\widehat{\mu}^{(\nu)}_{A+,p}(h_n)-\mu^{(\nu)}_{A+} = O_p \left( h_n^{p+1-\nu} + \sqrt{\frac{1}{nh_n^{1+2\nu}}} \right )
		\end{align*}
		Since $h_n \to 0$, $nh_n^{1+2\nu} \to \infty$, and $0 \leq \nu \leq p$, both terms in this bound converge to zero. In particular,
		\begin{align*}
			\widehat{\mu}^{(\nu)}_{D+,p}(h_n) \stackrel{p}{\to} \mu^{(\nu)}_{D+} \neq 0
		\end{align*}
		Thus, by the Continuous Mapping Theorem,
		\begin{align*}
			\frac{1}{\widehat{\mu}^{(\nu)}_{D+,p} (h_n)} = O_p (1)
		\end{align*}
		Furthermore, the definition of $\widetilde{\eta}_{+,\nu,p} \left ( h_n \right )$ multiplies the estimator for $W$ by $\frac{1}{\mu^{(\nu)}_{D+}}$ and subtracts the estimator for $D$ multiplied by $\frac{\mu^{(\nu)}_{W+}}{\left ( \mu^{(\nu)}_{D+} \right )^2}$. Both errors satisfy the bound established above. In addition, both multipliers are finite deterministic constants. Multiplying the estimators by these constants and taking their difference therefore yields
		\begin{align*}
			\widetilde{\eta}_{+,\nu,p} \left ( h_n \right )
			= O_p \left ( h_n^{p+1-\nu} + \sqrt{\frac{1}{n h_n^{1+2\nu}}} \right )
		\end{align*}
		Using the exact remainder defined in Section \ref{app_linearization},
		\begin{align*}
			R_{+,n} & \equiv - \frac{\widehat{\mu}^{(\nu)}_{D+,p} \left ( h_n \right ) - \mu^{(\nu)}_{D+}}{\widehat{\mu}^{(\nu)}_{D+,p} \left ( h_n \right )}
			\widetilde{\eta}_{+,\nu,p} \left ( h_n \right ) \notag \\
			& = O_p \left ( \left ( h_n^{p+1-\nu} + \sqrt{\frac{1}{n h_n^{1+2\nu}}} \right )^2 \right ) = O_p \left ( h_n^{2(p+1-\nu)} + \frac{1}{n h_n^{1+2\nu}} \right )
		\end{align*}
		The last equality follows by bounding the cross-product by the sum of the squared terms. The proof for $R_{-,n}$ is symmetric. Since $R_n \equiv R_{+,n} - R_{-,n}$, the desired result follows.
		
	\end{enumerate}
	
	\begin{enumerate}[(B)]
		\item Let $A \in \left \{ Y,W,D \right \}$. By part (B) of Lemma S.A.3 in \cite{cct2014supp}, as $h_n \to 0$,
		\begin{align*}
			\E \left [ \widehat{\mu}^{(\nu)}_{A+,p} \left ( h_n \right ) - \mu^{(\nu)}_{A+} \Big | \bm{R} \right ] & = h^{p+1-\nu}_{n} \textsc{B}_{A+,\nu,p,p+1} \left ( h_n \right ) + h^{p+2-\nu}_{n} \textsc{B}_{A+,\nu,p,p+2} \left ( h_n \right ) + o_p \left ( h^{p+2-\nu}_{n} \right )
		\end{align*}
		The expression for the conditional bias of the lower-side estimator is symmetric. Then, by the definition of $\widetilde{\eta}_{+,\nu,p} \left ( h_n \right )$ and the linearity of the conditional expectation,
		\begin{align*}
			\E \left [ \widetilde{\eta}_{+,\nu,p} \left ( h_n \right ) | \bm{R} \right ] & = \frac{1}{\mu^{(\nu)}_{D+}} \left ( h^{p+1-\nu}_{n} \textsc{B}_{W+,\nu,p,p+1} \left ( h_n \right ) + h^{p+2-\nu}_{n} \textsc{B}_{W+,\nu,p,p+2} \left ( h_n \right ) \right ) \\
			& - \frac{\mu^{(\nu)}_{W+}}{\left ( \mu^{(\nu)}_{D+} \right )^2} \left ( h^{p+1-\nu}_{n} \textsc{B}_{D+,\nu,p,p+1} \left ( h_n \right ) + h^{p+2-\nu}_{n} \textsc{B}_{D+,\nu,p,p+2} \left ( h_n \right ) \right ) \\
			& + o_p \left ( h^{p+2-\nu}_{n} \right ) \\
			& = h_n^{p+1-\nu} \widetilde{\textsc{B}}_{+,\nu,p,p+1} \left ( h_n \right )
			+ h_n^{p+2-\nu} \widetilde{\textsc{B}}_{+,\nu,p,p+2} \left ( h_n \right )
			+ o_p \left ( h_n^{p+2-\nu} \right )
		\end{align*}
		The first equality substitutes the expressions for the conditional bias of $W$ and $D$. The second follows from the definition of $\widetilde{\textsc{B}}_{+,\nu,p,q} \left ( h_n \right )$. The argument for $\widetilde{\eta}_{-,\nu,p} \left ( h_n \right )$ is symmetric. Combining these expressions yields the conditional expectation of $\widetilde{\theta}_{\nu,p} \left ( h_n \right )$. Using the definition of $\textsc{B}_{\nu,p,q} \left ( h_n \right )$ gives
		\begin{align*}
			\E \left [ \widetilde{\theta}_{\nu,p} \left ( h_n \right ) \Big | \bm{R} \right ] & = \mu^{(\nu)}_{Y+} - \mu^{(\nu)}_{Y-} + h^{p+1-\nu}_{n} \textsc{B}_{\nu,p,p+1} \left ( h_n \right ) + h^{p+2-\nu}_{n} \textsc{B}_{\nu,p,p+2} \left ( h_n \right ) + o_p \left ( h^{p+2-\nu}_{n} \right )
		\end{align*}
		which completes the proof.
	\end{enumerate}
	
	\begin{enumerate}[(V)]
		\item Conditional on $\bm{R}$, the upper- and lower-side estimators use disjoint sets of independent observations. Thus,
		\begin{align*}
			\textsc{V}_{\nu,p} \left ( h_n \right )
			& = \Var \left [ \widehat{\mu}^{(\nu)}_{Y+,p} \left ( h_n \right ) + \widetilde{\eta}_{+,\nu,p} \left ( h_n \right ) \Big | \bm{R} \right ] + \Var \left [ \widehat{\mu}^{(\nu)}_{Y-,p} \left ( h_n \right ) + \widetilde{\eta}_{-,\nu,p} \left ( h_n \right ) \Big | \bm{R} \right ]
		\end{align*}
		By the definition of $\widetilde{\eta}_{+,\nu,p} \left ( h_n \right )$ and the bilinearity of the conditional covariance,
		\begin{align*}
			\widetilde{\textsc{V}}_{+,\nu,p} \left ( h_n \right )
			& \equiv \Var \left [ \widetilde{\eta}_{+,\nu,p} \left ( h_n \right ) \Big | \bm{R} \right ] \\
			& = \frac{1}{\left ( \mu^{(\nu)}_{D+} \right )^2} \mathcal{V}_{WW+,\nu,p} \left ( h_n \right ) + \frac{\left ( \mu^{(\nu)}_{W+} \right )^2}{\left ( \mu^{(\nu)}_{D+} \right )^4} \mathcal{V}_{DD+,\nu,p} \left ( h_n \right ) - \frac{2 \mu^{(\nu)}_{W+}}{\left ( \mu^{(\nu)}_{D+} \right )^3} \mathcal{V}_{WD+,\nu,p} \left ( h_n \right )
		\end{align*}
		Similarly,
		\begin{align*}
			\widetilde{\textsc{V}}_{Y+,\nu,p} \left ( h_n \right )
			& \equiv \Cov \left [ \widehat{\mu}^{(\nu)}_{Y+,p} \left ( h_n \right ), \widetilde{\eta}_{+,\nu,p} \left ( h_n \right ) \Big | \bm{R} \right ] \\
			& = \frac{1}{\mu^{(\nu)}_{D+}} \mathcal{V}_{YW+,\nu,p} \left ( h_n \right )
			- \frac{\mu^{(\nu)}_{W+}}{\left ( \mu^{(\nu)}_{D+} \right )^2} \mathcal{V}_{YD+,\nu,p} \left ( h_n \right )
		\end{align*}
		The corresponding expressions for the lower-side estimators follow by replacing the $+$ subscripts with $-$ subscripts throughout. Finally, expanding the two conditional variances in the first display gives
		\begin{align*}
			\textsc{V}_{\nu,p} \left ( h_n \right )
			& = \mathcal{V}_{YY+,\nu,p} \left ( h_n \right )
			+ \mathcal{V}_{YY-,\nu,p} \left ( h_n \right ) \\
			& + \widetilde{\textsc{V}}_{+,\nu,p} \left ( h_n \right )
			+ \widetilde{\textsc{V}}_{-,\nu,p} \left ( h_n \right ) \\
			& + 2 \left ( \widetilde{\textsc{V}}_{Y+,\nu,p} \left ( h_n \right )
			+ \widetilde{\textsc{V}}_{Y-,\nu,p} \left ( h_n \right ) \right ) \\
			& = \textsc{V}_{Y,\nu,p} \left ( h_n \right )
			+ \widetilde{\textsc{V}}_{\nu,p} \left ( h_n \right )
			+ 2 \widetilde{\textsc{V}}_{Y,\nu,p} \left ( h_n \right )
		\end{align*}
		The last equality follows from the definitions in the statement. This completes the proof.
	\end{enumerate}
	
	\begin{enumerate}[(D)]
		\item Local polynomial estimators are linear in the outcome. Thus, the upper-side component of $\widetilde{\theta}_{\nu,p} \left ( h_n \right )$ is obtained using the outcome
		\begin{align*}
			Y + \frac{1}{\mu^{(\nu)}_{D+}} W
			- \frac{\mu^{(\nu)}_{W+}}{\left ( \mu^{(\nu)}_{D+} \right )^2} D
		\end{align*}
 		The corresponding expression for the lower-side component follows by replacing the $+$ subscripts with $-$ subscripts throughout. These transformed outcomes inherit bounded fourth moments and one-sided smoothness from Assumption \ref{ass_moments}. Their conditional variances have finite one-sided limits.These limits are strictly positive by part (f) of Assumption \ref{ass_moments}, because the coefficient of $Y$ in each transformed outcome is one. Thus, the argument used to prove part (D) of Lemma A.1 in \cite{cct2014supp} applies here. Using the bias and variance expressions from parts (B) and (V), and letting $n h_n^{2p+5} \to 0$,
		\begin{align*}
			\frac{
				\widetilde{\theta}_{\nu,p} \left ( h_n \right )
				- \left ( \mu^{(\nu)}_{Y+} - \mu^{(\nu)}_{Y-} \right )
				- h_n^{p+1-\nu} \textsc{B}_{\nu,p,p+1} \left ( h_n \right )
			}{
				\sqrt{\textsc{V}_{\nu,p} \left ( h_n \right )}
			}
			\stackrel{d}{\to} \mathcal{N} \left ( 0,1 \right )
		\end{align*}
		Furthermore, Assumption \ref{ass_moments} and the local polynomial expansion of the variance imply that
		\begin{align*}
			\frac{1}{\sqrt{\textsc{V}_{\nu,p} \left ( h_n \right )}}
			= O_p \left ( \sqrt{n h_n^{1+2\nu}} \right )
		\end{align*}
		Combining this bound with the one derived in part (R) yields
		\begin{align*}
			\frac{R_n}{\sqrt{\textsc{V}_{\nu,p} \left ( h_n \right )}}
			& = O_p \left (
				\sqrt{n h_n^{1+2\nu}} h_n^{2(p+1-\nu)}
				+ \frac{1}{\sqrt{n h_n^{1+2\nu}}}
			\right ) \\
			& = O_p \left (
				\sqrt{n h_n^{2p+5}} h_n^{p-\nu}
				+ \frac{1}{\sqrt{n h_n^{1+2\nu}}}
			\right )
			= o_p \left ( 1 \right )
		\end{align*}
		The last equality follows from $p \geq \nu$ and the bandwidth conditions in the statement. Finally, by the identity established in Section \ref{app_linearization},
		\begin{align*}
			T^{\textsc{bc}}_{\nu,p} \left ( h_n \right )
			& = \frac{
				\widetilde{\theta}_{\nu,p} \left ( h_n \right )
				- \left ( \mu^{(\nu)}_{Y+} - \mu^{(\nu)}_{Y-} \right )
				- h_n^{p+1-\nu} \textsc{B}_{\nu,p,p+1} \left ( h_n \right )
			}{
				\sqrt{\textsc{V}_{\nu,p} \left ( h_n \right )}
			}
			+ \frac{R_n}{\sqrt{\textsc{V}_{\nu,p} \left ( h_n \right )}} \stackrel{d}{\to} \mathcal{N} \left ( 0,1 \right )
		\end{align*}
		The convergence follows from Slutsky's Theorem. This completes the proof.
	\end{enumerate}

\appsubsection{Robust Bias-Corrected Confidence Intervals}

In this section, I derive the asymptotic distribution of the robust bias-corrected estimator for the target parameter $\theta_\nu$. This estimator is defined as the difference between $\widehat{\theta}_{\nu,p} \left ( h_n \right )$ and its estimated (rather than true) bias. Before stating the main proposition, it is convenient to define a few additional bias- and variance-related terms.

\appsubsubsection{Definitions of Bias Terms}\label{appsec_bc_bias}

Let $q \geq p+1$ denote the polynomial degree used with the pilot bandwidth $b_n$. Suppose that Assumption \ref{ass_moments} holds with $\delta \geq q+1$. The bias correction uses estimates of the $(p+1)$st derivatives. These estimates may themselves be biased. For any random variable $A$, define the following coefficients to characterize how this bias affects the corrected estimator:
\begin{align}
	\textsc{B}^{\textsc{bc}}_{A+,\nu,p,q} \left ( h_n, b_n \right ) & \equiv \frac{\mu^{(q+1)}_{A+}}{\left ( q+1 \right )!} \mathcal{B}_{+,p+1,q,q+1} \left ( b_n \right ) \frac{\mathcal{B}_{+,\nu,p,p+1} \left ( h_n \right )}{\left ( p+1 \right )!} \\
	\textsc{B}^{\textsc{bc}}_{A-,\nu,p,q} \left ( h_n, b_n \right ) & \equiv \frac{\mu^{(q+1)}_{A-}}{\left ( q+1 \right )!} \mathcal{B}_{-,p+1,q,q+1} \left ( b_n \right ) \frac{\mathcal{B}_{-,\nu,p,p+1} \left ( h_n \right )}{\left ( p+1 \right )!}
\end{align}
Natural estimators for the leading bias coefficients $\textsc{B}_{A+,\nu,p,p+1} \left ( h_n \right )$ and $\textsc{B}_{A-,\nu,p,p+1} \left ( h_n \right )$ are
\begin{align}
	\widehat{\textsc{B}}_{A+,\nu,p,q} \left ( h_n, b_n \right ) & \equiv \frac{\widehat{\mu}^{(p+1)}_{A+,q} \left ( b_n \right )}{\left ( p+1 \right )!} \mathcal{B}_{+,\nu,p,p+1} \left ( h_n \right )\\
	\widehat{\textsc{B}}_{A-,\nu,p,q} \left ( h_n, b_n \right ) & \equiv \frac{\widehat{\mu}^{(p+1)}_{A-,q} \left ( b_n \right )}{\left ( p+1 \right )!} \mathcal{B}_{-,\nu,p,p+1} \left ( h_n \right )
\end{align}
with $h_n$ and $b_n$ denoting the main and pilot bandwidths, respectively. Thus, an estimator for $\textsc{B}_{Y,\nu,p,p+1} \left ( h_n \right )$ is
\begin{align}
	\widehat{\textsc{B}}_{Y,\nu,p,q} \left ( h_n, b_n \right ) & \equiv \widehat{\textsc{B}}_{Y+,\nu,p,q} \left ( h_n, b_n \right ) - \widehat{\textsc{B}}_{Y-,\nu,p,q} \left ( h_n, b_n \right )
\end{align}
In addition, consider the leading bias coefficient of $\widetilde{\eta}_{\nu,p} \left ( h_n \right )$. First, define an infeasible estimator that retains the population coefficients in the linearization:
\begin{align}
	\widecheck{\widetilde{\textsc{B}}}_{\nu,p,q} \left ( h_n, b_n \right ) & \equiv \widecheck{\widetilde{\textsc{B}}}_{+,\nu,p,q} \left ( h_n, b_n \right ) - \widecheck{\widetilde{\textsc{B}}}_{-,\nu,p,q} \left ( h_n, b_n \right )
\end{align}
with
\begin{align}
	\widecheck{\widetilde{\textsc{B}}}_{+,\nu,p,q} \left ( h_n, b_n \right ) & \equiv \frac{1}{\mu^{(\nu)}_{D+}} \widehat{\textsc{B}}_{W+,\nu,p,q} \left ( h_n, b_n \right ) - \frac{\mu^{(\nu)}_{W+}}{\left ( \mu^{(\nu)}_{D+} \right )^2} \widehat{\textsc{B}}_{D+,\nu,p,q} \left ( h_n, b_n \right ) \\
	\widecheck{\widetilde{\textsc{B}}}_{-,\nu,p,q} \left ( h_n, b_n \right ) & \equiv \frac{1}{\mu^{(\nu)}_{D-}} \widehat{\textsc{B}}_{W-,\nu,p,q} \left ( h_n, b_n \right ) - \frac{\mu^{(\nu)}_{W-}}{\left ( \mu^{(\nu)}_{D-} \right )^2} \widehat{\textsc{B}}_{D-,\nu,p,q} \left ( h_n, b_n \right )
\end{align}
Second, replace the population coefficients with their estimated counterparts. This gives the feasible estimator:
\begin{align}
	\widehat{\widetilde{\textsc{B}}}_{\nu,p,q} \left ( h_n, b_n \right ) & \equiv \widehat{\widetilde{\textsc{B}}}_{+,\nu,p,q} \left ( h_n, b_n \right ) - \widehat{\widetilde{\textsc{B}}}_{-,\nu,p,q} \left ( h_n, b_n \right )
\end{align}
with
\begin{align}
	\widehat{\widetilde{\textsc{B}}}_{+,\nu,p,q} \left ( h_n, b_n \right ) & \equiv \frac{1}{\widehat{\mu}^{(\nu)}_{D+,p} \left ( h_n \right )} \widehat{\textsc{B}}_{W+,\nu,p,q} \left ( h_n, b_n \right ) - \frac{\widehat{\mu}^{(\nu)}_{W+,p} \left ( h_n \right )}{\left ( \widehat{\mu}^{(\nu)}_{D+,p} \left ( h_n \right ) \right )^2} \widehat{\textsc{B}}_{D+,\nu,p,q} \left ( h_n, b_n \right ) \\
	\widehat{\widetilde{\textsc{B}}}_{-,\nu,p,q} \left ( h_n, b_n \right ) & \equiv \frac{1}{\widehat{\mu}^{(\nu)}_{D-,p} \left ( h_n \right )} \widehat{\textsc{B}}_{W-,\nu,p,q} \left ( h_n, b_n \right ) - \frac{\widehat{\mu}^{(\nu)}_{W-,p} \left ( h_n \right )}{\left ( \widehat{\mu}^{(\nu)}_{D-,p} \left ( h_n \right ) \right )^2} \widehat{\textsc{B}}_{D-,\nu,p,q} \left ( h_n, b_n \right )
\end{align}
Adding the estimated bias coefficient for $Y$ yields the following infeasible and feasible estimators of $\textsc{B}_{\nu,p,p+1} \left ( h_n \right )$, respectively:
\begin{align}
	\widecheck{\textsc{B}}_{\nu,p,q} \left ( h_n, b_n \right ) & \equiv \widehat{\textsc{B}}_{Y,\nu,p,q} \left ( h_n, b_n \right ) + \widecheck{\widetilde{\textsc{B}}}_{\nu,p,q} \left ( h_n, b_n \right ) \\
	\widehat{\textsc{B}}_{\nu,p,q} \left ( h_n, b_n \right ) & \equiv \widehat{\textsc{B}}_{Y,\nu,p,q} \left ( h_n, b_n \right ) + \widehat{\widetilde{\textsc{B}}}_{\nu,p,q} \left ( h_n, b_n \right )
\end{align}

\appsubsubsection{Definitions of Bias-Corrected Estimators}\label{appsec_bc_estim}

The bias-corrected estimator for the target parameter $\theta_{\nu}$ is
\begin{align}
	\widehat{\theta}^{\textsc{bc}}_{\nu,p,q} \left ( h_n, b_n \right ) \equiv \widehat{\theta}_{\nu,p} \left ( h_n \right ) - h_n^{p+1-\nu} \widehat{\textsc{B}}_{\nu,p,q} \left ( h_n, b_n \right )
\end{align}
Analogously, define the bias-corrected linearized estimator as
\begin{align}
	\widetilde{\theta}^{\textsc{bc}}_{\nu,p,q} \left ( h_n, b_n \right ) \equiv \widetilde{\theta}_{\nu,p} \left ( h_n \right ) - h_n^{p+1-\nu} \widecheck{\textsc{B}}_{\nu,p,q} \left ( h_n, b_n \right )
\end{align}
Recall from Section \ref{app_linearization} that
\begin{align}
	\widehat{\theta}_{\nu,p} \left ( h_n \right ) - \theta_\nu
	= \widetilde{\theta}_{\nu,p} \left ( h_n \right )
	- \left ( \mu^{(\nu)}_{Y+} - \mu^{(\nu)}_{Y-} \right ) + R_n
\end{align}
Both $\widehat{\textsc{B}}_{\nu,p,q} \left ( h_n, b_n \right )$ and $\widecheck{\textsc{B}}_{\nu,p,q} \left ( h_n, b_n \right )$ contain $\widehat{\textsc{B}}_{Y,\nu,p,q} \left ( h_n,b_n \right )$. This term cancels when taking the difference between these two bias estimators. Define
\begin{align}
	R^{\textsc{bc}}_n
	\equiv h_n^{p+1-\nu}
	\left (
		\widehat{\widetilde{\textsc{B}}}_{\nu,p,q} \left ( h_n,b_n \right )
		- \widecheck{\widetilde{\textsc{B}}}_{\nu,p,q} \left ( h_n,b_n \right )
	\right )
\end{align}
Then, by the definitions of the bias-corrected estimators,
\begin{align}
	\widehat{\theta}^{\textsc{bc}}_{\nu,p,q} \left ( h_n,b_n \right ) - \theta_\nu
	& = \widetilde{\theta}^{\textsc{bc}}_{\nu,p,q} \left ( h_n,b_n \right )
	- \left ( \mu^{(\nu)}_{Y+} - \mu^{(\nu)}_{Y-} \right ) + R_n \notag \\
	& - h_n^{p+1-\nu}
	\left (
		\widehat{\textsc{B}}_{\nu,p,q} \left ( h_n,b_n \right )
		- \widecheck{\textsc{B}}_{\nu,p,q} \left ( h_n,b_n \right )
	\right ) \\
	& = \widetilde{\theta}^{\textsc{bc}}_{\nu,p,q} \left ( h_n,b_n \right )
	- \left ( \mu^{(\nu)}_{Y+} - \mu^{(\nu)}_{Y-} \right )
	+ R_n - R^{\textsc{bc}}_n
\end{align}

\appsubsubsection{Definitions of Variance Terms}\label{appsec_bc_var}

Consider the estimator of the $\nu$th derivative of the conditional mean of $A$ with bandwidth $h_n$. Consider also the estimator of the $(p+1)$st derivative of the conditional mean of $B$ with bandwidth $b_n$. Define the conditional covariance of these two estimators as
\begin{align}
	\mathcal{C}_{AB+,\nu,p,q} \left ( h_n,b_n \right )
	& \equiv \Cov \left [ \widehat{\mu}^{(\nu)}_{A+,p} \left ( h_n \right ), \widehat{\mu}^{(p+1)}_{B+,q} \left ( b_n \right ) \Big | \bm{R} \right ] \notag \\
	& = \frac{\nu! \left ( p+1 \right )!}{n h_n^\nu b_n^{p+1}} e_\nu' \Gamma^{-1}_{+,p} \left ( h_n \right ) \Psi_{AB+,p,q} \left ( h_n,b_n \right ) \Gamma^{-1}_{+,q} \left ( b_n \right ) e_{p+1}
\end{align}
The equality follows from the matrix representations in Section \ref{appsec_notation}, using the definition of $\Psi_{AB+,p,q} \left ( h_n,b_n \right )$ and the identities $e_\nu' H_p \left ( h_n \right ) = h_n^{-\nu} e_\nu'$ and $H_q \left ( b_n \right ) e_{p+1} = b_n^{-(p+1)} e_{p+1}$. Define $\mathcal{C}_{AB-,\nu,p,q} \left ( h_n,b_n \right )$ analogously, replacing the $+$ subscripts with $-$ subscripts throughout.

\appsubsubsection{General Theorem: Statement}

\begin{theorem}\label{thm_asydist_bias}
	Suppose that Assumptions \ref{ass_moments} and \ref{ass_kernel} hold with $\delta \geq q+1$, and $n \min \left \{ h_n, b_n \right \} \to \infty$.
	
	\begin{enumerate}[(R)]
		\item If $h_n \to 0$ and $n h_n^{1+2\nu} \to \infty$, and provided that $\kappa b_n < \kappa_0$, then
		\begin{align*}
			R^{\textsc{bc}}_{n} = O_p \left ( h_n^{2 (p+1-\nu)} + \frac{h_n^{p+1-\nu}}{\sqrt{n h_n^{1 + 2\nu}}} \right ) O_p \left ( 1 + \sqrt{\frac{1}{n b_n^{2p+3}}} \right )
		\end{align*}
	\end{enumerate}
	
	\begin{enumerate}[(B)]
		\item If $\max \left \{ h_n, b_n \right \} \to 0$, then
		\begin{align*}
			\E \left [ \widetilde{\theta}^{\textsc{bc}}_{\nu,p,q} \left ( h_n,b_n \right ) \Big | \bm{R} \right ]
			& = \mu^{(\nu)}_{Y+} - \mu^{(\nu)}_{Y-}
			+ h_n^{p+2-\nu} \textsc{B}_{\nu,p,p+2} \left ( h_n \right ) \\
			& - h_n^{p+1-\nu} b_n^{q-p}
			\textsc{B}^{\textsc{bc}}_{\nu,p,q} \left ( h_n,b_n \right ) \\
			& + o_p \left (
				h_n^{p+2-\nu} + h_n^{p+1-\nu} b_n^{q-p}
			\right )
		\end{align*}
		where
		\begin{align*}
			\textsc{B}^{\textsc{bc}}_{\nu,p,q} \left ( h_n, b_n \right ) & \equiv \textsc{B}^{\textsc{bc}}_{Y,\nu,p,q} \left ( h_n, b_n \right ) + \widetilde{\textsc{B}}^{\textsc{bc}}_{\nu,p,q} \left ( h_n, b_n \right )
		\end{align*}
		with
		\begin{align*}
			\textsc{B}^{\textsc{bc}}_{Y,\nu,p,q} \left ( h_n, b_n \right ) & \equiv \textsc{B}^{\textsc{bc}}_{Y+,\nu,p,q} \left ( h_n, b_n \right ) - \textsc{B}^{\textsc{bc}}_{Y-,\nu,p,q} \left ( h_n, b_n \right ) \\
			\widetilde{\textsc{B}}^{\textsc{bc}}_{\nu,p,q} \left ( h_n, b_n \right ) & \equiv \widetilde{\textsc{B}}^{\textsc{bc}}_{+,\nu,p,q} \left ( h_n, b_n \right ) - \widetilde{\textsc{B}}^{\textsc{bc}}_{-,\nu,p,q} \left ( h_n, b_n \right )
		\end{align*}
		and
		\begin{align*}
			\widetilde{\textsc{B}}^{\textsc{bc}}_{+,\nu,p,q} \left ( h_n, b_n \right ) & \equiv \frac{1}{\mu^{(\nu)}_{D+}} \textsc{B}^{\textsc{bc}}_{W+,\nu,p,q} \left ( h_n, b_n \right ) - \frac{\mu^{(\nu)}_{W+}}{\left ( \mu^{(\nu)}_{D+} \right )^2} \textsc{B}^{\textsc{bc}}_{D+,\nu,p,q} \left ( h_n, b_n \right ) \\
			\widetilde{\textsc{B}}^{\textsc{bc}}_{-,\nu,p,q} \left ( h_n, b_n \right ) & \equiv \frac{1}{\mu^{(\nu)}_{D-}} \textsc{B}^{\textsc{bc}}_{W-,\nu,p,q} \left ( h_n, b_n \right ) - \frac{\mu^{(\nu)}_{W-}}{\left ( \mu^{(\nu)}_{D-} \right )^2} \textsc{B}^{\textsc{bc}}_{D-,\nu,p,q} \left ( h_n, b_n \right )
		\end{align*}
		
	\end{enumerate}
	
	\begin{enumerate}[(V)]
		\item Whenever the local polynomial estimators are well defined,
		\begin{align*}
			\textsc{V}^{\textsc{bc}}_{\nu,p,q} \left ( h_n, b_n \right ) \equiv \Var \left [ \widetilde{\theta}^{\textsc{bc}}_{\nu,p,q} \left ( h_n, b_n \right ) \Big | \bm{R} \right ] = \textsc{V}^{\textsc{bc}}_{+,\nu,p,q} \left ( h_n, b_n \right ) + \textsc{V}^{\textsc{bc}}_{-,\nu,p,q} \left ( h_n, b_n \right )
		\end{align*}
		where
		\begin{align*}
			\textsc{V}^{\textsc{bc}}_{+,\nu,p,q} \left ( h_n, b_n \right ) & \equiv \textsc{V}_{+,\nu,p} \left ( h_n \right ) + h_n^{2(p+1-\nu)} \textsc{V}_{+,p+1,q} \left ( b_n \right ) \frac{\mathcal{B}^2_{+,\nu,p,p+1} \left ( h_n \right )}{\left ( p+1 \right )!^2} \notag \\
			& - 2 h_n^{p+1-\nu} \textsc{C}_{+,\nu,p,q} \left ( h_n, b_n \right ) \frac{\mathcal{B}_{+,\nu,p,p+1} \left ( h_n \right )}{\left ( p+1 \right )!} \\
			\textsc{V}^{\textsc{bc}}_{-,\nu,p,q} \left ( h_n, b_n \right ) & \equiv \textsc{V}_{-,\nu,p} \left ( h_n \right ) + h_n^{2(p+1-\nu)} \textsc{V}_{-,p+1,q} \left ( b_n \right ) \frac{\mathcal{B}^2_{-,\nu,p,p+1} \left ( h_n \right )}{\left ( p+1 \right )!^2} \notag \\
			& - 2 h_n^{p+1-\nu} \textsc{C}_{-,\nu,p,q} \left ( h_n, b_n \right ) \frac{\mathcal{B}_{-,\nu,p,p+1} \left ( h_n \right )}{\left ( p+1 \right )!}
		\end{align*}
		with
		\begin{align*}
			\textsc{V}_{+,u,v} \left ( x_n \right ) & \equiv \Var \left [ \widehat{\mu}^{(u)}_{Y+,v} \left ( x_n \right ) + \frac{1}{\mu^{(\nu)}_{D+}} \widehat{\mu}^{(u)}_{W+,v} \left ( x_n \right ) - \frac{\mu^{(\nu)}_{W+}}{\left ( \mu^{(\nu)}_{D+} \right )^2} \widehat{\mu}^{(u)}_{D+,v} \left ( x_n \right ) \Bigg | \bm{R} \right ] \\
			\textsc{V}_{-,u,v} \left ( x_n \right ) & \equiv \Var \left [ \widehat{\mu}^{(u)}_{Y-,v} \left ( x_n \right ) + \frac{1}{\mu^{(\nu)}_{D-}} \widehat{\mu}^{(u)}_{W-,v} \left ( x_n \right ) - \frac{\mu^{(\nu)}_{W-}}{\left ( \mu^{(\nu)}_{D-} \right )^2} \widehat{\mu}^{(u)}_{D-,v} \left ( x_n \right ) \Bigg | \bm{R} \right ] \\
			\textsc{C}_{+,\nu,p,q} \left ( h_n, b_n \right ) & \equiv \widetilde{\textsc{C}}_{Y+,\nu,p,q} \left ( h_n, b_n \right ) + \widetilde{\textsc{C}}_{+,\nu,p,q} \left ( h_n, b_n \right ) \\
			\textsc{C}_{-,\nu,p,q} \left ( h_n, b_n \right ) & \equiv \widetilde{\textsc{C}}_{Y-,\nu,p,q} \left ( h_n, b_n \right ) + \widetilde{\textsc{C}}_{-,\nu,p,q} \left ( h_n, b_n \right )
		\end{align*}
		and
		\begin{align*}
			\widetilde{\textsc{C}}_{Y+,\nu,p,q} \left ( h_n, b_n \right ) & \equiv \mathcal{C}_{YY+,\nu,p,q} \left ( h_n, b_n \right ) + \frac{1}{\mu^{(\nu)}_{D+}} \mathcal{C}_{YW+,\nu,p,q} \left ( h_n, b_n \right ) - \frac{\mu^{(\nu)}_{W+}}{\left ( \mu^{(\nu)}_{D+} \right )^2} \mathcal{C}_{YD+,\nu,p,q} \left ( h_n, b_n \right ) \\
			\widetilde{\textsc{C}}_{Y-,\nu,p,q} \left ( h_n, b_n \right ) & \equiv \mathcal{C}_{YY-,\nu,p,q} \left ( h_n, b_n \right ) + \frac{1}{\mu^{(\nu)}_{D-}} \mathcal{C}_{YW-,\nu,p,q} \left ( h_n, b_n \right ) - \frac{\mu^{(\nu)}_{W-}}{\left ( \mu^{(\nu)}_{D-} \right )^2} \mathcal{C}_{YD-,\nu,p,q} \left ( h_n, b_n \right )
		\end{align*}
		and
		\begin{align*}
			\widetilde{\textsc{C}}_{+,\nu,p,q} \left ( h_n, b_n \right ) & \equiv \frac{1}{\mu^{(\nu)}_{D+}} \mathcal{C}_{WY+,\nu,p,q} \left ( h_n, b_n \right ) - \frac{\mu^{(\nu)}_{W+}}{\left ( \mu^{(\nu)}_{D+} \right )^2} \mathcal{C}_{DY+,\nu,p,q} \left ( h_n, b_n \right ) \\
			& + \frac{1}{\left ( \mu^{(\nu)}_{D+} \right )^2} \mathcal{C}_{WW+,\nu,p,q} \left ( h_n, b_n \right ) - \frac{\mu^{(\nu)}_{W+}}{\left ( \mu^{(\nu)}_{D+} \right )^3} \mathcal{C}_{DW+,\nu,p,q} \left ( h_n, b_n \right ) \\
			& - \frac{\mu^{(\nu)}_{W+}}{\left ( \mu^{(\nu)}_{D+} \right )^3} \mathcal{C}_{WD+,\nu,p,q} \left ( h_n, b_n \right ) + \frac{\left ( \mu^{(\nu)}_{W+} \right )^2}{\left ( \mu^{(\nu)}_{D+} \right )^4} \mathcal{C}_{DD+,\nu,p,q} \left ( h_n, b_n \right ) \\
			\widetilde{\textsc{C}}_{-,\nu,p,q} \left ( h_n, b_n \right ) & \equiv \frac{1}{\mu^{(\nu)}_{D-}} \mathcal{C}_{WY-,\nu,p,q} \left ( h_n, b_n \right ) - \frac{\mu^{(\nu)}_{W-}}{\left ( \mu^{(\nu)}_{D-} \right )^2} \mathcal{C}_{DY-,\nu,p,q} \left ( h_n, b_n \right ) \\
			& + \frac{1}{\left ( \mu^{(\nu)}_{D-} \right )^2} \mathcal{C}_{WW-,\nu,p,q} \left ( h_n, b_n \right ) - \frac{\mu^{(\nu)}_{W-}}{\left ( \mu^{(\nu)}_{D-} \right )^3} \mathcal{C}_{DW-,\nu,p,q} \left ( h_n, b_n \right ) \\
			& - \frac{\mu^{(\nu)}_{W-}}{\left ( \mu^{(\nu)}_{D-} \right )^3} \mathcal{C}_{WD-,\nu,p,q} \left ( h_n, b_n \right ) + \frac{\left ( \mu^{(\nu)}_{W-} \right )^2}{\left ( \mu^{(\nu)}_{D-} \right )^4} \mathcal{C}_{DD-,\nu,p,q} \left ( h_n, b_n \right )
		\end{align*}
	\end{enumerate}
	
	\begin{enumerate}[(D)]
		\item Suppose that $h_n \to 0$ and $n h_n^{1+2\nu} \to \infty$. Suppose also that $n \min \left \{ h_n^{2p+3}, b_n^{2p+3} \right \} \max \left \{ h_n^2, b_n^{2(q-p)} \right \} \to 0$ and $\kappa \max \left \{ h_n,b_n \right \} < \kappa_0$. Then
		\begin{align*}
			T^{\textsc{rbc}}_{\nu,p,q} \left ( h_n, b_n \right ) \equiv \frac{\widehat{\theta}^{\textsc{bc}}_{\nu,p,q} \left ( h_n, b_n \right ) - \theta_{\nu}}{\sqrt{\textsc{V}^{\textsc{bc}}_{\nu,p,q} \left ( h_n, b_n \right )}} \stackrel{d}{\to} \mathcal{N} \left ( 0,1 \right ) 
		\end{align*}
	\end{enumerate}
	
\end{theorem}

\appsubsubsection{General Theorem: Proof}

	\begin{enumerate}[(R)]
		\item By the definitions in Sections \ref{appsec_bc_bias} and \ref{appsec_bc_estim},
		\begin{align*}
			R^{\textsc{bc}}_{n} & = h_n^{p+1-\nu} \left ( \frac{1}{\widehat{\mu}^{(\nu)}_{D+,p} \left ( h_n \right )} -  \frac{1}{\mu^{(\nu)}_{D+}} \right ) \widehat{\textsc{B}}_{W+,\nu,p,q} \left ( h_n, b_n \right ) \\
			& - h_n^{p+1-\nu} \left ( \frac{\widehat{\mu}^{(\nu)}_{W+,p} \left ( h_n \right )}{\left ( \widehat{\mu}^{(\nu)}_{D+,p} \left ( h_n \right ) \right )^2} -  \frac{\mu^{(\nu)}_{W+}}{\left ( \mu^{(\nu)}_{D+} \right )^2} \right ) \widehat{\textsc{B}}_{D+,\nu,p,q} \left ( h_n, b_n \right ) \\
			& - h_n^{p+1-\nu} \left ( \frac{1}{\widehat{\mu}^{(\nu)}_{D-,p} \left ( h_n \right )} -  \frac{1}{\mu^{(\nu)}_{D-}} \right ) \widehat{\textsc{B}}_{W-,\nu,p,q} \left ( h_n, b_n \right ) \\
			& + h_n^{p+1-\nu} \left ( \frac{\widehat{\mu}^{(\nu)}_{W-,p} \left ( h_n \right )}{\left ( \widehat{\mu}^{(\nu)}_{D-,p} \left ( h_n \right ) \right )^2} -  \frac{\mu^{(\nu)}_{W-}}{\left ( \mu^{(\nu)}_{D-} \right )^2} \right ) \widehat{\textsc{B}}_{D-,\nu,p,q} \left ( h_n, b_n \right )
		\end{align*}
		For $A \in \left \{ W,D \right \}$, the conditional mean of $\widehat{\mu}^{(p+1)}_{A+,q} \left ( b_n \right )$ is $O_p \left ( 1 \right )$. Its conditional variance is $O_p \left ( 1 \slash \left ( n b_n^{2p+3} \right ) \right )$. Under the stated regularity conditions, these bounds hold provided that $n b_n \to \infty$ and $\kappa b_n < \kappa_0$. The conditional Chebyshev inequality therefore gives
		\begin{align*}
			e_{p+1}' \widehat{\beta}_{A+,q} \left ( b_n \right ) = \frac{\widehat{\mu}^{(p+1)}_{A+,q} \left ( b_n \right )}{\left ( p+1 \right )!} = O_p \left ( 1 + \sqrt{\frac{1}{n b_n^{2p+3}}} \right )
		\end{align*}
		Since $\mathcal{B}_{+,\nu,p,p+1} \left ( h_n \right ) = O_p \left ( 1 \right )$, the definition in Section
		\ref{appsec_bc_bias} implies
		\begin{align*}
			\widehat{\textsc{B}}_{A+,\nu,p,q}
			\left ( h_n,b_n \right )
			= O_p \left (
				1 + \frac{1}{\sqrt{n b_n^{2p+3}}}
			\right )
		\end{align*}
		In addition, the proof of part (R) of Theorem \ref{thm_asydist} yields
		\begin{align*}
			\widehat{\mu}^{(\nu)}_{A+,p} \left ( h_n \right )
			- \mu^{(\nu)}_{A+}
			= O_p \left (
				h_n^{p+1-\nu}
				+ \frac{1}{\sqrt{n h_n^{1+2\nu}}}
			\right )
		\end{align*}
		for $A \in \left \{ W,D \right \}$. Since $h_n \to 0$ and $n h_n^{1+2\nu} \to \infty$, the estimator of $\mu^{(\nu)}_{D+}$ converges in probability to its nonzero population value. The reciprocal and the ratio with the squared denominator are continuously differentiable near these population values. A first-order Taylor expansion therefore gives
		\begin{align*}
			\frac{1}{\widehat{\mu}^{(\nu)}_{D+,p} \left ( h_n \right )} -  \frac{1}{\mu^{(\nu)}_{D+}} & = O_p \left ( h_n^{p+1-\nu} + \sqrt{\frac{1}{n h_n^{1 + 2\nu}}} \right ) \\
			\frac{\widehat{\mu}^{(\nu)}_{W+,p} \left ( h_n \right )}{\left ( \widehat{\mu}^{(\nu)}_{D+,p} \left ( h_n \right ) \right )^2} -  \frac{\mu^{(\nu)}_{W+}}{\left ( \mu^{(\nu)}_{D+} \right )^2} & = O_p \left ( h_n^{p+1-\nu} + \sqrt{\frac{1}{n h_n^{1 + 2\nu}}} \right )
		\end{align*}
		The argument for the lower-side estimators is symmetric. Substituting these bounds into the expression for
		$R^{\textsc{bc}}_n$ gives
		\begin{align*}
			R^{\textsc{bc}}_{n} & = h_n^{p+1-\nu} O_p \left ( h_n^{p+1-\nu} + \sqrt{\frac{1}{n h_n^{1 + 2\nu}}} \right ) O_p \left ( 1 + \sqrt{\frac{1}{n b_n^{2p+3}}} \right ) \\
			& = O_p \left ( h_n^{2 (p+1-\nu)} + \frac{h_n^{p+1-\nu}}{\sqrt{n h_n^{1 + 2\nu}}} \right ) O_p \left ( 1 + \sqrt{\frac{1}{n b_n^{2p+3}}} \right )
		\end{align*}
		which completes the proof.
	\end{enumerate}
	
	\begin{enumerate}[(B)]
		\item The conditional expectation of the bias-corrected linearized estimator $\widetilde{\theta}^{\textsc{bc}}_{\nu,p,q} \left ( h_n, b_n \right )$ is
		\begin{align*}
			\E \left [ \widetilde{\theta}^{\textsc{bc}}_{\nu,p,q} \left ( h_n, b_n \right ) \Big | \bm{R} \right ] & = \E \left [ \widetilde{\theta}_{\nu,p} \left ( h_n \right ) - h_n^{p+1-\nu} \widecheck{\textsc{B}}_{\nu,p,q} \left ( h_n, b_n \right ) | \bm{R} \right ] = \textsc{B}_1 - h_n^{p+1-\nu} \textsc{B}_2
		\end{align*}
		First, define
		\begin{align*}
			\textsc{B}_1 & \equiv \E \left [ \widetilde{\theta}_{\nu,p} \left ( h_n \right ) - h_n^{p+1-\nu}
				\textsc{B}_{\nu,p,p+1} \left ( h_n \right ) \Big | \bm{R} \right ] \\
			& = \mu^{(\nu)}_{Y+} - \mu^{(\nu)}_{Y-} + h_n^{p+2-\nu} \textsc{B}_{\nu,p,p+2} \left ( h_n \right )
			+ o_p \left ( h_n^{p+2-\nu} \right )
		\end{align*}
		The equality follows from part (B) of Theorem \ref{thm_asydist}. The two terms involving $h_n^{p+1-\nu} \textsc{B}_{\nu,p,p+1} \left ( h_n \right )$ cancel. Second, define
		\begin{align*}
			\textsc{B}_2
			& \equiv \E \left [ \widecheck{\textsc{B}}_{\nu,p,q} \left ( h_n,b_n \right ) - \textsc{B}_{\nu,p,p+1} \left ( h_n \right ) \Big | \bm{R} \right ]
		\end{align*}
		Recall that
		\begin{align*}
			\E \left [ \widehat{\textsc{B}}_{Y,\nu,p,q} \left ( h_n, b_n \right ) \Big | \bm{R} \right ] & = \E \Big [ e_{p+1}' \widehat{\beta}_{Y+,q} \left ( b_n \right ) \mathcal{B}_{+,\nu,p,p+1} \left ( h_n \right ) \\
			& - e_{p+1}' \widehat{\beta}_{Y-,q} \left ( b_n \right ) \mathcal{B}_{-,\nu,p,p+1} \left ( h_n \right ) \Big | \bm{R} \Big ] \\
			\textsc{B}_{Y,\nu,p,p+1} \left ( h_n \right ) & \equiv e_{p+1}' \beta_{Y+,q} \mathcal{B}_{+,\nu,p,p+1} \left ( h_n \right ) - e_{p+1}' \beta_{Y-,q} \mathcal{B}_{-,\nu,p,p+1} \left ( h_n \right )
		\end{align*}
		which implies that
		\begin{align*}
			& \E \left [ \widehat{\textsc{B}}_{Y,\nu,p,q} \left ( h_n, b_n \right ) \Big | \bm{R} \right ] - \textsc{B}_{Y,\nu,p,p+1} \left ( h_n \right ) \\
			& = e_{p+1}' \left ( \E \left [ \widehat{\beta}_{Y+,q} \left ( b_n \right ) \Big | \bm{R} \right ] - \beta_{Y+,q} \right ) \mathcal{B}_{+,\nu,p,p+1} \left ( h_n \right ) \\
			& - e_{p+1}' \left ( \E \left [ \widehat{\beta}_{Y-,q} \left ( b_n \right ) \Big | \bm{R} \right ] - \beta_{Y-,q} \right ) \mathcal{B}_{-,\nu,p,p+1} \left ( h_n \right )
		\end{align*}
		Similarly,
		\begin{align*}
			\E \left [ \widecheck{\widetilde{\textsc{B}}}_{+,\nu,p,q} \left ( h_n, b_n \right ) \Big | \bm{R} \right ] & = \E \left [ \frac{1}{\mu^{(\nu)}_{D+}} \widehat{\textsc{B}}_{W+,\nu,p,q} \left ( h_n, b_n \right ) - \frac{\mu^{(\nu)}_{W+}}{\left ( \mu^{(\nu)}_{D+} \right )^2} \widehat{\textsc{B}}_{D+,\nu,p,q} \left ( h_n, b_n \right ) \Bigg | \bm{R} \right ] \\
			& = \frac{1}{\mu^{(\nu)}_{D+}} \E \left [ \widehat{\textsc{B}}_{W+,\nu,p,q} \left ( h_n, b_n \right ) \Big | \bm{R} \right ] \\
			& - \frac{\mu^{(\nu)}_{W+}}{\left ( \mu^{(\nu)}_{D+} \right )^2} \E \left [ \widehat{\textsc{B}}_{D+,\nu,p,q} \left ( h_n, b_n \right ) \Big | \bm{R} \right ] \\
			& = \frac{1}{\mu^{(\nu)}_{D+}} \E \left [ e_{p+1}' \widehat{\beta}_{W+,q} \left ( b_n \right ) \mathcal{B}_{+,\nu,p,p+1} \left ( h_n \right ) \Big | \bm{R} \right ] \\
			& - \frac{\mu^{(\nu)}_{W+}}{\left ( \mu^{(\nu)}_{D+} \right )^2} \E \left [ e_{p+1}' \widehat{\beta}_{D+,q} \left ( b_n \right ) \mathcal{B}_{+,\nu,p,p+1} \left ( h_n \right ) \Big | \bm{R} \right ] \\
			\widetilde{\textsc{B}}_{+,\nu,p,p+1} \left ( h_n \right ) & \equiv \frac{1}{\mu^{(\nu)}_{D+}} \textsc{B}_{W+,\nu,p,p+1} \left ( h_n \right ) - \frac{\mu^{(\nu)}_{W+}}{\left ( \mu^{(\nu)}_{D+} \right )^2} \textsc{B}_{D+,\nu,p,p+1} \left ( h_n \right ) \\
			& = \frac{1}{\mu^{(\nu)}_{D+}} e_{p+1}' \beta_{W+,q} \mathcal{B}_{+,\nu,p,p+1} \left ( h_n \right ) \\
			& - \frac{\mu^{(\nu)}_{W+}}{\left ( \mu^{(\nu)}_{D+} \right )^2} e_{p+1}' \beta_{D+,q} \mathcal{B}_{+,\nu,p,p+1} \left ( h_n \right )
		\end{align*}
		which implies that
		\begin{align*}
			& \E \left [ \widecheck{\widetilde{\textsc{B}}}_{+,\nu,p,q} \left ( h_n, b_n \right ) \Big | \bm{R} \right ] - \widetilde{\textsc{B}}_{+,\nu,p,p+1} \left ( h_n \right ) \\
			& = \frac{1}{\mu^{(\nu)}_{D+}} e_{p+1}' \left ( \E \left [ \widehat{\beta}_{W+,q} \left ( b_n \right ) \Big | \bm{R} \right ] - \beta_{W+,q} \right ) \mathcal{B}_{+,\nu,p,p+1} \left ( h_n \right ) \\
			& - \frac{\mu^{(\nu)}_{W+}}{\left ( \mu^{(\nu)}_{D+} \right )^2} e_{p+1}' \left ( \E \left [ \widehat{\beta}_{D+,q} \left ( b_n \right ) \Big | \bm{R} \right ] - \beta_{D+,q} \right ) \mathcal{B}_{+,\nu,p,p+1} \left ( h_n \right )
		\end{align*}
		The corresponding expression for the lower side follows by symmetry. Combining these derivations,
		\begin{align*}
			\textsc{B}_2 & \equiv e_{p+1}' \left ( \E \left [ \widehat{\beta}_{Y+,q} \left ( b_n \right ) \Big | \bm{R} \right ] - \beta_{Y+,q} \right ) \mathcal{B}_{+,\nu,p,p+1} \left ( h_n \right ) \\
			& - e_{p+1}' \left ( \E \left [ \widehat{\beta}_{Y-,q} \left ( b_n \right ) \Big | \bm{R} \right ] - \beta_{Y-,q} \right ) \mathcal{B}_{-,\nu,p,p+1} \left ( h_n \right ) \\
			& + \frac{1}{\mu^{(\nu)}_{D+}} e_{p+1}' \left ( \E \left [ \widehat{\beta}_{W+,q} \left ( b_n \right ) \Big | \bm{R} \right ] - \beta_{W+,q} \right ) \mathcal{B}_{+,\nu,p,p+1} \left ( h_n \right ) \\
			& - \frac{\mu^{(\nu)}_{W+}}{\left ( \mu^{(\nu)}_{D+} \right )^2} e_{p+1}' \left ( \E \left [ \widehat{\beta}_{D+,q} \left ( b_n \right ) \Big | \bm{R} \right ] - \beta_{D+,q} \right ) \mathcal{B}_{+,\nu,p,p+1} \left ( h_n \right ) \\
			& - \frac{1}{\mu^{(\nu)}_{D-}} e_{p+1}' \left ( \E \left [ \widehat{\beta}_{W-,q} \left ( b_n \right ) \Big | \bm{R} \right ] - \beta_{W-,q} \right ) \mathcal{B}_{-,\nu,p,p+1} \left ( h_n \right ) \\
			& + \frac{\mu^{(\nu)}_{W-}}{\left ( \mu^{(\nu)}_{D-} \right )^2} e_{p+1}' \left ( \E \left [ \widehat{\beta}_{D-,q} \left ( b_n \right ) \Big | \bm{R} \right ] - \beta_{D-,q} \right ) \mathcal{B}_{-,\nu,p,p+1} \left ( h_n \right )
		\end{align*}
		Furthermore, by several applications of Lemma S.A.3 in \cite{cct2014supp}, for any random variable $A \in \left \{ Y, W, D \right \}$,
		\begin{align*}
			\E \left [ \left ( p+1 \right )! e_{p+1}' \widehat{\beta}_{A+,q} \left ( b_n \right ) \Big | \bm{R} \right ] & = \left ( p+1 \right )! e_{p+1}' \beta_{A+,q} + b_n^{q-p} \frac{\mu^{(q+1)}_{A+}}{\left ( q+1 \right )!} \mathcal{B}_{+,p+1,q,q+1} \left ( b_n \right ) + o_p \left ( b_n^{q-p} \right ) \\
			\E \left [ \left ( p+1 \right )! e_{p+1}' \widehat{\beta}_{A-,q} \left ( b_n \right ) \Big | \bm{R} \right ] & = \left ( p+1 \right )! e_{p+1}' \beta_{A-,q} + b_n^{q-p} \frac{\mu^{(q+1)}_{A-}}{\left ( q+1 \right )!} \mathcal{B}_{-,p+1,q,q+1} \left ( b_n \right ) + o_p \left ( b_n^{q-p} \right )
		\end{align*}
		Thus,
		\begin{align*}
			\textsc{B}_2 & = \left ( b_n^{q-p} \frac{\mu^{(q+1)}_{Y+}}{\left ( q+1 \right )!} \mathcal{B}_{+,p+1,q,q+1} \left ( b_n \right ) + o_p \left ( b_n^{q-p} \right ) \right ) \frac{\mathcal{B}_{+,\nu,p,p+1} \left ( h_n \right )}{\left ( p+1 \right )!} \\
			& - \left ( b_n^{q-p} \frac{\mu^{(q+1)}_{Y-}}{\left ( q+1 \right )!} \mathcal{B}_{-,p+1,q,q+1} \left ( b_n \right ) + o_p \left ( b_n^{q-p} \right ) \right ) \frac{\mathcal{B}_{-,\nu,p,p+1} \left ( h_n \right )}{\left ( p+1 \right )!} \\
			& + \frac{1}{\mu^{(\nu)}_{D+}} \left ( b_n^{q-p} \frac{\mu^{(q+1)}_{W+}}{\left ( q+1 \right )!} \mathcal{B}_{+,p+1,q,q+1} \left ( b_n \right ) + o_p \left ( b_n^{q-p} \right ) \right ) \frac{\mathcal{B}_{+,\nu,p,p+1} \left ( h_n \right )}{\left ( p+1 \right )!} \\
			& - \frac{\mu^{(\nu)}_{W+}}{\left ( \mu^{(\nu)}_{D+} \right )^2} \left ( b_n^{q-p} \frac{\mu^{(q+1)}_{D+}}{\left ( q+1 \right )!} \mathcal{B}_{+,p+1,q,q+1} \left ( b_n \right ) + o_p \left ( b_n^{q-p} \right ) \right ) \frac{\mathcal{B}_{+,\nu,p,p+1} \left ( h_n \right )}{\left ( p+1 \right )!} \\
			& - \frac{1}{\mu^{(\nu)}_{D-}} \left ( b_n^{q-p} \frac{\mu^{(q+1)}_{W-}}{\left ( q+1 \right )!} \mathcal{B}_{-,p+1,q,q+1} \left ( b_n \right ) + o_p \left ( b_n^{q-p} \right ) \right ) \frac{\mathcal{B}_{-,\nu,p,p+1} \left ( h_n \right )}{\left ( p+1 \right )!} \\
			& + \frac{\mu^{(\nu)}_{W-}}{\left ( \mu^{(\nu)}_{D-} \right )^2} \left ( b_n^{q-p} \frac{\mu^{(q+1)}_{D-}}{\left ( q+1 \right )!} \mathcal{B}_{-,p+1,q,q+1} \left ( b_n \right ) + o_p \left ( b_n^{q-p} \right ) \right ) \frac{\mathcal{B}_{-,\nu,p,p+1} \left ( h_n \right )}{\left ( p+1 \right )!}
		\end{align*}
		As a consequence,
		\begin{align*}
			\textsc{B}_2 & = b_n^{q-p} \textsc{B}^{\textsc{bc}}_{\nu,p,q} \left ( h_n, b_n \right ) + o_p \left ( b_n^{q-p} \right )
		\end{align*}
		Taken together, the conditional expectation of the linearized bias-corrected estimator $\widetilde{\theta}^{\textsc{bc}}_{\nu,p,q} \left ( h_n, b_n \right )$ is
		\begin{align*}
			\E \left [ \widetilde{\theta}^{\textsc{bc}}_{\nu,p,q} \left ( h_n, b_n \right ) \Big | \bm{R} \right ] & = \textsc{B}_1 - h_n^{p+1-\nu} \textsc{B}_2 \\
			& = \mu^{(\nu)}_{Y+} - \mu^{(\nu)}_{Y-} + h_n^{p+2-\nu} \textsc{B}_{\nu,p,p+2} \left ( h_n \right ) - h_n^{p+1-\nu} b_n^{q-p} \textsc{B}^{\textsc{bc}}_{\nu,p,q} \left ( h_n,b_n \right ) \\
			& + o_p \left ( h_n^{p+2-\nu} + h_n^{p+1-\nu} b_n^{q-p} \right )
		\end{align*}
		which completes the proof.
	\end{enumerate}
	
	\begin{enumerate}[(V)]
		\item The conditional variance of the linearized bias-corrected estimator $\widetilde{\theta}^{\textsc{bc}}_{\nu,p,q} \left ( h_n, b_n \right )$ is
		\begin{align*}
			& \Var \left [ \widetilde{\theta}^{\textsc{bc}}_{\nu,p,q} \left ( h_n, b_n \right ) \Big | \bm{R} \right ] \\
			& = \Var \left [ \widehat{\mu}^{(\nu)}_{Y+,p} \left ( h_n \right ) + \widetilde{\eta}_{+,\nu,p} \left ( h_n \right ) - h_n^{p+1-\nu} \left ( \widehat{\textsc{B}}_{Y+,\nu,p,q} \left ( h_n, b_n \right ) + \widecheck{\widetilde{\textsc{B}}}_{+,\nu,p,q} \left ( h_n, b_n \right ) \right ) \Big | \bm{R} \right ] \\
			& + \Var \left [ \widehat{\mu}^{(\nu)}_{Y-,p} \left ( h_n \right ) + \widetilde{\eta}_{-,\nu,p} \left ( h_n \right ) - h_n^{p+1-\nu} \left ( \widehat{\textsc{B}}_{Y-,\nu,p,q} \left ( h_n, b_n \right ) + \widecheck{\widetilde{\textsc{B}}}_{-,\nu,p,q} \left ( h_n, b_n \right ) \right ) \Big | \bm{R} \right ]
		\end{align*}
		where the equality follows from the bilinearity of the covariance and the conditional independence of observations on either side of the cutoff $c$. For compactness, define
		\begin{align*}
			\textsc{V}^{\textsc{bc}}_{+,\nu,p,q} \left ( h_n, b_n \right ) & \equiv \Var \bigg [ \widehat{\mu}^{(\nu)}_{Y+,p} \left ( h_n \right ) + \widetilde{\eta}_{+,\nu,p} \left ( h_n \right ) \\
			& - h_n^{p+1-\nu} \left ( \widehat{\textsc{B}}_{Y+,\nu,p,q} \left ( h_n, b_n \right ) + \widecheck{\widetilde{\textsc{B}}}_{+,\nu,p,q} \left ( h_n, b_n \right ) \right ) \Big | \bm{R} \bigg ] \\
			& = \Var \left [ \widehat{\mu}^{(\nu)}_{Y+,p} \left ( h_n \right ) + \widetilde{\eta}_{+,\nu,p} \left ( h_n \right ) \Big | \bm{R} \right ] \\
			& + h_n^{2 (p+1-\nu)} \Var \left [ \widehat{\textsc{B}}_{Y+,\nu,p,q} \left ( h_n, b_n \right ) + \widecheck{\widetilde{\textsc{B}}}_{+,\nu,p,q} \left ( h_n, b_n \right ) \Big | \bm{R} \right ] \\
			& - 2 h_n^{p+1-\nu} \Cov \left [ \widehat{\mu}^{(\nu)}_{Y+,p} \left ( h_n \right ), \widehat{\textsc{B}}_{Y+,\nu,p,q} \left ( h_n, b_n \right ) + \widecheck{\widetilde{\textsc{B}}}_{+,\nu,p,q} \left ( h_n, b_n \right ) \Big | \bm{R} \right ] \\
			& - 2 h_n^{p+1-\nu} \Cov \left [ \widetilde{\eta}_{+,\nu,p} \left ( h_n \right ), \widehat{\textsc{B}}_{Y+,\nu,p,q} \left ( h_n, b_n \right ) + \widecheck{\widetilde{\textsc{B}}}_{+,\nu,p,q} \left ( h_n, b_n \right ) \Big | \bm{R} \right ]
		\end{align*}
		First, leveraging the bilinearity of the covariance and the definitions from part (V) of Theorem \ref{thm_asydist},
		\begin{align*}
			\textsc{V}_{+,\nu,p} \left ( h_n \right ) \equiv \Var \left [ \widehat{\mu}^{(\nu)}_{Y+,p} \left ( h_n \right ) + \widetilde{\eta}_{+,\nu,p} \left ( h_n \right ) \Big | \bm{R} \right ] = \mathcal{V}_{YY+,\nu,p} \left ( h_n \right ) + \widetilde{\textsc{V}}_{+,\nu,p} \left ( h_n \right ) + 2 \widetilde{\textsc{V}}_{Y+,\nu,p} \left ( h_n \right )
		\end{align*}
		Second, the definitions of the estimated bias coefficients give
		\begin{align*}
			& \Var \left [
				\widehat{\textsc{B}}_{Y+,\nu,p,q}
				\left ( h_n,b_n \right )
				+ \widecheck{\widetilde{\textsc{B}}}_{+,\nu,p,q}
				\left ( h_n,b_n \right )
				\Big | \bm{R}
			\right ] \\
			& = \frac{ \mathcal{B}^2_{+,\nu,p,p+1} \left ( h_n \right )}{\left ( p+1 \right )!^2} \Var \left [ \widehat{\mu}^{(p+1)}_{Y+,q} \left ( b_n \right ) + \frac{1}{\mu^{(\nu)}_{D+}} \widehat{\mu}^{(p+1)}_{W+,q} \left ( b_n \right ) - \frac{\mu^{(\nu)}_{W+}}{\left ( \mu^{(\nu)}_{D+} \right )^2} \widehat{\mu}^{(p+1)}_{D+,q} \left ( b_n \right ) \Bigg | \bm{R} \right ] \\
			& = \frac{\mathcal{B}^2_{+,\nu,p,p+1} \left ( h_n \right )}{\left ( p+1 \right )!^2} \textsc{V}_{+,p+1,q} \left ( b_n \right )
		\end{align*}
		The factor involving $\mathcal{B}_{+,\nu,p,p+1} \left ( h_n \right )$ depends only on $\bm{R}$ and can therefore be factored out of the conditional variance. The last equality follows from the definition of $\textsc{V}_{+,p+1,q} \left ( b_n \right )$ in the statement. Third, expand the two conditional covariances using bilinearity. Then applying the definitions in the statement gives
		\begin{align*}
			\Cov \left [ \widehat{\mu}^{(\nu)}_{Y+,p} \left ( h_n \right ), \widehat{\textsc{B}}_{Y+,\nu,p,q} \left ( h_n, b_n \right ) + \widecheck{\widetilde{\textsc{B}}}_{+,\nu,p,q} \left ( h_n, b_n \right ) \Big | \bm{R} \right ] & = \widetilde{\textsc{C}}_{Y+,\nu,p,q} \left ( h_n, b_n \right ) \frac{\mathcal{B}_{+,\nu,p,p+1} \left ( h_n \right )}{\left ( p+1 \right )!} \\
			\Cov \left [ \widetilde{\eta}_{+,\nu,p} \left ( h_n \right ), \widehat{\textsc{B}}_{Y+,\nu,p,q} \left ( h_n, b_n \right ) + \widecheck{\widetilde{\textsc{B}}}_{+,\nu,p,q} \left ( h_n, b_n \right ) \Big | \bm{R} \right ] & = \widetilde{\textsc{C}}_{+,\nu,p,q} \left ( h_n, b_n \right ) \frac{\mathcal{B}_{+,\nu,p,p+1} \left ( h_n \right )}{\left ( p+1 \right )!}
		\end{align*}
		By the definition of $\textsc{C}_{+,\nu,p,q} \left ( h_n,b_n \right )$, adding these two expressions yields
		\begin{align*}
			\textsc{C}_{+,\nu,p,q} \left ( h_n,b_n \right ) \frac{\mathcal{B}_{+,\nu,p,p+1} \left ( h_n \right )}{\left ( p+1 \right )!}
		\end{align*}
		Combining these derivations,
		\begin{align*}
			\textsc{V}^{\textsc{bc}}_{+,\nu,p,q} \left ( h_n, b_n \right ) & = \textsc{V}_{+,\nu,p} \left ( h_n \right ) + h_n^{2 (p+1-\nu)} \textsc{V}_{+,p+1,q} \left ( b_n \right ) \frac{\mathcal{B}^2_{+,\nu,p,p+1} \left ( h_n \right )}{\left ( p+1 \right )!^2} \\
			& - 2 h_n^{p+1-\nu} \textsc{C}_{+,\nu,p,q} \left ( h_n, b_n \right ) \frac{\mathcal{B}_{+,\nu,p,p+1} \left ( h_n \right )}{\left ( p+1 \right )!}
		\end{align*}
		The expression for $\textsc{V}^{\textsc{bc}}_{-,\nu,p,q} \left ( h_n,b_n \right )$ follows by replacing the $+$ subscripts with $-$ subscripts throughout. To conclude, the conditional variance of the linearized bias-corrected estimator $\widetilde{\theta}^{\textsc{bc}}_{\nu,p,q} \left ( h_n, b_n \right )$ is
		\begin{align*}
			\textsc{V}^{\textsc{bc}}_{\nu,p,q} \left ( h_n, b_n \right ) & \equiv \Var \left [ \widetilde{\theta}^{\textsc{bc}}_{\nu,p,q} \left ( h_n, b_n \right ) \Big | \bm{R} \right ] = \textsc{V}^{\textsc{bc}}_{+,\nu,p,q} \left ( h_n, b_n \right ) + \textsc{V}^{\textsc{bc}}_{-,\nu,p,q} \left ( h_n, b_n \right )
		\end{align*}
	\end{enumerate}
	
	\begin{enumerate}[(D)]
		\item Local polynomial estimators are linear in the outcome. Thus, the upper-side component of
		$\widetilde{\theta}^{\textsc{bc}}_{\nu,p,q} \left ( h_n,b_n \right )$ is obtained by applying local polynomial estimation and bias correction to the outcome
		\begin{align*}
			Y + \frac{1}{\mu^{(\nu)}_{D+}} W
			- \frac{\mu^{(\nu)}_{W+}}
			{\left ( \mu^{(\nu)}_{D+} \right )^2} D
		\end{align*}
		The corresponding lower-side outcome follows by replacing the $+$ subscripts with $-$ subscripts throughout. By Assumption \ref{ass_moments}, both transformed outcomes have bounded fourth moments. Their conditional means have the required one-sided smoothness. The coefficient of $Y$ in each transformed outcome is one. Part (f) of Assumption \ref{ass_moments} therefore implies strictly positive one-sided limits for their conditional variances. By continuity, their conditional variances are bounded away from zero sufficiently close to $c$, on their respective sides. Conditional on $\bm{R}$, the variance of $\widetilde{\theta}^{\textsc{bc}}_{\nu,p,q}\left ( h_n,b_n \right )$ is a sum of squared estimation weights multiplied by the conditional variances of the transformed outcomes. Restricting this sum to observations sufficiently close to $c$ gives a lower bound for $\textsc{V}^{\textsc{bc}}_{\nu,p,q} \left ( h_n,b_n \right )$. Applying the kernel-weight argument in the proof of part (D) of Lemma S.A.4 in \cite{cct2014supp} then gives
		\begin{align*}
			\frac{1}{\textsc{V}^{\textsc{bc}}_{\nu,p,q} \left ( h_n,b_n \right )}
			= O_p \left ( \min \left \{ n h_n^{1+2\nu}, \frac{n b_n^{2p+3}}{h_n^{2(p+1-\nu)}} \right \} \right )
		\end{align*}
		In addition, by part (R) of Theorem \ref{thm_asydist},
		\begin{align*}
			R^2_n = O_p \left ( h_n^{4 (p+1-\nu)} + \frac{1}{n^2 h_n^{2+4\nu}} \right )
		\end{align*}
		Combining this expression with the inverse variance bound yields
		\begin{align*}
			\frac{R^2_n}{\textsc{V}^{\textsc{bc}}_{\nu,p,q} \left ( h_n, b_n \right )} = O_p \left ( n h_n^{2(p+1-\nu)} \min \left \{ h_n^{2p+3},b_n^{2p+3} \right \} + \frac{1}{n h_n^{1+2\nu}} \right ) = o_p \left ( 1 \right )
		\end{align*}
		Since $p \geq \nu$ and $h_n \to 0$, for sufficiently large $n$,
		\begin{align*}
			0 \leq n h_n^{2(p+1-\nu)} \min \left \{ h_n^{2p+3},b_n^{2p+3} \right \} \leq n h_n^2 \min \left \{ h_n^{2p+3},b_n^{2p+3} \right \} \to 0
		\end{align*}
		The convergence follows from the bandwidth condition in the statement. The other rate in the bound for
		$R_n^2$ divided by the conditional variance is $1 \slash \left ( n h_n^{1+2\nu} \right )$. This also converges to zero by assumption. Next, consider $R_n^{\textsc{bc}}$, the remainder arising from replacing the population coefficients in the bias correction with their estimators. By part (R) of this theorem, squaring and expanding the product gives
		\begin{align*}
			\left ( R_n^{\textsc{bc}} \right )^2 = O_p \Bigg ( h_n^{4(p+1-\nu)} + \frac{h_n^{2(p+1-\nu)}}{n h_n^{1+2\nu}} + \frac{h_n^{4(p+1-\nu)}}{n b_n^{2p+3}} + \frac{h_n^{2(p+1-\nu)}}{n^2 h_n^{1+2\nu} b_n^{2p+3}} \Bigg )
		\end{align*}
		Apply the inverse variance bound separately to the four terms:
		\begin{align*}
			\frac{h_n^{4(p+1-\nu)}}{\textsc{V}^{\textsc{bc}}_{\nu,p,q} \left ( h_n,b_n \right )} & = O_p \left ( n h_n^{2(p+1-\nu)} \min \left \{ h_n^{2p+3},b_n^{2p+3} \right \} \right ) \\
			\frac{ h_n^{2(p+1-\nu)}}{ n h_n^{1+2\nu} \textsc{V}^{\textsc{bc}}_{\nu,p,q} \left ( h_n,b_n \right )}
			& = O_p \left ( h_n^{2(p+1-\nu)} \right ) \\
			\frac{h_n^{4(p+1-\nu)}}{n b_n^{2p+3} \textsc{V}^{\textsc{bc}}_{\nu,p,q} \left ( h_n,b_n \right )}
			& = O_p \left ( h_n^{2(p+1-\nu)} \right ) \\
			\frac{h_n^{2(p+1-\nu)}}{n^2 h_n^{1+2\nu} b_n^{2p+3} \textsc{V}^{\textsc{bc}}_{\nu,p,q} \left ( h_n,b_n \right )}
			& = O_p \left ( \frac{1}{n h_n^{1+2\nu}} \right )
		\end{align*}
		Consequently,
		\begin{align*}
			\frac{\left ( R_n^{\textsc{bc}} \right )^2}{\textsc{V}^{\textsc{bc}}_{\nu,p,q} \left ( h_n,b_n \right )} = O_p \Bigg ( n h_n^{2(p+1-\nu)} \min \left \{ h_n^{2p+3},b_n^{2p+3} \right \} + h_n^{2(p+1-\nu)} + \frac{1}{n h_n^{1+2\nu}} \Bigg ) = o_p \left ( 1 \right )
		\end{align*}
		The rate $n h_n^{2(p+1-\nu)} \min \left \{ h_n^{2p+3},b_n^{2p+3} \right \}$ also appears in the bound for $R_n^2$ divided by the conditional variance. Its convergence to zero was established above. Furthermore, $h_n^{2(p+1-\nu)} \to 0$ because $h_n \to 0$ and $p+1-\nu \geq 1$. Finally, $1 \slash \left ( n h_n^{1+2\nu} \right ) \to 0$ by assumption. The argument used to prove part (D) of Theorem A.1 in \cite{cct2014supp} applies to the transformed outcomes defined above under the stated bandwidth conditions. It gives
		\begin{align*}
			\frac{\widetilde{\theta}^{\textsc{bc}}_{\nu,p,q} \left ( h_n,b_n \right ) - \left ( \mu^{(\nu)}_{Y+} - \mu^{(\nu)}_{Y-} \right )}{\sqrt{\textsc{V}^{\textsc{bc}}_{\nu,p,q} \left ( h_n,b_n \right )}}\stackrel{d}{\to} \mathcal{N} \left ( 0,1 \right )
		\end{align*}
		The bounds established above imply that $\left ( R_n-R_n^{\textsc{bc}} \right )$ divided by the conditional standard deviation converges in probability to zero. The robust bias-corrected test statistic for the estimator $\widehat{\theta}^{\textsc{bc}}_{\nu,p,q} \left ( h_n, b_n \right )$ is
		\begin{align*}
			T^{\textsc{rbc}}_{\nu,p,q} \left ( h_n, b_n \right ) & \equiv \frac{\widehat{\theta}^{\textsc{bc}}_{\nu,p,q} \left ( h_n, b_n \right ) - \theta_{\nu}}{\sqrt{\textsc{V}^{\textsc{bc}}_{\nu,p,q} \left ( h_n, b_n \right )}} \\
			& = \frac{\widetilde{\theta}^{\textsc{bc}}_{\nu,p,q} \left ( h_n, b_n \right ) - \left ( \mu^{(\nu)}_{Y+} - \mu^{(\nu)}_{Y-} \right )}{\sqrt{\textsc{V}^{\textsc{bc}}_{\nu,p,q} \left ( h_n, b_n \right )}} + \frac{R_n - R^{\textsc{bc}}_n}{\sqrt{\textsc{V}^{\textsc{bc}}_{\nu,p,q} \left ( h_n, b_n \right )}} \\
			& = \frac{\widetilde{\theta}^{\textsc{bc}}_{\nu,p,q} \left ( h_n, b_n \right ) - \left ( \mu^{(\nu)}_{Y+} - \mu^{(\nu)}_{Y-} \right )}{\sqrt{\textsc{V}^{\textsc{bc}}_{\nu,p,q} \left ( h_n, b_n \right )}} + o_p \left ( 1 \right ) \\
			& \stackrel{d}{\to} \mathcal{N} \left ( 0,1 \right )
		\end{align*}
		The conclusion therefore follows from Slutsky's Theorem.
	\end{enumerate}

\appsubsubsection{Proof of Proposition \ref{prop_confint}}\label{app_prop_confint}

The result follows by specializing parts (V) and (D) of Theorem \ref{thm_asydist_bias} to $\nu=0$, $p=1$, and $q=2$. In particular, the conditional variance of the bias-corrected linearized estimator $\widetilde{\theta}^{\textsc{bc}} \left ( h_n, b_n \right )$ is
\begin{align*}
	\textsc{V}^{\textsc{bc}} \left ( h_n, b_n \right ) \equiv \Var \left [ \widetilde{\theta}^{\textsc{bc}} \left ( h_n, b_n \right ) \Big | \bm{R} \right ] = \textsc{V}^{\textsc{bc}}_{+,0,1,2} \left ( h_n, b_n \right ) + \textsc{V}^{\textsc{bc}}_{-,0,1,2} \left ( h_n, b_n \right )
\end{align*}

\appsubsection{MSE-Optimal Bandwidths}

In this section, I first derive the asymptotic bandwidth $h_{\textsc{mse},\nu,p}$ that minimizes the leading terms of the conditional mean squared error (MSE) of the linearized estimator $\widetilde{\theta}_{\nu,p} \left ( h_n \right )$. I then specialize this result to obtain the main bandwidth. To select the pilot bandwidth, I minimize the leading terms of the conditional MSE of an estimator of the leading bias of $\widetilde{\theta} \left ( h_n \right )$. Before stating the general theorem, it is convenient to define additional recurring bias- and variance-related terms.

\appsubsubsection{Definitions of Bias Terms}\label{appsec_bw_bias}

Define the following population bias terms:
\begin{align}
	\dot{\mathcal{B}}_{+,\nu,p,q} \equiv \nu! e_{\nu}' \Gamma^{-1}_{p} \vartheta_{p,q} \qquad \dot{\mathcal{B}}_{-,\nu,p,q} \equiv \left ( -1 \right )^{\nu+q} \nu! e_{\nu}' \Gamma^{-1}_{p} \vartheta_{p,q}
\end{align}
Under the conditions of the theorem below, part (B) of Lemma S.A.3 in \cite{cct2014supp} gives
\begin{align}
	\mathcal{B}_{+,\nu,p,q} \left ( h_n \right ) = \dot{\mathcal{B}}_{+,\nu,p,q} + o_p \left ( 1 \right ) \qquad \mathcal{B}_{-,\nu,p,q} \left ( h_n \right ) = \dot{\mathcal{B}}_{-,\nu,p,q} + o_p \left ( 1 \right )
\end{align}
In turn, for any random variable $A$, define
\begin{align}
	\dot{\textsc{B}}_{A+,\nu,p,q} \equiv \frac{\mu^{(q)}_{A+}}{q!} \dot{\mathcal{B}}_{+,\nu,p,q} \qquad	\dot{\textsc{B}}_{A-,\nu,p,q} \equiv \frac{\mu^{(q)}_{A-}}{q!} \dot{\mathcal{B}}_{-,\nu,p,q}
\end{align}

\appsubsubsection{Definitions of Variance Terms}\label{appsec_bw_var}

For any pair of random variables $A$ and $B$, define the following population covariance terms:
\begin{align}
	\dot{\mathcal{V}}_{AB+,\nu,p} \equiv \frac{\sigma^2_{AB+}}{f_R \left ( c \right )} \nu!^2 e'_{\nu} \Gamma^{-1}_{p} \Psi_p \Gamma^{-1}_{p} e_{\nu} \\
	\dot{\mathcal{V}}_{AB-,\nu,p} \equiv \frac{\sigma^2_{AB-}}{f_R \left ( c \right )} \nu!^2 e'_{\nu} \Gamma^{-1}_{p} \Psi_p \Gamma^{-1}_{p} e_{\nu}
\end{align}
Under the conditions of the theorem below, part (V) of Lemma S.A.3 in \cite{cct2014supp} gives
\begin{align}
    \mathcal{V}_{AB+,\nu,p} \left ( h_n \right )
    & = \frac{1}{n h_n^{1+2\nu}} \left ( \dot{\mathcal{V}}_{AB+,\nu,p} + o_p \left ( 1 \right ) \right ) \\
    \mathcal{V}_{AB-,\nu,p} \left ( h_n \right )
    & = \frac{1}{n h_n^{1+2\nu}} \left ( \dot{\mathcal{V}}_{AB-,\nu,p} + o_p \left ( 1 \right ) \right )
\end{align}

\appsubsubsection{General Theorem: Statement}

\begin{theorem}\label{thm_optbandwidth}
	Suppose that Assumptions \ref{ass_moments} and \ref{ass_kernel} hold with $\delta \geq p+1$ and $0 \leq \nu \leq p$. Let $\mu^{(\nu)}_{D+} \neq 0$ and $\mu^{(\nu)}_{D-} \neq 0$. If $h_n \to 0$ and $n h_n \to \infty$, then the conditional mean squared error of the linearized estimator $\widetilde{\theta}_{\nu,p} \left ( h_n \right )$ is
	\begin{align*}
		\text{MSE} \left ( \widetilde{\theta}_{\nu,p} \left ( h_n \right ) \right ) & = h^{2(p+1-\nu)}_{n} \left ( \dot{\textsc{B}}^2_{\nu,p,p+1} + o_p \left ( 1 \right ) \right ) + \frac{1}{n h_n^{1+2\nu}} \left ( \dot{\textsc{V}}_{\nu,p} + o_p \left ( 1 \right ) \right )
	\end{align*}
	First,
	\begin{align*}
		\dot{\textsc{B}}_{\nu,p,q} \equiv \dot{\textsc{B}}_{Y,\nu,p,q} + \dot{\widetilde{\textsc{B}}}_{\nu,p,q}
	\end{align*}
	with
	\begin{align*}
		\dot{\textsc{B}}_{Y,\nu,p,q} & \equiv \dot{\textsc{B}}_{Y+,\nu,p,q} - \dot{\textsc{B}}_{Y-,\nu,p,q} \\
		\dot{\widetilde{\textsc{B}}}_{\nu,p,q} & \equiv \dot{\widetilde{\textsc{B}}}_{+,\nu,p,q} - \dot{\widetilde{\textsc{B}}}_{-,\nu,p,q}
	\end{align*}
	and
	\begin{align*}
		\dot{\widetilde{\textsc{B}}}_{+,\nu,p,q} & \equiv \frac{1}{\mu^{(\nu)}_{D+}} \dot{\textsc{B}}_{W+,\nu,p,q} - \frac{\mu^{(\nu)}_{W+}}{\left ( \mu^{(\nu)}_{D+} \right )^2} \dot{\textsc{B}}_{D+,\nu,p,q} \\
		\dot{\widetilde{\textsc{B}}}_{-,\nu,p,q} & \equiv \frac{1}{\mu^{(\nu)}_{D-}} \dot{\textsc{B}}_{W-,\nu,p,q} - \frac{\mu^{(\nu)}_{W-}}{\left ( \mu^{(\nu)}_{D-} \right )^2} \dot{\textsc{B}}_{D-,\nu,p,q}
	\end{align*}
	Second, 
	\begin{align*}
		\dot{\textsc{V}}_{\nu,p} \equiv \dot{\textsc{V}}_{Y,\nu,p} + \dot{\widetilde{\textsc{V}}}_{\nu,p} + 2 \dot{\widetilde{\textsc{V}}}_{Y,\nu,p}
	\end{align*}
	where
	\begin{align*}
		\dot{\textsc{V}}_{Y,\nu,p} & \equiv \dot{\mathcal{V}}_{YY+,\nu,p} + \dot{\mathcal{V}}_{YY-,\nu,p} \\
		\dot{\widetilde{\textsc{V}}}_{\nu,p} & \equiv \dot{\widetilde{\textsc{V}}}_{+,\nu,p} + \dot{\widetilde{\textsc{V}}}_{-,\nu,p} \\
		\dot{\widetilde{\textsc{V}}}_{Y,\nu,p} & \equiv \dot{\widetilde{\textsc{V}}}_{Y+,\nu,p} + \dot{\widetilde{\textsc{V}}}_{Y-,\nu,p}
	\end{align*}
	with
	\begin{align*}
		\dot{\widetilde{\textsc{V}}}_{+,\nu,p} & \equiv \frac{1}{\left ( \mu^{(\nu)}_{D+} \right )^2} \dot{\mathcal{V}}_{WW+,\nu,p} + \frac{\left ( \mu^{(\nu)}_{W+} \right )^2}{\left ( \mu^{(\nu)}_{D+} \right )^4} \dot{\mathcal{V}}_{DD+,\nu,p} - \frac{2 \mu^{(\nu)}_{W+}}{\left ( \mu^{(\nu)}_{D+} \right )^3} \dot{\mathcal{V}}_{WD+,\nu,p} \\
		\dot{\widetilde{\textsc{V}}}_{-,\nu,p} & \equiv \frac{1}{\left ( \mu^{(\nu)}_{D-} \right )^2} \dot{\mathcal{V}}_{WW-,\nu,p} + \frac{\left ( \mu^{(\nu)}_{W-} \right )^2}{\left ( \mu^{(\nu)}_{D-} \right )^4} \dot{\mathcal{V}}_{DD-,\nu,p} - \frac{2 \mu^{(\nu)}_{W-}}{\left ( \mu^{(\nu)}_{D-} \right )^3} \dot{\mathcal{V}}_{WD-,\nu,p}
	\end{align*}
	and
	\begin{align*}
		\dot{\widetilde{\textsc{V}}}_{Y+,\nu,p} & \equiv \frac{1}{\mu^{(\nu)}_{D+}} \dot{\mathcal{V}}_{YW+,\nu,p} - \frac{\mu^{(\nu)}_{W+}}{\left ( \mu^{(\nu)}_{D+} \right )^2} \dot{\mathcal{V}}_{YD+,\nu,p} \\
		\dot{\widetilde{\textsc{V}}}_{Y-,\nu,p} & \equiv \frac{1}{\mu^{(\nu)}_{D-}} \dot{\mathcal{V}}_{YW-,\nu,p} - \frac{\mu^{(\nu)}_{W-}}{\left ( \mu^{(\nu)}_{D-} \right )^2} \dot{\mathcal{V}}_{YD-,\nu,p}
	\end{align*}
	Finally, if $\dot{\textsc{B}}_{\nu,p,p+1} \neq 0$, the bandwidth that minimizes the leading terms of the conditional MSE is
	\begin{align*}
		h_{\textsc{mse},\nu,p} = \left ( \frac{\left ( 1+2\nu \right ) \dot{\textsc{V}}_{\nu,p}}{2(p+1-\nu) \dot{\textsc{B}}^2_{\nu,p,p+1}} \right )^{\frac{1}{2p+3}} n^{-\frac{1}{2p+3}}
	\end{align*}
\end{theorem}

\appsubsubsection{General Theorem: Proof}

The conditional mean squared error of $\widetilde{\theta}_{\nu,p} \left ( h_n \right )$ is
\begin{align*}
	\text{MSE} \left ( \widetilde{\theta}_{\nu,p} \left ( h_n \right ) \right ) & \equiv \E \left [ \left ( \widetilde{\theta}_{\nu,p} \left ( h_n \right ) - \left ( \mu^{(\nu)}_{Y+} - \mu^{(\nu)}_{Y-} \right ) \right )^2 \Big | \bm{R} \right ] \\
	& = \left ( \E \left [ \widetilde{\theta}_{\nu,p} \left ( h_n \right ) \Big | \bm{R} \right ] - \left ( \mu^{(\nu)}_{Y+} - \mu^{(\nu)}_{Y-} \right ) \right )^2 + \Var \left [ \widetilde{\theta}_{\nu,p} \left ( h_n \right ) \Big | \bm{R} \right ] 
\end{align*}
First, expand the conditional means of $Y$, $W$, and $D$ separately on each side of $c$ through order $p+1$. This follows the argument in the proof of part (B) of Theorem \ref{thm_asydist} and requires only $\delta \geq p+1$. The leading conditional bias is $h_n^{p+1-\nu} \textsc{B}_{\nu,p,p+1} \left ( h_n \right )$. By Section \ref{appsec_bw_bias}, the quantities $\mathcal{B}_{+,\nu,p,p+1} \left ( h_n \right )$ and $\mathcal{B}_{-,\nu,p,p+1} \left ( h_n \right )$ converge to $\dot{\mathcal{B}}_{+,\nu,p,p+1}$ and $\dot{\mathcal{B}}_{-,\nu,p,p+1}$, respectively. Their coefficients in $\textsc{B}_{\nu,p,p+1} \left ( h_n \right )$ are fixed population quantities. Thus, by the definitions in the statement,
\begin{align*}
    \textsc{B}_{\nu,p,p+1} \left ( h_n \right ) = \dot{\textsc{B}}_{\nu,p,p+1} + o_p \left ( 1 \right )
\end{align*}
Substituting this expression for the coefficient of the leading conditional bias yields
\begin{align*}
    \E \left [ \widetilde{\theta}_{\nu,p} \left ( h_n \right ) \Big | \bm{R} \right ] = \mu^{(\nu)}_{Y+} - \mu^{(\nu)}_{Y-} + h_n^{p+1-\nu} \left ( \dot{\textsc{B}}_{\nu,p,p+1} + o_p \left ( 1 \right ) \right )
\end{align*}
Hence, the squared conditional bias is
\begin{align*}
    \left ( \E \left [ \widetilde{\theta}_{\nu,p} \left ( h_n \right ) \Big | \bm{R} \right ] - \left ( \mu^{(\nu)}_{Y+} - \mu^{(\nu)}_{Y-} \right ) \right )^2 = h_n^{2(p+1-\nu)} \left ( \dot{\textsc{B}}^2_{\nu,p,p+1}
    + o_p \left ( 1 \right ) \right )
\end{align*}
Next, part (V) of Theorem \ref{thm_asydist} expresses the conditional variance as
\begin{align*}
    \Var \left [ \widetilde{\theta}_{\nu,p} \left ( h_n \right ) \Big | \bm{R} \right ] = \textsc{V}_{Y,\nu,p} \left ( h_n \right ) + \widetilde{\textsc{V}}_{\nu,p} \left ( h_n \right ) + 2 \widetilde{\textsc{V}}_{Y,\nu,p} \left ( h_n \right )
\end{align*}
Each term on the right-hand side is a linear combination of the conditional covariances $\mathcal{V}_{AB+,\nu,p} \left ( h_n \right )$ and $\mathcal{V}_{AB-,\nu,p} \left ( h_n \right )$. Substitute their expansions from Section \ref{appsec_bw_var} into these linear combinations. Factoring out $1 \slash \left ( n h_n^{1+2\nu} \right )$ leaves the population coefficient $\dot{\textsc{V}}_{\nu,p}$ defined in the statement. Therefore,
\begin{align*}
    \Var \left [ \widetilde{\theta}_{\nu,p} \left ( h_n \right ) \Big | \bm{R} \right ] = \frac{1}{n h_n^{1+2\nu}}
    \left ( \dot{\textsc{V}}_{\nu,p} + o_p \left ( 1 \right ) \right )
\end{align*}
Adding the squared conditional bias and the conditional variance gives the MSE expansion in the statement. Finally, part (f) of Assumption \ref{ass_moments} requires positive definite one-sided limits of the conditional covariance matrix of $Y$, $W$, and $D$. The linearized estimator combines estimates of the conditional derivatives of these variables. The coefficient multiplying the estimate for $Y$ is one on the upper side and minus one on the lower side. Thus, the limiting variance coefficient
$\dot{\textsc{V}}_{\nu,p}$ is strictly positive. If $\dot{\textsc{B}}_{\nu,p,p+1} \neq 0$,
the optimal bandwidth is chosen by minimizing
\begin{align*}
    h^{2(p+1-\nu)} \dot{\textsc{B}}^2_{\nu,p,p+1} + \frac{1}{n h^{1+2\nu}} \dot{\textsc{V}}_{\nu,p}
\end{align*}
with respect to $h > 0$. Differentiating with respect to $h$ and setting the derivative equal to zero yields
\begin{align*}
    h_{\textsc{mse},\nu,p}^{2p+3} = \frac{(1+2\nu)\dot{\textsc{V}}_{\nu,p}}{2(p+1-\nu)\dot{\textsc{B}}^2_{\nu,p,p+1}} \times n^{-1}
\end{align*}
This equation has a unique positive solution. The leading MSE criterion diverges as the bandwidth approaches zero or infinity. Thus, this solution is its unique minimizer and yields the bandwidth in the statement.

\appsubsubsection{Proof of Proposition \ref{prop_optbandwidth}}\label{app_prop_bandwidth}

For the main bandwidth, specialize Theorem \ref{thm_optbandwidth} to $\nu=0$ and $p=1$. Define
\begin{align*}
    \dot{\textsc{B}}_h
    & \equiv \dot{\textsc{B}}_{0,1,2}
    \qquad
    \dot{\textsc{V}}_h
    \equiv \dot{\textsc{V}}_{0,1}
\end{align*}
The theorem gives the first MSE expansion. If $\dot{\textsc{B}}_h \neq 0$, the resulting main bandwidth is
\begin{align*}
	h_{\textsc{mse}} = \left ( \frac{\dot{\textsc{V}}_{0,1}}{4 \dot{\textsc{B}}^2_{0,1,2}} \right )^{1 \slash 5} n^{- 1 \slash 5}
\end{align*}
For the pilot bandwidth, let $\widetilde{\textsc{B}}_h \left ( b_n \right )$ be an estimator of the leading bias coefficient $\dot{\textsc{B}}_h$. Since $\dot{\mathcal{B}}_{-,0,1,2} = \dot{\mathcal{B}}_{+,0,1,2}$, define
\begin{align*}
	\widetilde{\textsc{B}}_h \left ( b_n \right ) & \equiv \frac{\dot{\mathcal{B}}_{+,0,1,2}}{2} \Bigg [ \widehat{\mu}^{(2)}_{Y+,2} \left ( b_n \right ) - \widehat{\mu}^{(2)}_{Y-,2} \left ( b_n \right ) \\
	& + \frac{1}{\mu_{D+}} \widehat{\mu}^{(2)}_{W+,2} \left ( b_n \right ) - \frac{\mu_{W+}}{\mu_{D+}^2} \widehat{\mu}^{(2)}_{D+,2} \left ( b_n \right ) - \frac{1}{\mu_{D-}} \widehat{\mu}^{(2)}_{W-,2} \left ( b_n \right ) + \frac{\mu_{W-}}{\mu_{D-}^2} \widehat{\mu}^{(2)}_{D-,2} \left ( b_n \right ) \Bigg ]
\end{align*}
Starting from this definition, $\dot{\textsc{B}}_h$ can be obtained by replacing the estimators of second derivatives with their population counterparts. Thus, $\widetilde{\textsc{B}}_h \left ( b_n \right )$ estimates the leading bias coefficient using population coefficients from the original linearization of $\widehat{\theta} \left ( h_n \right )$ and limiting kernel constants. Define the coefficient of the leading conditional bias of $\widetilde{\textsc{B}}_h \left ( b_n \right )$ as
\begin{align*}
	\dot{\textsc{B}}_b & \equiv \frac{\dot{\mathcal{B}}_{+,0,1,2}}{2} \Bigg [ \dot{\textsc{B}}_{Y+,2,2,3} - \dot{\textsc{B}}_{Y-,2,2,3} \\
	& + \frac{1}{\mu_{D+}} \dot{\textsc{B}}_{W+,2,2,3} - \frac{\mu_{W+}}{\mu_{D+}^2} \dot{\textsc{B}}_{D+,2,2,3} - \frac{1}{\mu_{D-}} \dot{\textsc{B}}_{W-,2,2,3} + \frac{\mu_{W-}}{\mu_{D-}^2} \dot{\textsc{B}}_{D-,2,2,3} \Bigg ]
\end{align*}
Define the coefficient of the leading conditional variance as
\begin{align*}
	\dot{\textsc{V}}_b & \equiv \frac{\left ( \dot{\mathcal{B}}_{+,0,1,2} \right )^2}{4} \bigg [ \dot{\mathcal{V}}_{YY+,2,2} + \frac{1}{\mu_{D+}^2} \dot{\mathcal{V}}_{WW+,2,2} + \frac{\mu_{W+}^2}{\mu_{D+}^4} \dot{\mathcal{V}}_{DD+,2,2} \\
	& - \frac{2 \mu_{W+}}{\mu_{D+}^3} \dot{\mathcal{V}}_{WD+,2,2} + \frac{2}{\mu_{D+}} \dot{\mathcal{V}}_{YW+,2,2} - \frac{2 \mu_{W+}}{\mu_{D+}^2} \dot{\mathcal{V}}_{YD+,2,2} \\
	& + \dot{\mathcal{V}}_{YY-,2,2} + \frac{1}{\mu_{D-}^2} \dot{\mathcal{V}}_{WW-,2,2} + \frac{\mu_{W-}^2}{\mu_{D-}^4} \dot{\mathcal{V}}_{DD-,2,2} \\
	& - \frac{2 \mu_{W-}}{\mu_{D-}^3} \dot{\mathcal{V}}_{WD-,2,2} + \frac{2}{\mu_{D-}} \dot{\mathcal{V}}_{YW-,2,2} - \frac{2 \mu_{W-}}{\mu_{D-}^2} \dot{\mathcal{V}}_{YD-,2,2} \bigg ]
\end{align*}
Compute the conditional expectation and variance of $\widetilde{\textsc{B}}_h \left ( b_n \right )$ using the conditional bias and covariance expansions for the individual estimators of second-order derivatives. Collecting the resulting terms according to the definitions of $\dot{\textsc{B}}_b$ and $\dot{\textsc{V}}_b$ yields
\begin{align*}
	\E \left [ \widetilde{\textsc{B}}_h \left ( b_n \right ) \Big | \bm{R} \right ] & = \dot{\textsc{B}}_h + b_n \left ( \dot{\textsc{B}}_b + o_p \left ( 1 \right ) \right ) \\
	\Var \left [ \widetilde{\textsc{B}}_h \left ( b_n \right ) \Big | \bm{R} \right ] & = \frac{1}{n b_n^5} \left ( \dot{\textsc{V}}_b + o_p \left ( 1 \right ) \right )
\end{align*}
The conditional MSE is the sum of squared conditional bias and conditional variance:
\begin{align*}
	\text{MSE} \left ( \widetilde{\textsc{B}}_h \left ( b_n \right ) \right ) & \equiv \E \left [ \left ( \widetilde{\textsc{B}}_h \left ( b_n \right ) - \dot{\textsc{B}}_h \right )^2 \Big | \bm{R} \right ] = b_n^2 \left ( \dot{\textsc{B}}_b^2 + o_p \left ( 1 \right ) \right ) + \frac{1}{n b_n^5} \left ( \dot{\textsc{V}}_b + o_p \left ( 1 \right ) \right )
\end{align*}
If $\dot{\textsc{B}}_b \neq 0$, differentiating the leading MSE criterion with respect to the bandwidth gives
\begin{align*}
	2 b_n \dot{\textsc{B}}_b^2 - \frac{5 \dot{\textsc{V}}_b}{n b_n^6} = 0
\end{align*}
Thus, the optimal pilot bandwidth is
\begin{align*}
	b_{\textsc{mse}} = \left ( \frac{5 \dot{\textsc{V}}_b}{2 \dot{\textsc{B}}_b^2} \right )^{1 \slash 7} n^{-1 \slash 7}
\end{align*}
Both the main bandwidth and pilot bandwidth derivations require only $\delta \geq 3$.

\appsubsection{Assessing the Plausibility of Local Common Trends}\label{app_testct}

In this section, I construct a joint test of the pre-period restrictions used to assess the plausibility of Assumption \ref{ass_ct1}. As in standard difference-in-differences designs, the tests use outcomes observed before the focal round of treatment assignment. Consider a treatment history $h_{t-1}=\left ( h_{t-k-1},0_k \right )$. Units with this history are untreated in the $k$ periods preceding the focal round of treatment assignment. Then, based on Assumption \ref{ass_ct1}, an empirical researcher may specify a variety of conjectures depending on context and application. Let $t-k \leq u < v < t$ and define $\Delta_{u,v} \equiv Y_v - Y_u$.

Let $\ell = 1, \dots, \overline{\ell}$ index a fixed collection of comparisons. For each comparison, $G_\ell$ indicates membership in a selected group of nonzero subsequent treatment paths. The upper-limit restriction for comparison $\ell$ is
\begin{align}
	& \lim_{r \downarrow c} \E \left [ \Delta_{u,v} \bigg | H_{t-1} = h_{t-1}, R_t = r, \bigcap_{s=1}^{\tau} \left \{ D_{t+s} = 0 \right \} \right ] \notag \\
	& = \lim_{r \downarrow c} \E \left [ \Delta_{u,v} \big | H_{t-1} = h_{t-1}, R_t = r, G_\ell = 1 \right ]
\end{align}
The lower limit is defined symmetrically. In Section \ref{sec_check_ct}, three main possibilities of $G_\ell$ were considered.

\paragraph{Aggregate Comparisons} To pool all nonzero subsequent treatment paths, define
\begin{align}
	G_\ell \equiv \mathbb{I} \left [ \bigcup_{s=1}^{\tau} \left \{ D_{t+s} \neq 0 \right \} \right ]
\end{align}
The indicator $G_\ell$ equals one whenever at least one subsequent treatment state is nonzero.

\paragraph{Comparisons with Individual Treatment Paths} To compare the all-zero path with an individual nonzero
subsequent treatment path, define
\begin{align}
	G_\ell \equiv \mathbb{I} \left [ \left ( D_{t+1}, \dots, D_{t+\tau} \right ) = \left ( d_{t+1}, \dots, d_{t+\tau} \right ) \right ]
\end{align}
for any $\left ( d_{t+1}, \dots, d_{t+\tau} \right ) \neq 0_\tau$.

\paragraph{Intermediate Groupings} The grouping can be chosen to reflect features of subsequent treatment assignment relevant to the empirical application. For example, to pool paths with exactly one subsequent treated period, define
\begin{align}
	G_\ell \equiv \mathbb{I} \left [ \sum_{s=1}^{\tau} D_{t+s} = 1 \right ]
\end{align}
Alternatively, to pool paths involving treatment in period $t+1$, define
\begin{align}
	G_\ell \equiv \mathbb{I} \left [ D_{t+1} = 1 \right ]
\end{align}
Hereafter, to keep notation compact, I omit the event $H_{t-1} = h_{t-1}$ from the conditioning set and define the following random variables:
\begin{align}
	R \equiv R_t \qquad D \equiv \prod_{s=1}^{\tau} \left ( 1 - D_{t+s} \right ) \qquad J \equiv \Delta_{u,v} \times D \qquad K_\ell \equiv \Delta_{u,v} \times G_\ell
\end{align}
For all $\ell=1,\dots,\overline{\ell}$, suppose that $\mu_{D+}$, $\mu_{D-}$, $\mu_{G_\ell+}$, and $\mu_{G_\ell-}$ are strictly positive. By the Law of Iterated Expectations and the properties of limits, the upper- and lower-limit restrictions can then be written as
\begin{align}\label{eq:hypotheses}
    \pi_{\ell,+} & \equiv \frac{\mu_{J+}}{\mu_{D+}} - \frac{\mu_{K_\ell+}}{\mu_{G_\ell+}} = 0
    \qquad
    \pi_{\ell,-} \equiv \frac{\mu_{J-}}{\mu_{D-}} - \frac{\mu_{K_\ell-}}{\mu_{G_\ell-}} = 0
\end{align}
For consistency with previous sections, I consider more general parameters that involve derivatives of arbitrary order $\nu$. For $\nu>0$, suppose that $\mu^{(\nu)}_{D+}$, $\mu^{(\nu)}_{D-}$, $\mu^{(\nu)}_{G_\ell+}$, and $\mu^{(\nu)}_{G_\ell-}$ are nonzero.
\begin{align}
	\pi_{\ell,+,\nu} \equiv \frac{\mu^{(\nu)}_{J+}}{\mu^{(\nu)}_{D+}} - \frac{\mu^{(\nu)}_{K_\ell+}}{\mu^{(\nu)}_{G_\ell+}} \qquad \qquad \pi_{\ell,-,\nu} \equiv \frac{\mu^{(\nu)}_{J-}}{\mu^{(\nu)}_{D-}} - \frac{\mu^{(\nu)}_{K_\ell-}}{\mu^{(\nu)}_{G_\ell-}}
\end{align}
Setting $\nu=0$ recovers the pre-period restrictions defined above. In the remainder of this section, I construct a statistic to test $\pi_{\ell,+,\nu} = \pi_{\ell,-,\nu}=0$ jointly for all $\ell=1,\dots,\overline{\ell}$. The linearization and bias correction arguments follow those used for $\widehat{\theta}^{\textsc{bc}}_{\nu,p,q} \left ( h_n,b_n \right )$. In addition, the joint test requires the covariances of the estimated contrasts.

\appsubsubsection{Definitions of Estimators}

Given a bandwidth sequence $h_n$, natural estimators for the target parameters are
\begin{align}
	\widehat{\pi}_{\ell,+,\nu,p} \left ( h_n \right ) \equiv \frac{\widehat{\mu}^{(\nu)}_{J+,p} \left ( h_n \right )}{\widehat{\mu}^{(\nu)}_{D+,p} \left ( h_n \right )} - \frac{\widehat{\mu}^{(\nu)}_{K_\ell+,p} \left ( h_n \right )}{\widehat{\mu}^{(\nu)}_{G_\ell+,p} \left ( h_n \right )} \qquad \widehat{\pi}_{\ell,-,\nu,p} \left ( h_n \right ) \equiv \frac{\widehat{\mu}^{(\nu)}_{J-,p} \left ( h_n \right )}{\widehat{\mu}^{(\nu)}_{D-,p} \left ( h_n \right )} - \frac{\widehat{\mu}^{(\nu)}_{K_\ell-,p} \left ( h_n \right )}{\widehat{\mu}^{(\nu)}_{G_\ell-,p} \left ( h_n \right )}
\end{align}
A first-order Taylor expansion of $\widehat{\pi}_{\ell,+,\nu,p} \left ( h_n \right ) - \pi_{\ell,+,\nu}$ yields the following linearized estimator:
\begin{align}
	\widetilde{\pi}_{\ell,+,\nu,p} \left ( h_n \right ) & \equiv \frac{1}{\mu^{(\nu)}_{D+}} \left ( \widehat{\mu}^{(\nu)}_{J+,p} \left ( h_n \right ) - \mu^{(\nu)}_{J+} \right ) - \frac{\mu^{(\nu)}_{J+}}{\left ( \mu^{(\nu)}_{D+} \right )^2} \left ( \widehat{\mu}^{(\nu)}_{D+,p} \left ( h_n \right ) - \mu^{(\nu)}_{D+} \right ) \notag \\
	& - \frac{1}{\mu^{(\nu)}_{G_\ell+}} \left ( \widehat{\mu}^{(\nu)}_{K_\ell+,p} \left ( h_n \right ) - \mu^{(\nu)}_{K_\ell+} \right ) + \frac{\mu^{(\nu)}_{K_\ell+}}{\left ( \mu^{(\nu)}_{G_\ell+} \right )^2} \left ( \widehat{\mu}^{(\nu)}_{G_\ell+,p} \left ( h_n \right ) - \mu^{(\nu)}_{G_\ell+} \right )
\end{align}
The expression for $\widetilde{\pi}_{\ell,-,\nu,p} \left ( h_n \right )$ is symmetric.

\appsubsubsection{Bias of Linearized Estimators}

Suppose that the conditions in parts (a)-(c) of Assumption \ref{ass_moments} hold for $R$, $J$, $D$, $K_\ell$, and $G_\ell$, with $\delta \geq p+2$. Suppose also that Assumption \ref{ass_kernel} holds. If $h_n \to 0$ and $n h_n \to \infty$, the conditional expectation of the linearized estimator is
\begin{align}
	\E \left [ \widetilde{\pi}_{\ell,+,\nu,p} \left ( h_n \right ) \Big | \bm{R} \right ] & = h^{p+1-\nu}_{n} \textsc{B}_{\ell,+,\nu,p,p+1} \left ( h_n \right ) + h^{p+2-\nu}_{n} \textsc{B}_{\ell,+,\nu,p,p+2} \left ( h_n \right ) + o_p \left ( h^{p+2-\nu}_{n} \right )
\end{align}
where
\begin{align}
    \textsc{B}_{\ell,+,\nu,p,q} \left ( h_n \right ) & \equiv \textsc{B}_{1+,\nu,p,q} \left ( h_n \right ) - \textsc{B}_{\ell,2+,\nu,p,q} \left ( h_n \right )
\end{align}
with
\begin{align}
	\textsc{B}_{1+,\nu,p,q} \left ( h_n \right ) & \equiv \frac{1}{\mu^{(\nu)}_{D+}} \textsc{B}_{J+,\nu,p,q} \left ( h_n \right ) - \frac{\mu^{(\nu)}_{J+}}{\left ( \mu^{(\nu)}_{D+} \right )^2} \textsc{B}_{D+,\nu,p,q} \left ( h_n \right ) \\
	\textsc{B}_{\ell,2+,\nu,p,q} \left ( h_n \right ) & \equiv \frac{1}{\mu^{(\nu)}_{G_\ell +}} \textsc{B}_{K_\ell +,\nu,p,q} \left ( h_n \right ) - \frac{\mu^{(\nu)}_{K_\ell +}}{\left ( \mu^{(\nu)}_{G_\ell +} \right )^2} \textsc{B}_{G_\ell +,\nu,p,q} \left ( h_n \right )
\end{align}
The conditional expectation of the linearized estimator $\widetilde{\pi}_{\ell,-,\nu,p} \left ( h_n \right )$ is defined symmetrically.

\appsubsubsection{Variance of Linearized Estimators}

Whenever the local polynomial estimators are well defined, the conditional variance of the linearized estimator $\widetilde{\pi}_{\ell,+,\nu,p} \left ( h_n \right )$ is
\begin{align}
	\textsc{V}_{\ell,+,\nu,p} \left ( h_n \right ) \equiv \Var \left [ \widetilde{\pi}_{\ell,+,\nu,p} \left ( h_n \right ) \Big | \bm{R} \right ] = \textsc{V}_{11+,\nu,p} \left ( h_n \right ) + \textsc{V}_{\ell,22+,\nu,p} \left ( h_n \right ) - 2 \textsc{V}_{\ell,12+,\nu,p} \left ( h_n \right )
\end{align}
where
\begin{align}
	\textsc{V}_{11+,\nu,p} \left ( h_n \right ) & \equiv \frac{1}{\left ( \mu^{(\nu)}_{D+} \right )^2} \mathcal{V}_{JJ+,\nu,p} \left ( h_n \right ) + \frac{\left ( \mu^{(\nu)}_{J+} \right )^2}{\left ( \mu^{(\nu)}_{D+} \right )^4} \mathcal{V}_{DD+,\nu,p} \left ( h_n \right ) - \frac{2 \mu^{(\nu)}_{J+}}{\left ( \mu^{(\nu)}_{D+} \right )^3} \mathcal{V}_{JD+,\nu,p} \left ( h_n \right ) \\
	\textsc{V}_{\ell,22+,\nu,p} \left ( h_n \right ) & \equiv \frac{1}{\left ( \mu^{(\nu)}_{G_\ell+} \right )^2} \mathcal{V}_{K_\ell K_\ell +,\nu,p} \left ( h_n \right ) + \frac{\left ( \mu^{(\nu)}_{K_\ell +} \right )^2}{\left ( \mu^{(\nu)}_{G_\ell +} \right )^4} \mathcal{V}_{G_\ell G_\ell +,\nu,p} \left ( h_n \right ) - \frac{2 \mu^{(\nu)}_{K_\ell +}}{\left ( \mu^{(\nu)}_{G_\ell +} \right )^3} \mathcal{V}_{K_\ell G_\ell +,\nu,p} \left ( h_n \right )
\end{align}
and
\begin{align}
	\textsc{V}_{\ell,12+,\nu,p} \left ( h_n \right ) & \equiv \frac{1}{\mu^{(\nu)}_{D+} \mu^{(\nu)}_{G_\ell +}} \mathcal{V}_{J K_\ell +,\nu,p} \left ( h_n \right ) - \frac{\mu^{(\nu)}_{K_\ell +}}{\mu^{(\nu)}_{D+} \left ( \mu^{(\nu)}_{G_\ell +} \right )^2} \mathcal{V}_{J G_\ell +,\nu,p} \left ( h_n \right ) \notag \\
	& - \frac{\mu^{(\nu)}_{J+}}{\left ( \mu^{(\nu)}_{D+} \right )^2 \mu^{(\nu)}_{G_\ell +}} \mathcal{V}_{DK_\ell +,\nu,p} \left ( h_n \right ) + \frac{\mu^{(\nu)}_{J+} \mu^{(\nu)}_{K_\ell +}}{\left ( \mu^{(\nu)}_{D+} \mu^{(\nu)}_{G_\ell +} \right )^2} \mathcal{V}_{DG_\ell +,\nu,p} \left ( h_n \right ) 
\end{align}
The expression for the conditional variance of $\widetilde{\pi}_{\ell,-,\nu,p} \left ( h_n \right )$ follows by replacing the $+$ subscripts with $-$ subscripts throughout.

\appsubsubsection{Definitions of Bias-Corrected Estimators}

The bias-corrected estimator of $\pi_{\ell,+,\nu}$ is defined as
\begin{align}
	\widehat{\pi}^{\textsc{bc}}_{\ell,+,\nu,p,q} \left ( h_n, b_n \right ) & \equiv \widehat{\pi}_{\ell,+,\nu,p} \left ( h_n \right ) - h_n^{p+1-\nu} \widehat{\textsc{B}}_{\ell,+,\nu,p,q} \left ( h_n, b_n \right )
\end{align}
Here, $h_n^{p+1-\nu}\widehat{\textsc{B}}_{\ell,+,\nu,p,q} \left ( h_n,b_n \right )$ estimates the leading bias of
$\widetilde{\pi}_{\ell,+,\nu,p} \left ( h_n \right )$. Specifically,
\begin{align}
	\widehat{\textsc{B}}_{\ell,+,\nu,p,q} \left ( h_n, b_n \right ) & \equiv \widehat{\textsc{B}}_{1+,\nu,p,q} \left ( h_n, b_n \right ) - \widehat{\textsc{B}}_{\ell,2+,\nu,p,q} \left ( h_n, b_n \right )
\end{align}
with
\begin{align}
	\widehat{\textsc{B}}_{1+,\nu,p,q} \left ( h_n, b_n \right ) & \equiv \frac{1}{\widehat{\mu}^{(\nu)}_{D+,p} \left ( h_n \right )} \widehat{\textsc{B}}_{J+,\nu,p,q} \left ( h_n, b_n \right ) - \frac{\widehat{\mu}^{(\nu)}_{J+,p} \left ( h_n \right )}{\left ( \widehat{\mu}^{(\nu)}_{D+,p} \left ( h_n \right ) \right )^2} \widehat{\textsc{B}}_{D+,\nu,p,q} \left ( h_n, b_n \right ) \\
	\widehat{\textsc{B}}_{\ell,2+,\nu,p,q} \left ( h_n, b_n \right ) & \equiv \frac{1}{\widehat{\mu}^{(\nu)}_{G_\ell+,p} \left ( h_n \right )} \widehat{\textsc{B}}_{K_\ell+,\nu,p,q} \left ( h_n, b_n \right ) - \frac{\widehat{\mu}^{(\nu)}_{K_\ell+,p} \left ( h_n \right )}{\left ( \widehat{\mu}^{(\nu)}_{G_\ell+,p} \left ( h_n \right ) \right )^2} \widehat{\textsc{B}}_{G_\ell+,\nu,p,q} \left ( h_n, b_n \right )
\end{align}
In addition, the bias-corrected linearized estimator is defined as
\begin{align}
	\widetilde{\pi}^{\textsc{bc}}_{\ell,+,\nu,p,q} \left ( h_n, b_n \right ) & \equiv \widetilde{\pi}_{\ell,+,\nu,p} \left ( h_n \right ) - h_n^{p+1-\nu} \widecheck{\textsc{B}}_{\ell,+,\nu,p,q} \left ( h_n, b_n \right )
\end{align}
where
\begin{align}
	\widecheck{\textsc{B}}_{\ell,+,\nu,p,q} \left ( h_n, b_n \right ) & \equiv \widecheck{\textsc{B}}_{1+,\nu,p,q} \left ( h_n, b_n \right ) - \widecheck{\textsc{B}}_{\ell,2+,\nu,p,q} \left ( h_n, b_n \right )
\end{align}
with
\begin{align}
	\widecheck{\textsc{B}}_{1+,\nu,p,q} \left ( h_n, b_n \right ) & \equiv \frac{1}{\mu^{(\nu)}_{D+}} \widehat{\textsc{B}}_{J+,\nu,p,q} \left ( h_n, b_n \right ) - \frac{\mu^{(\nu)}_{J+}}{\left ( \mu^{(\nu)}_{D+} \right )^2} \widehat{\textsc{B}}_{D+,\nu,p,q} \left ( h_n, b_n \right ) \\
	\widecheck{\textsc{B}}_{\ell,2+,\nu,p,q} \left ( h_n, b_n \right ) & \equiv \frac{1}{\mu^{(\nu)}_{G_\ell+}} \widehat{\textsc{B}}_{K_\ell+,\nu,p,q} \left ( h_n, b_n \right ) - \frac{\mu^{(\nu)}_{K_\ell+}}{\left ( \mu^{(\nu)}_{G_\ell+} \right )^2} \widehat{\textsc{B}}_{G_\ell+,\nu,p,q} \left ( h_n, b_n \right )
\end{align}
The lower-limit estimators are defined by replacing the $+$ subscripts with $-$ subscripts throughout.

\appsubsubsection{Variance of Bias-Corrected Linearized Estimators}\label{appsec_ct_var_bc}

Whenever the local polynomial estimators are well defined, the conditional variance of the bias-corrected linearized estimator is
\begin{align}
    \textsc{V}^{\textsc{bc}}_{\ell,+,\nu,p,q} \left ( h_n,b_n \right )
    & \equiv \Var \left [ \widetilde{\pi}^{\textsc{bc}}_{\ell,+,\nu,p,q} \left ( h_n,b_n \right ) \Big | \bm{R} \right ] \notag \\
    & = \textsc{V}_{\ell,+,\nu,p} \left ( h_n \right ) + h_n^{2(p+1-\nu)} \textsc{V}_{\ell,+,p+1,q} \left ( b_n \right ) \frac{\mathcal{B}^2_{+,\nu,p,p+1} \left ( h_n \right )}{\left [ \left ( p+1 \right )! \right ]^2}
    \notag \\
    & - 2 h_n^{p+1-\nu} \textsc{C}_{\ell,+,\nu,p,q} \left ( h_n,b_n \right ) \frac{\mathcal{B}_{+,\nu,p,p+1}\left ( h_n \right )}{\left ( p+1 \right )!}
\end{align}
Here, the pilot variance retains the population coefficients
from the original linearization:
\begin{align}
    \textsc{V}_{\ell,+,p+1,q} \left ( b_n \right ) & \equiv \Var \Bigg [ \frac{1}{\mu^{(\nu)}_{D+}} \widehat{\mu}^{(p+1)}_{J+,q} \left ( b_n \right ) - \frac{\mu^{(\nu)}_{J+}}{\left ( \mu^{(\nu)}_{D+} \right )^2} \widehat{\mu}^{(p+1)}_{D+,q} \left ( b_n \right ) \\
	& - \frac{1}{\mu^{(\nu)}_{G_\ell+}} \widehat{\mu}^{(p+1)}_{K_\ell+,q} \left ( b_n \right ) + \frac{\mu^{(\nu)}_{K_\ell+}}{\left ( \mu^{(\nu)}_{G_\ell+} \right )^2} \widehat{\mu}^{(p+1)}_{G_\ell+,q} \left ( b_n \right ) \Bigg | \bm{R} \Bigg ]
\end{align}
The covariance term excludes the kernel bias factor and is defined as
\begin{align}
	& \textsc{C}_{\ell,+,\nu,p,q} \left ( h_n, b_n \right ) \notag \\
	& = \textsc{C}_{11+,\nu,p,q} \left ( h_n, b_n \right ) - \textsc{C}_{\ell,12+,\nu,p,q} \left ( h_n, b_n \right ) - \textsc{C}_{\ell,21+,\nu,p,q} \left ( h_n, b_n \right ) + \textsc{C}_{\ell,22+,\nu,p,q} \left ( h_n, b_n \right )
\end{align}
with
\begin{align}
	\textsc{C}_{11+,\nu,p,q} \left ( h_n, b_n \right ) & = \frac{1}{\left ( \mu^{(\nu)}_{D+} \right )^2} \mathcal{C}_{JJ+,\nu,p,q} \left ( h_n, b_n \right ) - \frac{\mu^{(\nu)}_{J+}}{\left ( \mu^{(\nu)}_{D+} \right )^3} \mathcal{C}_{JD+,\nu,p,q} \left ( h_n, b_n \right ) \notag \\
	& - \frac{\mu^{(\nu)}_{J+}}{\left ( \mu^{(\nu)}_{D+} \right )^3} \mathcal{C}_{DJ+,\nu,p,q} \left ( h_n, b_n \right ) + \frac{\left ( \mu^{(\nu)}_{J+} \right )^2}{\left ( \mu^{(\nu)}_{D+} \right )^4} \mathcal{C}_{DD+,\nu,p,q} \left ( h_n, b_n \right ) \\
	\textsc{C}_{\ell,12+,\nu,p,q} \left ( h_n, b_n \right ) & = \frac{1}{\mu^{(\nu)}_{D+} \mu^{(\nu)}_{G_\ell+}} \mathcal{C}_{JK_\ell+,\nu,p,q} \left ( h_n, b_n \right ) - \frac{\mu^{(\nu)}_{K_\ell+}}{\mu^{(\nu)}_{D+} \left ( \mu^{(\nu)}_{G_\ell+} \right )^2} \mathcal{C}_{JG_\ell+,\nu,p,q} \left ( h_n, b_n \right ) \notag \\
	& - \frac{\mu^{(\nu)}_{J+}}{\left ( \mu^{(\nu)}_{D+} \right )^2 \mu^{(\nu)}_{G_\ell+}} \mathcal{C}_{DK_\ell+,\nu,p,q} \left ( h_n, b_n \right ) + \frac{\mu^{(\nu)}_{J+} \mu^{(\nu)}_{K_\ell+}}{\left ( \mu^{(\nu)}_{D+} \right )^2 \left ( \mu^{(\nu)}_{G_\ell+} \right )^2} \mathcal{C}_{DG_\ell+,\nu,p,q} \left ( h_n, b_n \right ) \\
	\textsc{C}_{\ell,21+,\nu,p,q} \left ( h_n, b_n \right ) & = \frac{1}{\mu^{(\nu)}_{G_\ell+} \mu^{(\nu)}_{D+}} \mathcal{C}_{K_\ell J+,\nu,p,q} \left ( h_n, b_n \right ) - \frac{\mu^{(\nu)}_{J+}}{\mu^{(\nu)}_{G_\ell+} \left ( \mu^{(\nu)}_{D+} \right )^2} \mathcal{C}_{K_\ell D+,\nu,p,q} \left ( h_n, b_n \right ) \notag \\
	& - \frac{\mu^{(\nu)}_{K_\ell+}}{\left ( \mu^{(\nu)}_{G_\ell +} \right )^2 \mu^{(\nu)}_{D+}} \mathcal{C}_{G_\ell J+,\nu,p,q} \left ( h_n, b_n \right ) + \frac{\mu^{(\nu)}_{K_\ell +} \mu^{(\nu)}_{J+}}{\left ( \mu^{(\nu)}_{G_\ell +} \right )^2 \left ( \mu^{(\nu)}_{D+} \right )^2} \mathcal{C}_{G_\ell D+,\nu,p,q} \left ( h_n, b_n \right ) \\
	\textsc{C}_{\ell,22+,\nu,p,q} \left ( h_n, b_n \right ) & = \frac{1}{\left ( \mu^{(\nu)}_{G_\ell +} \right )^2} \mathcal{C}_{K_\ell K_\ell +,\nu,p,q} \left ( h_n, b_n \right ) - \frac{\mu^{(\nu)}_{K_\ell +}}{\left ( \mu^{(\nu)}_{G_\ell +} \right )^3} \mathcal{C}_{K_\ell G_\ell+,\nu,p,q} \left ( h_n, b_n \right ) \notag \\
	& - \frac{\mu^{(\nu)}_{K_\ell+}}{\left ( \mu^{(\nu)}_{G_\ell+} \right )^3} \mathcal{C}_{G_\ell K_\ell +,\nu,p,q} \left ( h_n, b_n \right ) + \frac{\left ( \mu^{(\nu)}_{K_\ell +} \right )^2}{\left ( \mu^{(\nu)}_{G_\ell +} \right )^4} \mathcal{C}_{G_\ell G_\ell +,\nu,p,q} \left ( h_n, b_n \right )
\end{align}
The expression for the conditional variance of the bias-corrected linearized estimator $\widetilde{\pi}^{\textsc{bc}}_{\ell,-,\nu,p,q} \left ( h_n,b_n \right )$ follows by replacing the $+$ subscripts with $-$ subscripts throughout.

\appsubsubsection{Covariances of Bias-Corrected Linearized Estimators}\label{appsec_ct_cov_bc}

To conduct joint tests, I derive the conditional covariance of $\widetilde{\pi}^{\textsc{bc}}_{\ell,+,\nu,p,q} \left ( h_n,b_n \right )$ and $\widetilde{\pi}^{\textsc{bc}}_{m,+,\nu,p,q} \left ( h_n,b_n \right )$, the bias-corrected linearized estimators for comparisons $\ell,m \in \left \{ 1,\dots,\overline{\ell} \right \}$. All comparisons use the same bandwidths $h_n$ and $b_n$ and polynomial degrees $p$ and $q$. Because the comparisons share the all-zero reference group, their estimators may be correlated even when the groups indicated by $G_\ell$ and $G_m$ do not overlap. The following identities hold whenever the local polynomial estimators are well defined. First, consider the conditional covariance of the original linearized estimators (prior to bias correction). Expanding the two linearizations gives
\begin{align}
    \textsc{V}_{\ell m,+,\nu,p} \left ( h_n \right ) & \equiv \Cov \left [ \widetilde{\pi}_{\ell,+,\nu,p} \left ( h_n \right ), \widetilde{\pi}_{m,+,\nu,p} \left ( h_n \right ) \Big | \bm{R} \right ] \notag \\
    & = \textsc{V}_{11+,\nu,p} \left ( h_n \right ) - \textsc{V}_{\ell,12+,\nu,p} \left ( h_n \right ) - \textsc{V}_{m,12+,\nu,p} \left ( h_n \right ) + \textsc{V}_{\ell m,22+,\nu,p} \left ( h_n \right )
\end{align}
The terms $\textsc{V}_{11+,\nu,p} \left ( h_n \right )$, $\textsc{V}_{\ell,12+,\nu,p} \left ( h_n \right )$, and $\textsc{V}_{m,12+,\nu,p} \left ( h_n \right )$ are defined above. The remaining term is the conditional covariance of the linearized estimators for the ratios $\mu^{(\nu)}_{K_\ell+} \slash \mu^{(\nu)}_{G_\ell+}$ and $\mu^{(\nu)}_{K_m+} \slash \mu^{(\nu)}_{G_m+}$. These ratios correspond to the groups indicated by $G_\ell$ and $G_m$, respectively:
\begin{align}
    \textsc{V}_{\ell m,22+,\nu,p} \left ( h_n \right ) & \equiv \frac{1}{\mu^{(\nu)}_{G_\ell+}\mu^{(\nu)}_{G_m+}} \mathcal{V}_{K_\ell K_m+,\nu,p} \left ( h_n \right ) - \frac{\mu^{(\nu)}_{K_m+}}{\mu^{(\nu)}_{G_\ell+} \left ( \mu^{(\nu)}_{G_m+} \right )^2} \mathcal{V}_{K_\ell G_m+,\nu,p} \left ( h_n \right ) \notag \\
    & - \frac{\mu^{(\nu)}_{K_\ell+}}{\left ( \mu^{(\nu)}_{G_\ell+} \right )^2 \mu^{(\nu)}_{G_m+}} \mathcal{V}_{G_\ell K_m+,\nu,p} \left ( h_n \right ) + \frac{\mu^{(\nu)}_{K_\ell+}\mu^{(\nu)}_{K_m+}}{\left ( \mu^{(\nu)}_{G_\ell+}\mu^{(\nu)}_{G_m+} \right )^2} \mathcal{V}_{G_\ell G_m+,\nu,p} \left ( h_n \right )
\end{align}
Second, consider the conditional covariance of the estimated bias coefficients $\widecheck{\textsc{B}}_{\ell,+,\nu,p,q} \left ( h_n,b_n \right )$ and $\widecheck{\textsc{B}}_{m,+,\nu,p,q} \left ( h_n,b_n \right )$. Both coefficients use estimates of the $(p+1)$st derivatives computed with bandwidth $b_n$ and polynomial degree $q$. Define $\textsc{V}_{\ell m,+,p+1,q} \left ( b_n \right )$ using the expansion of $\textsc{V}_{\ell m,+,\nu,p} \left ( h_n \right )$ above. In that expansion and the definitions of its components, replace every primitive covariance $\mathcal{V}_{AB+,\nu,p} \left ( h_n \right )$ with $\mathcal{V}_{AB+,p+1,q} \left ( b_n \right )$ and keep all population coefficients unchanged, evaluated at order $\nu$. Each estimated bias coefficient also contains the factor $\mathcal{B}_{+,\nu,p,p+1} \left ( h_n \right ) \slash \left ( p+1 \right )!$, which depends only on $\bm{R}$ and can thus be factored out of the conditional covariance, yielding
\begin{align}
    & \Cov \left [ \widecheck{\textsc{B}}_{\ell,+,\nu,p,q} \left ( h_n,b_n \right ), \widecheck{\textsc{B}}_{m,+,\nu,p,q} \left ( h_n,b_n \right ) \Big | \bm{R} \right ] = \textsc{V}_{\ell m,+,p+1,q} \left ( b_n \right ) \frac{\mathcal{B}^2_{+,\nu,p,p+1} \left ( h_n \right )}{\left [ \left ( p+1 \right )! \right ]^2}
\end{align}
Third, consider the conditional covariance of $\widetilde{\pi}_{\ell,+,\nu,p} \left ( h_n \right )$ and $\widecheck{\textsc{B}}_{m,+,\nu,p,q} \left ( h_n,b_n \right )$. I factor $\mathcal{B}_{+,\nu,p,p+1} \left ( h_n \right ) \slash \left ( p+1 \right )!$ out of this conditional covariance. Define the remaining covariance term as
\begin{align}
    & \textsc{C}_{\ell m,+,\nu,p,q} \left ( h_n,b_n \right ) \notag \\
	& \equiv \textsc{C}_{11+,\nu,p,q} \left ( h_n,b_n \right ) - \textsc{C}_{m,12+,\nu,p,q} \left ( h_n,b_n \right ) - \textsc{C}_{\ell,21+,\nu,p,q} \left ( h_n,b_n \right ) + \textsc{C}_{\ell m,22+,\nu,p,q} \left ( h_n,b_n \right )
\end{align}
The first three terms are defined above. The remaining term contains the covariances of the $\nu$th derivative estimates for $K_\ell$ and $G_\ell$ with the $(p+1)$st derivative estimates for $K_m$ and $G_m$:
\begin{align}
    \textsc{C}_{\ell m,22+,\nu,p,q} \left ( h_n,b_n \right ) & \equiv \frac{1}{\mu^{(\nu)}_{G_\ell+}\mu^{(\nu)}_{G_m+}} \mathcal{C}_{K_\ell K_m+,\nu,p,q} \left ( h_n,b_n \right ) - \frac{\mu^{(\nu)}_{K_m+}}{\mu^{(\nu)}_{G_\ell+} \left ( \mu^{(\nu)}_{G_m+} \right )^2} \mathcal{C}_{K_\ell G_m+,\nu,p,q} \left ( h_n,b_n \right ) \notag \\
    & - \frac{\mu^{(\nu)}_{K_\ell+}}{\left ( \mu^{(\nu)}_{G_\ell+} \right )^2 \mu^{(\nu)}_{G_m+}} \mathcal{C}_{G_\ell K_m+,\nu,p,q} \left ( h_n,b_n \right ) + \frac{\mu^{(\nu)}_{K_\ell+}\mu^{(\nu)}_{K_m+}}{\left ( \mu^{(\nu)}_{G_\ell+}\mu^{(\nu)}_{G_m+} \right )^2} \mathcal{C}_{G_\ell G_m+,\nu,p,q} \left ( h_n,b_n \right )
\end{align}
Multiplying $\textsc{C}_{\ell m,+,\nu,p,q} \left ( h_n,b_n \right )$ by the kernel bias factor gives the conditional covariance of the original linearized estimator and the estimated bias coefficient:
\begin{align}
    & \Cov \left [ \widetilde{\pi}_{\ell,+,\nu,p} \left ( h_n \right ), \widecheck{\textsc{B}}_{m,+,\nu,p,q} \left ( h_n,b_n \right ) \Big | \bm{R} \right ] = \textsc{C}_{\ell m,+,\nu,p,q} \left ( h_n,b_n \right ) \frac{\mathcal{B}_{+,\nu,p,p+1} \left ( h_n \right )}{\left ( p+1 \right )!}
\end{align}
Under independent sampling, only terms involving the same observation contribute to the conditional covariance of the derivative estimators. For a given observation, interchanging $A$ and $B$ leaves their conditional covariance unchanged. The product of the main and pilot estimation weights is also unchanged because these weights do not depend on the variable being estimated. Therefore, $\mathcal{C}_{AB+,\nu,p,q} \left ( h_n,b_n \right ) = \mathcal{C}_{BA+,\nu,p,q} \left ( h_n,b_n \right )$. The estimated bias coefficient retains the population coefficients from the original linearization. The common kernel bias factor is excluded from $\textsc{C}_{\ell m,+,\nu,p,q} \left ( h_n,b_n \right )$. Applying the symmetry of the primitive covariances therefore gives $\textsc{C}_{\ell m,+,\nu,p,q} \left ( h_n,b_n \right ) = \textsc{C}_{m\ell,+,\nu,p,q} \left ( h_n,b_n \right )$. Finally, expand the conditional covariance of the two bias-corrected linearized estimators using their definitions. The first step gives the covariance of the original linearized estimators. The second step gives the covariance of the estimated bias coefficients, which is multiplied by $h_n^{2(p+1-\nu)}$ in this expansion. The two remaining terms are
\begin{align}
    & - h_n^{p+1-\nu} \Cov \left [ \widetilde{\pi}_{\ell,+,\nu,p} \left ( h_n \right ), \widecheck{\textsc{B}}_{m,+,\nu,p,q} \left ( h_n,b_n \right ) \Big | \bm{R} \right ] \notag \\
    & - h_n^{p+1-\nu} \Cov \left [ \widetilde{\pi}_{m,+,\nu,p} \left ( h_n \right ), \widecheck{\textsc{B}}_{\ell,+,\nu,p,q} \left ( h_n,b_n \right ) \Big | \bm{R} \right ]
\end{align}
These two covariances are equal by the symmetry established above, and their sum produces the factor of two in the final term. Substituting the covariance expressions derived in the three steps gives
\begin{align}
    & \textsc{V}^{\textsc{bc}}_{\ell m,+,\nu,p,q} \left ( h_n,b_n \right ) \equiv \Cov \left [ \widetilde{\pi}^{\textsc{bc}}_{\ell,+,\nu,p,q} \left ( h_n,b_n \right ), \widetilde{\pi}^{\textsc{bc}}_{m,+,\nu,p,q} \left ( h_n,b_n \right ) \Big | \bm{R} \right ] \notag \\
    & = \textsc{V}_{\ell m,+,\nu,p} \left ( h_n \right ) \notag \\
    & + h_n^{2(p+1-\nu)} \textsc{V}_{\ell m,+,p+1,q} \left ( b_n \right ) \frac{\mathcal{B}^2_{+,\nu,p,p+1} \left ( h_n \right )}{\left [ \left ( p+1 \right )! \right ]^2} \notag \\
    & - 2 h_n^{p+1-\nu} \textsc{C}_{\ell m,+,\nu,p,q} \left ( h_n,b_n \right ) \frac{\mathcal{B}_{+,\nu,p,p+1} \left ( h_n \right )}{\left ( p+1 \right )!} \label{eq:ct_cov_bc}
\end{align}
Setting $m=\ell$ recovers the conditional variance derived in the previous section:
\begin{align}
    \textsc{V}^{\textsc{bc}}_{\ell\ell,+,\nu,p,q} \left ( h_n,b_n \right ) = \textsc{V}^{\textsc{bc}}_{\ell,+,\nu,p,q} \left ( h_n,b_n \right )
\end{align}
The lower-limit covariance expressions follow by replacing the $+$ subscripts with $-$ subscripts throughout. Under independent sampling, the upper- and lower-limit linearized estimators use disjoint observations conditional on $\bm{R}$, implying that their conditional covariance is zero:
\begin{align}
    \Cov \left [ \widetilde{\pi}^{\textsc{bc}}_{\ell,+,\nu,p,q} \left ( h_n,b_n \right ), \widetilde{\pi}^{\textsc{bc}}_{m,-,\nu,p,q} \left ( h_n,b_n \right ) \Big | \bm{R} \right ] = 0
\end{align}

\appsubsubsection{Joint Test}\label{appsec_ct_joint}

I first collect the conditional variances and covariances derived above into a matrix. For the upper-limit comparisons, define the $\overline{\ell} \times \overline{\ell}$ matrix
\begin{align}
    \bm{V}^{\textsc{bc}}_{+,\nu,p,q} \left ( h_n,b_n \right ) & \equiv \left [ \textsc{V}^{\textsc{bc}}_{\ell m,+,\nu,p,q} \left ( h_n,b_n \right ) \right ]_{\ell,m}
\end{align}
The diagonal entries are the conditional variances of the bias-corrected linearized estimators (Section \ref{appsec_ct_var_bc}). The off-diagonal entries are their conditional covariances (Section \ref{appsec_ct_cov_bc}). Define $\bm{V}^{\textsc{bc}}_{-,\nu,p,q} \left ( h_n,b_n \right )$ by replacing the $+$ subscripts with $-$ subscripts throughout. Under independent sampling, the upper- and lower-limit linearized estimators are uncorrelated conditional on $\bm{R}$. Thus, the full $2\overline{\ell} \times 2\overline{\ell}$ conditional covariance matrix is
\begin{align}
    \bm{V}^{\textsc{bc}}_{\nu,p,q} \left ( h_n,b_n \right ) & \equiv \begin{bmatrix} \bm{V}^{\textsc{bc}}_{+,\nu,p,q} \left ( h_n,b_n \right ) & \bm{0} \\ \bm{0} & \bm{V}^{\textsc{bc}}_{-,\nu,p,q} \left ( h_n,b_n \right ) \end{bmatrix}
\end{align}
The first $\overline{\ell}$ rows and columns correspond to the upper-limit comparisons. The remaining $\overline{\ell}$ rows and columns correspond to the lower-limit comparisons. Each $\bm{0}$ is an $\overline{\ell} \times \overline{\ell}$ matrix of zeros. Using the same ordering, collect the parameters and their bias-corrected estimators into the vectors
\begin{align}
    \bm{\pi}_{\nu} & \equiv \left [ \pi_{1,+,\nu},\dots,\pi_{\overline{\ell},+,\nu},\pi_{1,-,\nu},\dots,\pi_{\overline{\ell},-,\nu} \right ]' \\
    \widehat{\bm{\pi}}^{\textsc{bc}}_{\nu,p,q} \left ( h_n,b_n \right ) & \equiv \bigg [ \widehat{\pi}^{\textsc{bc}}_{1,+,\nu,p,q} \left ( h_n,b_n \right ),\dots,\widehat{\pi}^{\textsc{bc}}_{\overline{\ell},+,\nu,p,q} \left ( h_n,b_n \right ),\notag\\
	& \quad \ \ \, \widehat{\pi}^{\textsc{bc}}_{1,-,\nu,p,q} \left ( h_n,b_n \right ),\dots,\widehat{\pi}^{\textsc{bc}}_{\overline{\ell},-,\nu,p,q} \left ( h_n,b_n \right ) \bigg ]'
\end{align}
Letting $\bm{0}$ denote a vector of $2\overline{\ell}$ zeros, the joint null hypothesis is
\begin{align}
    H_0:\ \bm{\pi}_{\nu} = \bm{0}
\end{align}
This hypothesis imposes $\pi_{\ell,+,\nu}=\pi_{\ell,-,\nu}=0$ for every $\ell=1,\dots,\overline{\ell}$.

\appsubsubsection{Asymptotic Distribution of the Joint Test Statistics}\label{appsec_ct_joint_dis}

Let $\overline{\ell}$ be fixed, $0 \leq \nu \leq p$, and $q \geq p+1$. Suppose that parts (a)-(c) of Assumption \ref{ass_moments} hold for $R$, $J$, $D$, $K_\ell$, and $G_\ell$ for all $\ell=1,\dots,\overline{\ell}$, with $\delta \geq q+1$. Suppose also that Assumption \ref{ass_kernel} holds. Further assume that, for every pair of variables $A$ and $B$ used in the comparisons, $\Cov \left [ A,B | R=r \right ]$ is Lipschitz continuous in $r$ separately on either side of $c$ and has finite one-sided limits at $c$.

To conduct joint statistical inference, positive variances for the individual linearized estimators are not sufficient. In fact, their covariance matrix must also be nonsingular. To ensure this property, consider the following $\overline{\ell}$ random variables:
\begin{align}
    \frac{1}{\mu^{(\nu)}_{D+}} J - \frac{\mu^{(\nu)}_{J+}}{\left ( \mu^{(\nu)}_{D+} \right )^2} D - \frac{1}{\mu^{(\nu)}_{G_\ell+}} K_\ell + \frac{\mu^{(\nu)}_{K_\ell+}}{\left ( \mu^{(\nu)}_{G_\ell+} \right )^2} G_\ell
\end{align}
for $\ell=1,\dots,\overline{\ell}$. These variables use the population coefficients from the original linearizations. Suppose that the upper limit at $c$ of their conditional covariance matrix given $R=r$ is positive definite. Impose the corresponding lower-limit condition by replacing the $+$ subscripts with $-$ subscripts throughout.

In addition, suppose that the bandwidth sequences satisfy $\max \left \{ h_n,b_n \right \} \to 0$, $n \min \left \{ h_n,b_n \right \} \to \infty$, $n h_n^{1+2\nu} \to \infty$, and $n \min \left \{ h_n^{2p+3},b_n^{2p+3} \right \} \max \left \{ h_n^2,b_n^{2(q-p)} \right \} \to 0$. Under these conditions,
\begin{align}\label{app_test_infeasible}
    \left [ \bm{V}^{\textsc{bc}}_{\nu,p,q} \left ( h_n,b_n \right ) \right ]^{-1/2} \left [ \widehat{\bm{\pi}}^{\textsc{bc}}_{\nu,p,q} \left ( h_n,b_n \right ) - \bm{\pi}_{\nu} \right ] \stackrel{d}{\to} \mathcal{N} \left ( \bm{0}, \bm{I} \right )
\end{align}
Here, $\bm{I}$ is the identity matrix of dimension $2\overline{\ell}$. The argument in part (D) of Theorem \ref{thm_asydist_bias} applies to linear combinations of the bias-corrected linearized estimators. The stated bandwidth conditions make the remaining bias and the linearization remainders negligible after standardization. The joint normal limit then follows by the Cramér-Wold device and Slutsky's Theorem.

Since the covariance matrix in equation \eqref{app_test_infeasible} is unknown, I replace it with an estimate to implement the test. Let $\widehat{\bm{V}}^{\textsc{bc}}_{\nu,p,q} \left ( h_n,b_n \right )$ denote the estimator of the covariance matrix. Here I replace the population coefficients in each entry with their corresponding local polynomial estimates and estimate the conditional variances and covariances using the nearest-neighbor procedure described in Section \ref{sec:stderr}. Throughout, I use the same nearest-neighbor sets for all variables on each side of the cutoff and the same estimated population moments wherever they appear in different covariance entries. Applying the argument of Theorem A.3 in \cite{cct2014supp} to the finitely many variable pairs, and invoking consistency of the estimated population coefficients, yields
\begin{align}
    \left [ \bm{V}^{\textsc{bc}}_{\nu,p,q} \left ( h_n,b_n \right ) \right ]^{-1/2} \widehat{\bm{V}}^{\textsc{bc}}_{\nu,p,q} \left ( h_n,b_n \right ) \left [ \bm{V}^{\textsc{bc}}_{\nu,p,q} \left ( h_n,b_n \right ) \right ]^{-1/2} \stackrel{p}{\to} \bm{I}
\end{align}
Consequently, replacing the true covariance matrix with its estimator preserves the limiting distribution of the joint test statistic. Under $H_0$, the feasible joint Wald statistic is
\begin{align}
    W^{\textsc{rbc}}_{\nu,p,q} \left ( h_n,b_n \right ) & \equiv \widehat{\bm{\pi}}^{\textsc{bc}}_{\nu,p,q} \left ( h_n,b_n \right )' \left [ \widehat{\bm{V}}^{\textsc{bc}}_{\nu,p,q} \left ( h_n,b_n \right ) \right ]^{-1} \widehat{\bm{\pi}}^{\textsc{bc}}_{\nu,p,q} \left ( h_n,b_n \right ) \stackrel{d}{\to} \chi^2 \left ( 2\overline{\ell} \right )
\end{align}
At significance level $\alpha$, the joint null hypothesis $H_0$ is rejected when $W^{\textsc{rbc}}_{\nu,p,q} \left ( h_n,b_n \right )$ exceeds the $\left ( 1-\alpha \right )$ quantile of the $\chi^2 \left ( 2\overline{\ell} \right )$ distribution.

Setting $\nu=0$ gives the joint test of the pre-period restrictions in equation \eqref{eq:hypotheses}. For the aggregate comparison alone, $\overline{\ell}=1$ and the reference distribution is $\chi^2 \left ( 2 \right )$. Instead, if the test seeks to compare the all-zero path with every individual nonzero path, $\overline{\ell}=2^\tau-1$ and the asymptotic distribution is $\chi^2 \left ( 2 \left ( 2^\tau-1 \right ) \right )$. Finally, intermediate groupings imply $2\overline{\ell}$ restrictions.

\appsubsubsection{MSE-Optimal Bandwidths}\label{appsec_ct_bandwidths}

For each pre-treatment outcome change $\Delta_{u,v}$ and collection of comparison groups $\left \{ G_\ell \right \}_{\ell = 1}^{\overline{\ell}}$, I select a common main bandwidth and a common pilot bandwidth for all restrictions in the joint test.

The optimal main bandwidth minimizes the leading terms of the sum of conditional MSEs of the linearized estimators of the contrasts. This sum includes every comparison on both sides of the cutoff. Using the conditional bias and variance expressions derived above gives
\begin{align}
    & \sum_{\ell=1}^{\overline{\ell}} \left [ \text{MSE} \left ( \widetilde{\pi}_{\ell,+,\nu,p} \left ( h_n \right ) \right ) + \text{MSE} \left ( \widetilde{\pi}_{\ell,-,\nu,p} \left ( h_n \right ) \right ) \right ] \notag \\
    & = h_n^{2(p+1-\nu)} \sum_{\ell=1}^{\overline{\ell}} \left [ \textsc{B}_{\ell,+,\nu,p,p+1}^{2} \left ( h_n \right ) + \textsc{B}_{\ell,-,\nu,p,p+1}^{2} \left ( h_n \right ) \right ] \notag \\
    & + \sum_{\ell=1}^{\overline{\ell}} \left [ \textsc{V}_{\ell,+,\nu,p} \left ( h_n \right ) + \textsc{V}_{\ell,-,\nu,p} \left ( h_n \right ) \right ] + o_p \left ( h_n^{2(p+1-\nu)} \right )
\end{align}

The optimal pilot bandwidth minimizes the leading terms of the sum of conditional MSEs of the estimators of the leading bias coefficients. These estimators retain the population coefficients from the original linearizations, as in Section \ref{app_prop_bandwidth}.

For implementation, I use regularized direct plug-in selectors analogous to those described in Section \ref{sec:bandwidth}. For inference, I apply the same rule-of-thumb adjustment to the main bandwidth.

\appsubsection{Confidence Intervals for Horizon-Specific Parameters}\label{appsec_aggrconfint}

Section \ref{sec_aggreg} describes an approach to aggregate history-specific parameters within each focal period and horizon. Under Assumption \ref{ass_mark}, the ADTE is homogeneous across treatment histories whose last $k$ states are zero. It is therefore natural to estimate a single parameter for each focal period $t$ and horizon $\tau$ in the subpopulation satisfying $D_{t-k} = \dots = D_{t-1} = 0$. Earlier treatment histories are allowed to vary across units. However, because the ADTE is common across these histories, estimating history-specific aggregation weights is not necessary.

For a given horizon $\tau$, the period-specific effects can then be aggregated across focal periods, following the approach in equation \eqref{eq:aggr_reltime}. To construct confidence intervals for this aggregate target parameter, one must account for the fact that the same unit may contribute to estimates for several focal periods. Consequently, the estimators for different period-specific effects may be correlated. However, under the additional conditions stated below, these cross-period covariances are negligible to a first-order approximation.

Units are independently and identically sampled, conditional on any common calendar time components. Draws within the same unit are allowed to be dependent over time. Suppose that the number of periods is fixed and maintain the period-specific regularity and bandwidth conditions of Theorem \ref{thm_asydist_bias}, with $\max \left \{ h_{t,n},b_{t,n} \right \} \to 0$ and bounded ratios $h_{t,n} \slash b_{t,n}$. For every pair of distinct focal periods with overlapping eligible populations, assume that the joint density of their running variables, conditional on eligibility in both periods, is bounded near $\left ( c,c \right )$. For units eligible in both focal periods, consider any pair of residuals for $Y$, $W$, or $D$, one from each period. Assume that the conditional expectation of their absolute product, given $\left ( R_{it},R_{is} \right )$, is bounded near $\left ( c,c \right )$. Each residual is centered on its own period-specific conditional mean.

Cross-period covariances can be shown to be negligible following the localization argument for clustered local polynomial estimators discussed in Section 4.4.1 of \cite{noackolmarothe2026}. Intuitively, a unit contributes to two period-specific estimators only when its running variables belong to both estimation windows. Because the joint density is bounded at the cutoff, the probability of this event decreases with the product of the window widths. For the linearized estimators, the first-order cross-period covariances are $O \left ( 1 \slash \left ( n h_{t,n}^{\nu} h_{s,n}^{\nu} \right ) \right )$, while individual variances are of order $1 \slash \left ( n h_{t,n}^{1+2\nu} \right )$. Dividing each cross-period covariance by the product of the corresponding marginal standard deviations gives $O \left ( \sqrt{h_{t,n}h_{s,n}} \right )$, which converges to zero. Because the number of periods is fixed, their aggregate contribution is negligible to first order.

Let $\omega_{t\tau}$ denote fixed, nonnegative calendar time weights that sum to one over the eligible focal periods. Equal weighting assigns $\omega_{t\tau}=1 \slash \left ( \overline{t}-\tau-k \right )$. Define
\begin{align}
	\theta_{\tau,\nu} \equiv \sum_{t=k+1}^{\overline{t}-\tau} \omega_{t\tau} \theta_{t,\tau,\nu} \qquad
	\widehat{\theta}^{\textsc{bc}}_{\tau,\nu,p,q} \equiv \sum_{t=k+1}^{\overline{t}-\tau} \omega_{t\tau} \widehat{\theta}^{\textsc{bc}}_{t,\tau,\nu,p,q} \left ( h_{t,n},b_{t,n} \right )
\end{align}
A first-order approximation to the variance of the horizon-specific estimator is
\begin{align}
	\textsc{V}^{\textsc{bc}}_{\tau,\nu,p,q} & \equiv \sum_{t=k+1}^{\overline{t}-\tau} \omega_{t\tau}^2 \textsc{V}^{\textsc{bc}}_{t,\tau,\nu,p,q}
	\end{align}
where $\textsc{V}^{\textsc{bc}}_{t,\tau,\nu,p,q}$ denotes the variance of the period-specific linearized bias-corrected estimator derived in Section \ref{appsec_bc_var}. I estimate this expression using the period-specific nearest-neighbor approach:
	\begin{align}
	\widehat{\textsc{V}}^{\textsc{bc}}_{\tau,\nu,p,q} & \equiv \sum_{t=k+1}^{\overline{t}-\tau} \omega_{t\tau}^2 \widehat{\textsc{V}}^{\textsc{bc}}_{t,\tau,\nu,p,q}
\end{align}
The central limit argument in part (D) of Theorem \ref{thm_asydist_bias} applies to the weighted average of the period-specific linearized bias-corrected estimators. In addition, each period-specific estimator of the variance divided by its theoretical counterpart converges in probability to one. Because the number of periods is fixed, the same holds for the estimator of the variance of the horizon-specific estimator. Slutsky's Theorem therefore gives
\begin{align}
	\frac{\widehat{\theta}^{\textsc{bc}}_{\tau,\nu,p,q} - \theta_{\tau,\nu}}{\sqrt{\widehat{\textsc{V}}^{\textsc{bc}}_{\tau,\nu,p,q}}} & \stackrel{d}{\to} \mathcal{N} \left ( 0,1 \right )
\end{align}
Setting $\nu=0$, $p=1$, and $q=2$ yields the normal approximation for the horizon-specific ADTE estimated using local linear regression and local quadratic bias correction.

\appsection{Monte Carlo Simulation}\label{sec_simulation}

In this section, I conduct a Monte Carlo simulation to assess the performance of the proposed estimators and inference procedures. To do so, I employ a data generating process similar to that presented by \cite{bilaschon2023}\footnote{Specifically, the authors conduct a Monte Carlo simulation in Appendix B.}, which models both school districts' decisions to hold local referenda and the subsequent evolution of an outcome of interest based on referendum results.

\appsubsection{Data Generating Process}\label{appsec_mc_dgp}

Let $i \in \left \{ 1, \dots, n \right \}$ and $t \in \left \{ 1, \dots, \overline{t} \right \}$ index school districts and calendar years, respectively. Let $Q_{it}$ be an indicator for whether district $i$ holds a referendum in year $t$. When $Q_{it} = 1$, denote the approval vote margin by $R_{it}$ and define $D_{it} \equiv \mathbb{I} \left [ R_{it} > 0 \right ]$ as the indicator for referendum approval. When $Q_{it}=0$, the vote margin is not observed and $D_{it}=0$. In each period, a district's decision to hold a referendum is influenced, among other factors, by whether referenda were recently approved or rejected. Specifically, 
\begin{align}\label{sim:hold}
	Q_{it} = \mathbb{I} \left [ A^q_i + B^q_t + U^q_{it} + \sum_{s=1}^{\overline{s}} \rho^q_s D_{i,t-s} + \sum_{s=1}^{\overline{s}} \delta^q_s Q_{i,t-s} \left ( 1 - D_{i,t-s} \right ) > \overline{q} \right ]
\end{align}
where $A^q_i$ is a district-specific intercept, $B^q_t$ is a time-specific intercept, $U^q_{it}$ is an idiosyncratic shock, and $\overline{q}$ is a deterministic threshold that must be exceeded for a referendum to be held. If a referendum is held, the approval vote margin is determined according to a similar statistical model:
\begin{align}\label{sim:vote}
	R_{it} = A^r_i + B^r_t + U^r_{it} + \sum_{s=1}^{\overline{s}} \rho^r_s D_{i,t-s} + \sum_{s=1}^{\overline{s}} \delta^r_s Q_{i,t-s} \left ( 1 - D_{i,t-s} \right ) + \sum_{s=1}^{\overline{s}} \gamma_s \left ( 1 - Q_{i,t-s} \right )
\end{align} 
where $A^r_i$ and $B^r_t$ are district- and time-specific intercepts, respectively, and $U^r_{it}$ is an idiosyncratic shock. This specification allows the vote margin to depend on past referendum approvals, past referendum rejections, and periods in which no referendum was held. The outcome of interest is generated according to
\begin{align}\label{sim:outcome}
	Y_{it} = \sum_{g=1}^{t} \theta_{igt} D_{ig} + \eta \overline{R}_i + V_{it}
\end{align}
Here, $V_{it}$ is an idiosyncratic shock and $\overline{R}_i$ is the average approval vote margin across referenda held by district $i$ during the observation period. I set $\overline{R}_i=0$ for districts that hold no referenda. The summation allows the effects of multiple referendum approvals to accumulate additively. Importantly, the stochastic treatment effect $\theta_{igt}$ is determined by
\begin{align}\label{sim:effect}
	\theta_{igt} = \theta_{0g} + \theta_{1g} \left ( t - g \right ) + \phi R_{ig}
\end{align}
This formulation allows the effect of referendum approval to vary across districts, calendar times, and cohorts (with a cohort defined as the calendar time at which a referendum is held). Here, $\theta_{0g}$ is a cohort-specific random variable capturing the permanent effect of a referendum approved in calendar time $g$, and $\theta_{1g}$ is a cohort-specific random variable that measures the incremental effect per unit of time elapsed since approval, meaning that this additional effect is proportional to the relative time $t-g$. Finally, $\phi$ is a deterministic coefficient that quantifies the extent to which the treatment effect varies with the magnitude of the approval vote margin.

\appsubsection{Parameterization}\label{appsec_mc_params}

The distributional assumptions and parameter values largely mirror those in \cite{bilaschon2023}. In the referendum equation \eqref{sim:hold}, the unit intercept is assumed to be $A^q_i \sim \mathcal{N} \left ( 0, \sigma^2_q \right )$, the time intercept is $B^q_t \sim \mathcal{N} \left ( 0, \sigma^2_q \right )$, and the idiosyncratic shock is $U^q_{it} \sim \mathcal{N} \left ( 0, \sigma^2_{uq} \right )$. Similarly, in the vote margin equation \eqref{sim:vote}, $A^r_i \sim \mathcal{N} \left ( 0, \sigma^2_r \right )$, $B^r_t \sim \mathcal{N} \left ( 0, \sigma^2_r \right )$, and $U^r_{it} \sim \mathcal{N} \left ( 0, \sigma^2_{ur} \right )$. In the outcome equation \eqref{sim:outcome}, the idiosyncratic shock is given by $V_{it} \sim \mathcal{N} \left ( 0, \sigma^2_v \right )$. In the treatment effect equation \eqref{sim:effect}, the permanent effect is modeled as $\theta_{0g} \sim \mathcal{N} \left ( \overline{\theta}_{0} , \sigma^2_0 \right )$ and the additional, time-proportional effect is $\theta_{1g} \sim \mathcal{N} \left ( \overline{\theta}_{1} , \sigma^2_1 \right )$. All draws of the intercepts, shocks, and treatment effect coefficients are mutually independent. District-specific intercepts are drawn once per district, while time-specific intercepts and treatment effect coefficients are drawn once per calendar year and are common to all districts. Idiosyncratic shocks are drawn independently across districts and years.

I simulate data for $n = 10,000$ school districts over a period of $\overline{t} = 30$ years. To initialize treatment histories, I draw $Q_{i0} \sim \text{Binomial} \left ( 1, 0.1 \right )$ and, conditional on $Q_{i0}=1$, $R_{i0} \sim \mathcal{N} \left ( \mu_r,\sigma^2_r \right )$. I determine $D_{i0}$ using the approval rule defined in Section \ref{appsec_mc_dgp}. For $t<0$, I set $Q_{it}=D_{it}=0$. Observations in period zero initialize the assignment histories but do not contribute directly to the outcome sum.

The chosen parameter values are as follows: $\overline{q} = 1.7$; $\sigma^2_q = 0.5$; $\sigma^2_{uq} = 1$; $\sigma^2_r = 0.01$; $\sigma^2_{ur} = 0.02$; $\mu_r = 0.05$; $\sigma^2_v = 0.2$; $\eta = 0$; $\phi = 0.1$; $\overline{\theta}_{0} = 0.4$; $\overline{\theta}_{1} = -0.1$; $\sigma^2_{0} = 0.08^2$; $\sigma^2_{1} = 0.02^2$.

In addition, I restrict the treatment effect to change only during the first two years after approval. Specifically, I replace $\theta_{1g} \left ( t-g \right )$ in equation \eqref{sim:effect} with $\theta_{1g} \min \left \{ t-g,2 \right \}$. The effect of each referendum approval remains constant thereafter.

Finally, I set $\overline{s} = 3$ and $\left ( \rho^q_1, \rho^q_2, \rho^q_3 \right ) = \left ( -0.9, -0.5, -0.1 \right )$, $\left ( \delta^q_1, \delta^q_2, \delta^q_3 \right ) = \left ( 0.5, 0.3, 0.1 \right )$, $\left ( \rho^r_1, \rho^r_2, \rho^r_3 \right ) = \left ( -0.05, -0.03, -0.01 \right )$, $\left ( \delta^r_1, \delta^r_2, \delta^r_3 \right ) = \left ( 0.05, 0.03, 0.01 \right )$, and $\left ( \gamma_1, \gamma_2, \gamma_3 \right ) = \left ( 0.05, 0.03, 0.01 \right )$.

\appsubsection{Target Parameters}\label{appsec_mc_targetparam}

Under the parameterization in Section \ref{appsec_mc_params}, the effect of a referendum approved in period $g$ on the outcome observed $\tau$ periods later is $\theta_{0g} + \theta_{1g} \min \left \{ \tau,2 \right \} + \phi R_{ig}$. Evaluating this effect at the approval cutoff $R_{ig}=0$ gives
\begin{align}\label{eq_mc_theta}
	\mathrm{ADTE}_{g,\tau} = \theta_{0g} + \theta_{1g} \min \left \{ \tau,2 \right \}
\end{align}
Averaging the cohort-specific ADTEs over the draws of the treatment effect coefficients yields
\begin{align}
	\E \left [ \mathrm{ADTE}_{g,\tau} \right ] = \overline{\theta}_{0} + \overline{\theta}_{1} \min \left \{ \tau,2 \right \}
\end{align}
where the expectation is taken over the draws of $\theta_{0g}$ and $\theta_{1g}$. To preserve a sufficient sample size in each discontinuity window, I pool eligible observations across focal calendar years. The simulation evaluates the performance of the proposed estimator under this implementation, with one estimate at each horizon.

\appsubsection{Results}

I leverage Assumption \ref{ass_mark} by restricting attention to jurisdictions that held a referendum in period $g$ and did not approve a ballot measure in any of the preceding $k=4$ periods, so that $Q_{ig}=1$ and $D_{i,g-1} = D_{i,g-2} = D_{i,g-3} = D_{i,g-4} = 0$. For each horizon $\tau \in \left \{ 0,1,2,3,4,5 \right \}$, I compute the bias-corrected estimator using local linear regression and local quadratic bias estimation. Bandwidth selection and adjustment follow Section \ref{sec:bandwidth}. I conduct 1,000 replications. I first assess estimation and confidence interval coverage under the baseline DGP. I then introduce departures from local common trends to evaluate the power of the pre-trends tests.

\clearpage
\appsubsubsection{Estimation}

Table \ref{apptab_estim} reports the output of the simulation pertaining to estimation. Reassuringly, the point estimates capture the dynamic pattern implied by the data generating process. The estimator is essentially unbiased, and its median absolute error remains modest. Moreover, the range between the 10th and 90th percentiles of the estimates widens modestly at longer horizons. This pattern is consistent with fewer available focal calendar years and fewer observations with no additional treatment throughout the relevant horizon.

\vspace{2mm}

\begin{table}[H]
\setstretch{1}
	\centering
	\caption{Simulation Results: Estimation}
	\label{apptab_estim}
	\begingroup
\renewcommand{\arraystretch}{1.4}
\setlength{\tabcolsep}{5pt}
\begin{tabular}{*{5}{>{\centering\arraybackslash}m{1.6cm}}}
\hline\hline
$\tau$ & $\mathrm{ADTE}_{\tau}$ & Bias & MAE & p90$-$p10 \\
\midrule
$0$ & $0.4$ & $0.001$ & $0.022$ & $0.102$ \\
$1$ & $0.3$ & $0.002$ & $0.025$ & $0.115$ \\
$2$ & $0.2$ & $-0.001$ & $0.028$ & $0.129$ \\
$3$ & $0.2$ & $-0.002$ & $0.028$ & $0.128$ \\
$4$ & $0.2$ & $0.001$ & $0.034$ & $0.147$ \\
$5$ & $0.2$ & $-0.001$ & $0.033$ & $0.153$ \\
\hline\hline
\end{tabular}
\endgroup

	\par\medskip\medskip
		\begin{minipage}{\textwidth}
		\footnotesize\textsc{Notes:} This table reports results from 1,000 replications of the data generating process described in Section \ref{appsec_mc_dgp}, with the parameterization detailed in Section \ref{appsec_mc_params}. Target parameters are estimated with $\widehat{\theta}^{\textsc{bc}}$, the bias-corrected estimator defined in equation \eqref{eq_thetabc}. Estimation uses the bandwidth adjustment described in Section \ref{sec:bandwidth}. The first column reports the horizon $\tau$. The second reports $\mathrm{ADTE}_{\tau}$, the average effect implied by the data generating process. In each replication, the estimation error is the difference between the bias-corrected estimate and the population value of equation \eqref{apte_gen_lie} for the pooled sample. Bias is the average estimation error. MAE denotes the median absolute estimation error. The last column reports the difference between the 90th and 10th percentiles of the bias-corrected estimates across replications.
		\end{minipage}
\end{table}

\clearpage
\appsubsubsection{Statistical Inference}

Table \ref{apptab_inference} reports the output of the simulation pertaining to inference. Reassuringly, the bias-corrected confidence intervals have coverage close to the nominal level across all horizons. Average interval length increases with $\tau$, alongside the Monte Carlo standard deviation of estimation errors. The wider intervals at longer horizons are therefore consistent with greater sampling uncertainty.

\vspace{2mm}

\begin{table}[H]
\setstretch{1}
	\centering
	\caption{Simulation Results: Statistical Inference}
	\label{apptab_inference}
	\begingroup
\renewcommand{\arraystretch}{1.4}
\setlength{\tabcolsep}{5pt}
\begin{tabular}{*{4}{>{\centering\arraybackslash}m{1.6cm}}}
\hline\hline
$\tau$ & Coverage & AIL & MC SD \\
\midrule
$0$ & $0.956$ & $0.132$ & $0.035$ \\
$1$ & $0.951$ & $0.145$ & $0.039$ \\
$2$ & $0.949$ & $0.158$ & $0.043$ \\
$3$ & $0.961$ & $0.172$ & $0.044$ \\
$4$ & $0.945$ & $0.188$ & $0.052$ \\
$5$ & $0.943$ & $0.203$ & $0.053$ \\
\hline\hline
\end{tabular}
\endgroup

	\par\medskip\medskip
	\begin{minipage}{\textwidth}
		\footnotesize\textsc{Notes:} This table reports results from 1,000 replications of the data generating process described in Section \ref{appsec_mc_dgp}, with the parameterization detailed in Section \ref{appsec_mc_params}. Confidence intervals are constructed using $\widehat{\theta}^{\textsc{bc}}$, the bias-corrected estimator defined in equation \eqref{eq_thetabc}. Estimation uses the bandwidth adjustment described in Section \ref{sec:bandwidth}. The nominal coverage probability is $0.95$. The first column reports the horizon $\tau$. In each replication, the estimation error is the difference between the bias-corrected estimate and the population value of equation \eqref{apte_gen_lie} for the pooled sample. Coverage is the fraction of confidence intervals containing this population value. AIL denotes the average length of the confidence intervals. MC SD is the Monte Carlo standard deviation of the estimation errors across replications.
	\end{minipage}
\end{table}

\clearpage
\appsubsubsection{Assessing the Plausibility of Local Common Trends}

Section \ref{sec_check_ct} introduces tests for assessing the plausibility of local common trends. Section \ref{app_testct} develops the corresponding statistical theory. Here, I evaluate the empirical size and power of these tests. Motivated by \cite{roth2022}, I examine whether the tests detect departures from local common trends that induce economically meaningful bias in the estimated ADTE.

I consider two departures from local common trends, corresponding to the selection mechanisms discussed in Section \ref{sec_localct}. First, I allow persistent determinants of referendum participation to affect untreated outcome growth. I replace the outcome equation \eqref{sim:outcome} with
\begin{align}\label{eq_dgp_selchars}
	Y_{it} = \sum_{g=1}^{t} \theta_{igt} D_{ig} + \eta \overline{R}_i + V_{it} + \lambda t \frac{A^q_i}{\sigma_q}
\end{align}
The additional term contributes $\lambda A^q_i \slash \sigma_q$ to the annual growth in untreated potential outcomes. Since $A^q_i$ also affects whether a district holds subsequent referenda, the same district characteristic affects both selection into subsequent treatment and growth in untreated outcomes. Second, I consider selection on gains by replacing the outcome equation \eqref{sim:outcome} with
\begin{align}\label{eq_dgp_selgains}
	Y_{it} = \sum_{g=1}^{t} \theta_{igt} D_{ig} + \eta \overline{R}_i + V_{it} + \lambda t \frac{A^r_i}{\sigma_r}
\end{align}
The additional term contributes $\lambda A^r_i \slash \sigma_r$ to the annual growth in untreated potential outcomes. The same district characteristic affects future vote margins and gains from treatment through the component $\phi R_{ig}$ in equation \eqref{sim:effect}. This specification therefore relates growth in untreated outcomes to a determinant of both subsequent treatment assignment and treatment gains.

To make these departures economically interpretable, I calibrate $\lambda$ separately for each selection mechanism. I choose values that induce additional population biases in the five-period-ahead pooled estimator equal to 10, 20, 30, 40, and 50 percent of the DGP mean ADTE of $0.2$. For example, the 10 percent scenario induces an additional population bias of $0.02$. For each resulting DGP, I evaluate the aggregate, frequency-based, and timing-based comparisons discussed in Section \ref{sec_check_ct}.

Table \ref{apptab_pretrend_power} reports the empirical size and power of the joint pre-trends tests based on the outcome change $Y_{g-1}-Y_{g-3}$. The restriction to four preceding untreated periods, combined with the two-year cap, ensures that earlier approvals do not affect this pre-treatment outcome change. Under the baseline DGP, rejection rates are close to the nominal significance level of 5 percent. Power increases with the population bias induced by the departure from local common trends. Departures that induce a bias of 30 percent are detected in about 49-76 percent of replications. At a target bias of 50 percent, power reaches 97-98 percent under selection on persistent characteristics and 70-75 percent under selection on gains. Thus, power depends on the economic mechanism governing subsequent selection into treatment, even when the departures induce the same bias in ADTE estimation.

\vspace{2mm}

\begin{table}[H]
\setstretch{1}
\centering
\caption{Simulation Results: Size and Power of Pre-Trends Tests}
\label{apptab_pretrend_power}
\begingroup
\renewcommand{\arraystretch}{1.4}
\setlength{\tabcolsep}{5pt}
\begin{tabular}{@{\hspace{10pt}}*{4}{>{\centering\arraybackslash}m{1.8cm}}@{\hspace{10pt}}}
\hline\hline
\multirow{2}{1.8cm}[-1pt]{\centering Target Bias (\%)} & \multicolumn{3}{c}{Type of Comparison Group} \\
\cline{2-4}
 & \makebox[\linewidth][c]{Aggregate} & \makebox[\linewidth][c]{Frequency} & Timing \\
\midrule
\addlinespace[2pt]
\multicolumn{4}{@{}l@{}}{\textit{Panel A: Selection on Persistent Characteristics}} \\
\addlinespace[2pt]
$0$ (size) & $0.053$ & $0.061$ & $0.048$ \\
$10$ & $0.132$ & $0.128$ & $0.119$ \\
$20$ & $0.456$ & $0.400$ & $0.368$ \\
$30$ & $0.756$ & $0.725$ & $0.678$ \\
$40$ & $0.930$ & $0.916$ & $0.880$ \\
$50$ & $0.975$ & $0.978$ & $0.969$ \\
\midrule
\addlinespace[4pt]
\multicolumn{4}{@{}l@{}}{\textit{Panel B: Selection on Gains}} \\
\addlinespace[2pt]
$0$ (size) & $0.053$ & $0.061$ & $0.048$ \\
$10$ & $0.139$ & $0.135$ & $0.125$ \\
$20$ & $0.357$ & $0.353$ & $0.317$ \\
$30$ & $0.540$ & $0.558$ & $0.493$ \\
$40$ & $0.656$ & $0.674$ & $0.620$ \\
$50$ & $0.725$ & $0.747$ & $0.699$ \\
\hline\hline
\end{tabular}
\endgroup

\par\medskip\medskip
\begin{minipage}{\textwidth}
	\footnotesize\textsc{Notes:} This table reports rejection rates for the joint Wald tests described in Section \ref{app_testct}, using 1,000 replications. The significance level is $0.05$. The baseline data generating process is described in Section \ref{appsec_mc_dgp}, with the parameterization detailed in Section \ref{appsec_mc_params}. The first column reports the target additional population bias in the five-period-ahead pooled estimator, expressed as a percentage of the DGP mean ADTE of $0.2$. Rows with zero target bias report empirical size under the baseline DGP. Rows with positive target bias report power under the departures from local common trends specified in equation \eqref{eq_dgp_selchars} for Panel A and equation \eqref{eq_dgp_selgains} for Panel B. For each departure, $\lambda$ is chosen to generate the target population bias. All tests compare the pre-treatment outcome change $Y_{g-1}-Y_{g-3}$ across groups defined by subsequent treatment paths. The reference group follows the all-zero path $D_{g+1} = \dots = D_{g+5} = 0$. Aggregate comparisons pool all nonzero paths into a single comparison group. Frequency-based comparisons divide these paths into two groups: exactly one treatment and at least two treatments between $g+1$ and $g+5$. Timing-based comparisons divide these paths according to whether the first subsequent treatment occurs between $g+1$ and $g+2$ or between $g+3$ and $g+5$. Each Wald test jointly compares the reference group with the comparison groups on both sides of the cutoff. Thus, aggregate tests have two restrictions, while frequency- and timing-based tests have four.
\end{minipage}
\end{table}

% References cited in the Supplementary Appendix.
\clearpage
\begin{spacing}{1}
    \def\bibfont{\small}
    \renewcommand{\bibsection}{\section*{References}\addcontentsline{appsec}{section}{References}}
    \putbib
\end{spacing}

\clearpage
\end{bibunit}

\end{document}